\documentclass[12pt,authoryear]{elsarticle}

\usepackage{natbib}
\usepackage{amsmath,amssymb,amsthm,setspace}
\usepackage[utf8]{inputenc}
\usepackage[T1]{fontenc}
\usepackage{lmodern}
\usepackage{enumitem}
\usepackage{graphicx}
\usepackage{booktabs}
\usepackage{longtable}
\usepackage{array}
\usepackage{caption}
\usepackage{subcaption}
\usepackage{xcolor}
\usepackage{colortbl}
\usepackage{multirow}
\usepackage{accents}
\usepackage{float}

\usepackage{url}
\usepackage[colorlinks=true,linkcolor=blue,citecolor=blue,urlcolor=blue]{hyperref}

\newtheorem{assumption}{Assumption}
\newtheorem{proposition}{Proposition}

\makeatletter
\def\fps@figure{htbp}
\makeatother

\newcommand{\di}{\text{d}}
\newcommand{\bX}{\mathbf{X}}
\newcommand{\bD}{\mathbf{D}}

\newcommand{\bF}{\mathbf{F}}
\newcommand{\bK}{\mathbf{K}}
\newcommand{\bS}{\mathbf{S}}

\newcommand{\by}{\mathbf{y}}

\newcommand{\bu}{\mathbf{u}}
\newcommand{\bU}{\mathbf{U}}
\newcommand{\bY}{\mathbf{Y}}

\newcommand{\bR}{\mathbf{R}}
\newcommand{\bV}{\mathbf{V}}

\newcommand{\bE}{\mathbf{E}}
\newcommand{\ba}{\mathbf{a}}
\newcommand{\bb}{\mathbf{b}}
\newcommand{\bh}{\mathbf{h}}
\newcommand{\bo}{\mathbf{o}}
\newcommand{\bH}{\mathbf{H}}
\newcommand{\bI}{\mathbf{I}}

\newcommand{\bA}{\mathbf{A}}
\newcommand{\bB}{\mathbf{B}}
\newcommand{\bM}{\mathbf{M}}
\newcommand{\bQ}{\mathbf{Q}}

\newcommand{\bbf}{\mathbf{f}}

\newcommand{\cov}{\text{Cov}}

\newcommand{\bC}{\mathbf{C}}

\newcommand{\bv}{\mathbf{v}}

\renewcommand{\epsilon}{\varepsilon}

\newcommand{\Em}{\mathbb E}

\newcommand{\e}{\text{e}}
\newcommand{\Var}{\text{Var}}
\newcommand{\var}{\text{Var}}

\newcommand{\vect}[1]{\boldsymbol #1}

\newcommand{\vtheta}{\vect{\theta}}

\newcommand{\vOmega}{\vect{\Omega}}
\newcommand{\vomega}{\vect{\omega}}
\newcommand{\vepsilon}{\vect{\epsilon}}
\newcommand{\vSigma}{\vect{\Sigma}}
\newcommand{\vLambda}{\vect{\Lambda}}

\newcommand{\bL}{\vect{L}}
\newcommand{\vlambda}{\vect{\lambda}}

\newcommand{\vnu}{\vect{\nu}}

\newcommand{\vrho}{\vect{\rho}}
\newcommand{\vPsi}{\vect{\Psi}}
\newcommand{\vDelta}{\vect{\Delta}}
\newcommand{\vecf}{\text{vec}}
\newcommand{\tr}{\text{tr}}
\newcommand{\diag}{\text{diag}}

\newcommand{\distn}[1]{\mathcal{#1}}

\newcommand{\bone}{\mbox{$1 \hspace{-1.0mm} {\bf 1}$}}
\newcommand{\bzero}{\mathbf{0}}

\newcommand{\blank}{{{}\cdot{}}}
\newcommand\given[1][]{\:#1\vert\:}
\renewcommand{\tilde}{\widetilde}
\renewcommand{\leq}{\leqslant}
\renewcommand{\geq}{\geqslant}
\newcommand{\citetc}[1]{{\setcitestyle{citesep={,}}\citet{#1}}}
\newcommand{\citeps}[1]{{\setcitestyle{citesep={;}}\citep{#1}}}

\makeatletter
\let\save@mathaccent\mathaccent
\newcommand*\if@single[3]{%
  \setbox0\hbox{${\mathaccent"0362{#1}}^H$}%
  \setbox2\hbox{${\mathaccent"0362{\kern0pt#1}}^H$}%
  \ifdim\ht0=\ht2 #3\else #2\fi
  }
\newcommand*\rel@kern[1]{\kern#1\dimexpr\macc@kerna}
\newcommand*\widebar[1]{\@ifnextchar^{{\wide@bar{#1}{0}}}{\wide@bar{#1}{1}}}
\newcommand*\wide@bar[2]{\if@single{#1}{\wide@bar@{#1}{#2}{1}}{\wide@bar@{#1}{#2}{2}}}
\newcommand*\wide@bar@[3]{%
  \begingroup
  \def\mathaccent##1##2{%
    \let\mathaccent\save@mathaccent
    \if#32 \let\macc@nucleus\first@char \fi
    \setbox\z@\hbox{$\macc@style{\macc@nucleus}_{}$}%
    \setbox\tw@\hbox{$\macc@style{\macc@nucleus}{}_{}$}%
    \dimen@\wd\tw@
    \advance\dimen@-\wd\z@
    \divide\dimen@ 3
    \@tempdima\wd\tw@
    \advance\@tempdima-\scriptspace
    \divide\@tempdima 10
    \advance\dimen@-\@tempdima
    \ifdim\dimen@>\z@ \dimen@0pt\fi
    \rel@kern{0.6}\kern-\dimen@
    \if#31
      \overline{\rel@kern{-0.6}\kern\dimen@\macc@nucleus\rel@kern{0.4}\kern\dimen@}%
      \advance\dimen@0.4\dimexpr\macc@kerna
      \let\final@kern#2%
      \ifdim\dimen@<\z@ \let\final@kern1\fi
      \if\final@kern1 \kern-\dimen@\fi
    \else
      \overline{\rel@kern{-0.6}\kern\dimen@#1}%
    \fi
  }%
  \macc@depth\@ne
  \let\math@bgroup\@empty \let\math@egroup\macc@set@skewchar
  \mathsurround\z@ \frozen@everymath{\mathgroup\macc@group\relax}%
  \macc@set@skewchar\relax
  \let\mathaccentV\macc@nested@a
  \if#31
    \macc@nested@a\relax111{#1}%
  \else
    \def\gobble@till@marker##1\endmarker{}%
    \futurelet\first@char\gobble@till@marker#1\endmarker
    \ifcat\noexpand\first@char A\else
      \def\first@char{}%
    \fi
    \macc@nested@a\relax111{\first@char}%
  \fi
  \endgroup
}
\makeatother
\newcommand{\anon}{1}

\journal{Journal of Economic Dynamics and Control}

\begin{document}

\begin{frontmatter}

\title{Bayesian Dynamic Factor Models for High-Dimensional Matrix-Valued Time Series\if1\anon\tnoteref{ack}\fi}

\if1\anon
\tnotetext[ack]{We would like to thank Marta Ba\'{n}bura, Domenico
Giannone, Michele Lenza, Elena Bobeica, Catalina Mart\'{i}nez Hern\'{a}ndez, Danilo Leiva-Le\'{o}n,
and Carlos Montes-Gald\'{o}n for many constructive suggestions. We are also grateful
for insightful discussions with participants at the DG-Economics Internal Seminar at
the European Central Bank, the $94^{\text{th}}$ SEA Annual Meeting, the 2025
\"{O}rebro Workshop on Macro- and Financial Econometrics, and the 2025
CFE-CMStatistics Conference. All remaining errors are our own.}
\fi

\if1\anon
\author[purdue]{Joshua C. C. Chan}
\author[jhu]{Wei Zhang}
\address[purdue]{Department of Economics, Purdue University}
\address[jhu]{Department of Economics, Johns Hopkins University}
\fi
\if0\anon
\author{}
\address{}
\fi

\begin{abstract}
We introduce a class of Bayesian dynamic factor models for matrix-valued time series, with autoregressive factor dynamics and idiosyncratic components that allow stochastic volatility, outliers, and a Kronecker-structured covariance capturing cross-row and cross-column correlation. Exploiting the matrix structure, we make these richly parameterized models tractable in high dimensions and develop an efficient Gibbs sampler for estimation. For model comparison, we propose a unified approach based on the cross-entropy importance-sampling estimator of the marginal likelihood, which under a common criterion selects the factor dimension, a vector versus matrix structure, and the idiosyncratic specification. Monte Carlo experiments confirm that the estimator reliably recovers the true model. In an application to an OECD macroeconomic panel of 190 time series, the data favor both cross-sectional correlation and stochastic volatility, and the model delivers statistically significant out-of-sample forecast gains over a static matrix factor benchmark.
\end{abstract}


\end{frontmatter}

\newpage
\onehalfspacing

\section{Introduction}\label{sec:intro}
Dynamic factor models are a workhorse for summarizing the comovement in
large macroeconomic panels, and they underpin much of modern empirical
macroeconomics and forecasting \citeps{stock2002forecasting,
forni2000generalized, bernanke2005measuring, doz2012quasi}. An increasing
share of these panels is naturally matrix-valued: a set of macroeconomic
indicators is observed for a cross-section of countries over time, so that
each period delivers an indicator-by-country matrix. Such multi-country
panels are a leading application of factor analysis, where a
small number of common factors summarize the world, regional, and
country-specific sources of macroeconomic comovement
\citeps{kose2003international, del2008dynamic}. The standard practice
is to stack the matrix into a long vector and fit a vector dynamic factor
model or a large Bayesian vector autoregression \citeps{banbura2010large,
koop2013forecasting, carriero2019large, chan2023comparing}. This vectorization discards the
matrix structure of the data---the strong dependence among indicators
within a country and among countries for a given indicator---and the number
of loadings grows with the product of the two cross-sectional dimensions,
which strains estimation as the panel widens.

A parsimonious alternative is to exploit the matrix structure directly. \citet{wang2019factor} introduce a matrix factor model in which row and column loadings enter through a bilinear form, subsequently extended with linear constraints \citep{chen2020constrained}, thresholds \citep{liu2019helping}, and time-varying loadings \citep{chen2023time}. Modeling the row and column loadings separately sharply reduces the number of free loadings, making the matrix structure attractive in high dimensions. These models, however, omit two features essential for macroeconomic applications. First, they are static: the factor dynamics are left unspecified, so they cannot produce iterated multi-step forecasts. Second, the idiosyncratic errors are assumed homoskedastic, ruling out the time-varying volatility, heavy tails, and outliers pervasive in macroeconomic data \citeps{cogley2005drifts, justiniano2008time, stock2016core, chan2023comparing}. We take up the matrix structure as a parsimony device and build a fully Bayesian dynamic factor model that addresses both.

This paper makes two methodological contributions. First, we introduce a class of Bayesian dynamic factor models for matrix-valued time series, designed around the empirical features of macroeconomic and financial data. The latent factors follow autoregressive processes, which capture the persistent comovement characteristic of common components and support recursive multi-step forecasting.\footnote{See, e.g.,
\citetc{sargent1977business, stock2012dynamic, poncela2021factor} for autoregressive factor dynamics in the vector dynamic factor literature.} We allow rich dynamics and dependence in the idiosyncratic components. Along the time dimension, we incorporate stochastic volatility \citep{carriero2016common, kastner2017efficient}, heavy-tailed errors \citep{jacquier2004bayesian}, and outlier components to absorb large, infrequent extreme observations such as those around the COVID-19 pandemic \citep{carriero2024capturing, chan2025bayesian}. Along the cross-section, we adopt a Kronecker structure built from two full covariance matrices that captures the cross-row and cross-column correlations left unexplained by the common components.

Bringing these features together yields a flexible but heavily parameterized
model, and the matrix structure is what keeps estimation tractable. Exploiting
the Kronecker form of the likelihood, we develop an efficient Gibbs sampler
that scales to the dimensions encountered in macroeconomic panels. To identify
the factors and loadings, we extend the lower-triangular scheme of
\citet{bai2015identification} to the matrix setting by anchoring both the row
and column loading matrices, and we sample the constrained loadings using the
fast hyperplane-truncation algorithm of \citet{cong2017fast}.

The second contribution is a unified approach to model comparison. The matrix factor literature has relied on eigenvalue-ratio criteria to select the number of factors. Empirical practice, however, raises broader questions of model structure: whether the data support a matrix factor structure relative to a vector dynamic factor model (VDFM),\footnote{We denote a matrix dynamic factor model by MDFM and the vector counterpart by VDFM; the ``m'' and ``v'' distinguish the matrix and vector formulations.} whether the idiosyncratic component is homoskedastic or exhibits time-varying volatility, and whether weak cross-sectional correlation is present. We address these questions through the marginal likelihood, which we compute for each model using the cross-entropy importance-sampling estimator of \citet{chan2015marginal}. This estimator draws independently from a calibrated importance density---obtained by minimizing, within a given parametric family, the Kullback--Leibler divergence to the posterior---rather than from correlated Markov chain Monte Carlo samples. Casting the choice of the factor dimension, the matrix-versus-vector structure, and the idiosyncratic specification as a Bayesian model comparison renders these distinct questions directly comparable under a common criterion \citeps{geweke1996measuring, chan2018bayesian, fong2020marginal}.

We demonstrate the effectiveness of the cross-entropy importance-sampling estimator at distinguishing among competing specifications in a series of Monte Carlo experiments. Across all three comparison dimensions---the factor dimension, the matrix-versus-vector structure, and the idiosyncratic specification---the estimator reliably assigns the highest marginal likelihood to the data-generating model. The simulations also confirm that the factor estimates are close to their true values, with accuracy improving in longer samples.

We illustrate the usefulness of the proposed MDFM in an application with an OECD macroeconomic panel of $19$ countries and $10$ indicators, a total of $190$ time series. The data decisively favor specifications with both cross-sectional correlation and stochastic volatility---features absent from the static matrix factor model---and the estimated loadings display clear grouping patterns across countries and indicators. The estimated volatility paths spike during the Asian financial crisis, the Great Recession, and the COVID-19 pandemic, underscoring the empirical relevance of the very features that distinguish the MDFM from earlier matrix factor models. 

In an out-of-sample forecasting exercise benchmarked against the static matrix factor model of \citet{wang2019factor} with its eigenvalue-ratio rank-selection rule, the proposed MDFM with the Kronecker idiosyncratic covariance and stochastic volatility delivers statistically significant reductions in pooled root mean squared forecast error of $2.5\%$ at the one-quarter horizon and $4.3\%$ at the one-year horizon \citep{marcellino2006comparison}. The advantage is broadest at the policy-relevant short and medium horizons: the MDFM improves on the benchmark in $27$ of the $29$ pooled indicator-country series at $h=1$ and $28$ of $29$ at $h=4$, with the largest gains in real-activity indicators (real GDP, household consumption, and unemployment improve by roughly $6$--$10\%$) and in the continental European, Nordic, and Korean economies whose business cycles are tightly tied to the global factor. Beyond point forecasts, the fully Bayesian treatment yields posterior predictive densities that integrate over parameter, factor, and volatility uncertainty---a dimension of forecast quality emphasized in the Bayesian macroeconometric forecasting literature \citep{carriero2015bayesian} and central to the growth-at-risk agenda \citep{adrian2019vulnerable}. The predictive variance widens visibly during the 2008--2009 financial crisis and the 2020 COVID-19 contraction.

This paper connects several literatures. It contributes to the tradition of Bayesian dynamic factor and large Bayesian time-series models for macroeconomic and financial datasets \citeps{aguilar2000bayesian, west2003bayesian, lopes2004bayesian, del2008dynamic, banbura2010large, huber2020multi, chan2023large}, to which it adds a
matrix-valued formulation with flexible idiosyncratic dynamics. It is also
related to the growing statistical literature on matrix and tensor factor
models \citeps{wang2019factor, chen2022factor, yu2022projected, he2024matrix},
which is predominantly frequentist and focused on
estimation and asymptotic theory rather than the Bayesian inference,
forecasting, and model comparison that we emphasize. 

Two recent works are closest to ours. \citet{yu2024dynamic} and \citet{qin2025bayesian} both propose dynamic matrix factor models with matrix autoregressive factor evolution; \citet{yu2024dynamic} take a frequentist approach with iterative eigendecomposition and eigenvalue-ratio factor selection, while \citet{qin2025bayesian} adopt a Bayesian approach with a Kronecker-structured idiosyncratic covariance. Our framework differs in two key respects that mirror our two contributions. First, in modeling: neither accommodates the time-varying volatility, heavy tails, and outlier components that we introduce and that are empirically important for macroeconomic data. Second, in model comparison: we develop a unified approach that compares a wide range of competing specifications, which neither paper provides. 

The rest of this paper is organized as follows. Section~\ref{sec:model}
specifies the model, identification, priors, and Bayesian estimation.
Section~\ref{sec:selection} introduces the marginal likelihood estimator.
Section~\ref{sec:monte} presents results from the Monte Carlo experiments.
Section~\ref{sec:application} reports the empirical application to an OECD
macroeconomic panel and the out-of-sample forecasting exercise.
Section~\ref{sec:conclusion} concludes.

\section{A Matrix Dynamic Factor Model with Stochastic Volatility}\label{sec:model}
Building on the framework of \citet{wang2019factor}, we introduce a dynamic factor
model for matrix-valued time series. Following the spirit of the approximate factor
model of \citet{chamberlain1983arbitrage}, we allow cross-row and cross-column
correlations in the idiosyncratic components. We then incorporate time-varying
volatility and outlier adjustments, present identification restrictions, and outline
the Bayesian priors and posterior simulation. We close the section by sketching a
heteroskedastic extension.

\subsection{The Model}\label{sec:submodel}
Consider an $n\times k$ data matrix $\bY_t$ observed at time $t$. To fix ideas,
suppose $\bY_t$ is a panel of macroeconomic indicators where rows index $n$
countries and columns index $k$ variables. The matrix dynamic factor model is
\begin{align}
\bY_t &= \bA\bF_t\bB' + \bE_t,
\quad\vecf(\bE_t)\sim\distn{N}\!\left(\bzero,\,\omega_t\vSigma_c\otimes\vSigma_r\right),\label{eq:mdfm}\\
\vecf(\bF_t) &= \bH_{\vrho_1}\vecf(\bF_{t-1}) + \cdots + \bH_{\vrho_q}\vecf(\bF_{t-q}) + \bu_t,
\quad\bu_t\sim\distn{N}(\bzero,\vLambda),\label{eq:errorlags}
\end{align}
where $\bA$ ($n\times p_1$) and $\bB$ ($k\times p_2$) are the row and column loading
matrices, $\bF_t$ is a $p_1\times p_2$ latent factor matrix, $\vSigma_r$ ($n\times n$)
and $\vSigma_c$ ($k\times k$) are the row- and column-wise idiosyncratic covariance
matrices, and the $p_1p_2\times p_1p_2$ factor innovation covariance is
$\vLambda = \diag(\lambda_{1,1}^2,\ldots,\lambda_{p_1p_2,p_1p_2}^2)$.
$\bH_{\vrho_l}$, $l=1,\ldots,q$, are autoregressive coefficient matrices.

The bilinear form $\bA\bF_t\bB'$ captures cross-country and cross-variable
dependencies. The $i$-th row of $\bY_t$ is
$\bY_{i,\cdot,t} = \bA_{i,\cdot}\bF_t\bB' + \bE_{i,\cdot,t}$: the common component
$\bA_{i,\cdot}\bF_t\bB'$ is a linear combination of the rows of $\bF_t\bB'$ weighted
by the $i$-th row of $\bA$. Similarly, the $j$-th column has common component
formed by linear combinations of columns of $\bA\bF_t$ with weights from the $j$-th
column of $\bB'$. Hence rows of $\bA$ describe how countries load on latent factors,
columns of $\bB'$ describe how indicators respond to latent factors, and rows
(columns) of $\bF_t$ are interpreted as latent factors driving the country
(indicator) dimension.

\subsubsection{Important features of the model}
Relative to the static matrix factor model of \citet{wang2019factor}, our model
adds three features tailored to macroeconomic and financial data: autoregressive
factor dynamics, a Kronecker-structured idiosyncratic covariance that captures
cross-row and cross-column correlation, and idiosyncratic errors that accommodate
stochastic volatility and outliers.

\paragraph{Factor dynamics}
Modeling factor dynamics is particularly important in macroeconomic applications,
where latent state variables typically follow stable laws of motion that embody
business-cycle persistence. The autoregressive specification in
\eqref{eq:errorlags} captures persistent comovement and provides a natural
framework for recursive forecasting. Posterior simulation from the resulting dynamic system delivers predictive densities as standard byproducts of estimation; we exploit this in the forecasting exercise of Section~\ref{sec:application}, which compares the predictive performance of our specification against the static matrix factor benchmark.

\paragraph{Kronecker idiosyncratic covariance}
The Kronecker covariance of $\vecf(\bE_t)$ separates row-wise and column-wise
residual correlations. Specifically, for any row $i$, the conditional covariance
is $\cov(\bY_{i,\cdot,t}'\given\bA,\bF_t,\bB) = \omega_t\sigma_{r,i,i}^2\vSigma_c$,
and for any column $j$, $\cov(\bY_{\cdot,j,t}\given\bA,\bF_t,\bB) =
\omega_t\sigma_{c,j,j}^2\vSigma_r$. $\vSigma_r$ and $\vSigma_c$ thus capture
row-wise and column-wise residual covariances not explained by the common
component, making the model an approximate factor model in the sense of
\citet{chamberlain1983arbitrage} \citep[see also][]{stock2005implications,
	barigozzi2024dynamic,giglio2025test}.

The Kronecker structure also greatly reduces the number of free parameters relative to an unrestricted idiosyncratic covariance.\footnote{An unrestricted $nk\times nk$ covariance matrix contains $nk(nk+1)/2$ parameters. For the OECD panel considered below ($n=19$, $k=10$, $T=115$), this is $18{,}145$ parameters---more than 150 times the sample size---whereas the Kronecker specification $\vSigma_u=\vSigma_c\otimes\vSigma_r$ requires only $n(n+1)/2+k(k+1)/2-1=244$, while still allowing both within-row and within-column residual dependence.}  This dimensionality reduction is critical for the tractability of large Bayesian multivariate models and is analogous to the use of Kronecker-structured priors in large Bayesian VARs \citep{chan2018large}. The restriction is stronger than the approximate factor framework of \citet{chamberlain1983arbitrage}, which only requires bounded eigenvalues of the idiosyncratic covariance. In return, it yields conjugate matrix-normal--inverse-Wishart posteriors for $(\bA,\vSigma_r)$ and $(\bB,\vSigma_c)$, enabling joint rather than element-wise sampling of loadings and covariance matrices (Section~\ref{sec:bayesian}).

\paragraph{Stochastic volatility and outliers} For homoskedasticity, we can set $\omega_t = 1$ for all $t$. It is also well known that allowing for time-varying volatility is essential in modeling macroeconomic and financial data, especially after the COVID-19 pandemic.\footnote{For evidence in vector autoregressions, see \citet{cross2016forecasting}, \citet{chan2018bayesian}, and \citet{chan2023comparing}; for factor models, \citet{aguilar2000bayesian,chib2006analysis,kastner2017efficient,li2022leverage}.} We implement three popular specifications under the conditionally Gaussian framework in \eqref{eq:errorlags} through the scalar latent state $\omega_t$.


\textbf{Specification 1. Common stochastic volatility.}
Following \citet{carriero2015bayesian,carriero2016common}, let $\omega_t = \e^{h_t}$ with log-volatility $h_t$ following a stationary AR(1) process with zero mean:
\begin{equation}\label{eq:sv}
h_t = \phi h_{t-1} + u_t^h,\qquad u_t^h\sim\distn{N}(0,\sigma_h^2),
\end{equation}
for $t = 2,\ldots,T$, with $|\phi|<1$ and $h_1\sim\distn{N}(0,\sigma_h^2/(1-\phi^2))$.

\textbf{Specification 2. Explicit outliers.}
Following \citet{stock2016core}, $\omega_t = o_t^2$ where $o_t$ follows a mixture distribution distinguishing regular observations ($o_t = 1$) from outliers ($o_t \geq 2$). The outlier probability $p_o$ has a beta prior. This specification has become standard for handling the extreme observations of macroeconomic series during the onset of COVID-19 pandemic.

\textbf{Specification 3. Fat-tailed innovations.}
Following \citet{jacquier2004bayesian}, $\omega_t = q_t^2$ with $q_t^2\sim
\distn{IG}(l/2,l/2)$. The marginal distribution of $\vecf(\bE_t)$ is then a multivariate $t$ with zero mean, scale matrix $\vSigma_c\otimes\vSigma_r$, and degrees of freedom $l$.

\subsubsection{Relation to vectorized factor models}
A natural benchmark for the model in \eqref{eq:mdfm}--\eqref{eq:errorlags} is the
standard VDFM applied to the vectorized panel:
\begin{equation}\label{eq:dfm}
\begin{aligned}
\by_t &= \bM\bbf_t + \vepsilon_t,\\
\bbf_t &= \bH_{\rho}\bbf_{t-1} + \vnu_t,
\end{aligned}
\end{equation}
where $\by_t$ is the $nk\times 1$ vectorized panel, $\bM$ is an $nk\times p$
loading matrix, $\bbf_t$ is a $p\times 1$ vector of factors, and $\bH_\rho$ is the
factor AR coefficient. We set $p = p_1 \times p_2$ for direct comparison with the
MDFM.

The MDFM is a restricted version of \eqref{eq:dfm} in which the factor loadings
exhibit a Kronecker product structure:
\begin{equation}\label{eq:specialdfm}
	\vecf(\bY_t) = (\bB\otimes\bA)\vecf(\bF_t) + \vecf(\bE_t).
\end{equation}
This restriction reduces the loading parameter count from $nk \times p_1 p_2$ in the unrestricted VDFM to $np_1 + kp_2$ in the MDFM, with corresponding improvements in estimation accuracy. Under principal components analysis applied to the vectorized model, \citet{bai2003inferential} establishes that each row of $\bB \otimes \bA$ converges at rate $\min\{\sqrt{T}, nk\}^{-1}$, which is dominated by $T^{-1/2}$ in the empirically relevant regime where $T \ll nk$. The matrix structure delivers sharper rates by exploiting the fact that each column of $\bY_t$ is informative about $\bA$ and each row about $\bB$. \citet[Theorem 1]{chen2023statistical} show via $\alpha$-PCA that $\bA$ is recovered at rate $\min\{n, Tk\}^{-1/2}$ and $\bB$ at rate $\min\{k, Tn\}^{-1/2}$, so the effective sample size for $\bA$ scales with $Tk$ rather than $T$ alone, and symmetrically for $\bB$. This efficiency gain is valuable in the macro applications we target, where $T$ is short relative to $nk$.

Whether the data actually support the Kronecker structure is, of course, an empirical
question. \citet{he2024matrix} propose a family of randomized specification
tests for the matrix structure; we instead adopt a Bayesian model comparison
approach using marginal likelihoods (Section~\ref{sec:selection}), which has the
advantage of treating the VDFM-versus-MDFM choice on the same footing as the
choice of factor-matrix dimension and idiosyncratic covariance structure. The
Monte Carlo experiments in Section~\ref{sec:mc_structure} confirm that the
marginal likelihood estimator reliably distinguishes the two specifications in
either direction.

\subsection{Identification}\label{sec:identification}
The model in \eqref{eq:mdfm}--\eqref{eq:errorlags} is not identified without
further restrictions. The matrix factor literature commonly resolves this by
estimating only the column spaces of the loading matrices, which are uniquely
identifiable without further restrictions
\citep{wang2019factor,chen2020constrained,chen2021autoregressive,yu2022projected}.

The Bayesian dynamic factor model literature has adopted a range of
identification strategies that pin down loadings and factors uniquely. The
most common is the positive lower-triangular (PLT) restriction introduced by
\citet{geweke1996measuring} and developed further for dynamic factor and
stochastic volatility models by
\citet{aguilar2000bayesian},\cite{west2003bayesian} and \cite{lopes2004bayesian}; the formal
identification result for dynamic factor models is established in
\citet{bai2015identification}, and \citet{bai2013principal} provide the corresponding identification analysis for principal components estimators, including a triangular-block restriction (PC2) that is the frequentist analog of the lower-triangular Bayesian scheme used here. Closely related anchoring strategies that
combine sign and zero restrictions on selected loadings to identify regional
or group factors include \citet{del2008dynamic} and the related international
business cycle literature. Alternative schemes that avoid ordering altogether include
the order-invariant approach of \citet{chan2018invariant}, which restricts
the factor space rather than the ordering of individual variables; and the generalized
lower-triangular (GLT) representation of
\citet{fruhwirth2024sparse,fruhwirth2024gltcounts}, which treats the
leading-variable allocation as a parameter to be inferred. See also
\citet{kaufmann2019bayesian,leung2016order} for related order-invariant
proposals.

In this paper we adopt the lower-triangular identification of
\citet{bai2015identification}, extending it to the matrix factor setting by
anchoring both row and column loading matrices. The scheme pins down not only
the column spaces but also the levels and signs of the loadings and factors,
delivering economic interpretability for the rows and columns of $\bF_t$.
Proofs of the identification results below are given in \ref{app:proofs} of the
Online Supplemental Materials.

\begin{assumption}\label{assumption1}
	\rm Factors and idiosyncratic errors are independent of each other.
\end{assumption}
\begin{assumption}\label{assumption2}
	\rm Factor series are independent of each other; $\bH_{\vrho_l}$ is diagonal for
	$l = 1,\ldots,q$; and $\cov(\bu_t) = \bI_{p_1p_2}$.
\end{assumption}
\begin{assumption}\label{assumption3}
	\rm One of $\bA$ and $\bB$ is a lower-triangular matrix with ones on the diagonal,
	while the other is a lower-triangular matrix with strictly positive diagonal
	elements.
\end{assumption}
\begin{proposition}\label{prop1}
	Under Assumptions~\ref{assumption1}--\ref{assumption3}, the dynamic factor matrix
	$\bF_t$ and the loading matrices $\bA$ and $\bB$ in
	\eqref{eq:mdfm}--\eqref{eq:errorlags} are uniquely identified.
\end{proposition}

A second identification issue arises from the Kronecker structure of the
idiosyncratic covariance: $\vSigma_c$ and $\vSigma_r$ are identified only up
to a scalar, since for any $m \in \mathbb{R}\setminus\{0\}$,
$\vSigma_c \otimes \vSigma_r = (m\vSigma_c) \otimes (m^{-1}\vSigma_r)$.
We fix the scale by normalizing the $(1,1)$ element of $\vSigma_c$ to one.

An alternative identification scheme places the unit-diagonal restriction on
both $\bA$ and $\bB$ symmetrically and instead allows the factor innovation
covariance $\cov(\bu_t)$ to be a positive diagonal matrix rather than the
identity. Formally:

\begin{assumption}\label{assumption4}
	\rm Factor series are independent of each other; $\bH_{\vrho_l}$ is a
	diagonal matrix for $l = 1, \ldots, q$; and $\cov(\bu_t)$ is a positive
	definite diagonal matrix.
\end{assumption}
\begin{assumption}\label{assumption5}
	\rm The factor loading matrices $\bA$ and $\bB$ are both lower-triangular
	matrices with ones on the diagonal.
\end{assumption}
\begin{proposition}\label{prop2}
	Under Assumptions~\ref{assumption1}, \ref{assumption4}, and \ref{assumption5},
	the dynamic factor matrix $\bF_t$ and the loading matrices $\bA$ and $\bB$
	in \eqref{eq:mdfm}--\eqref{eq:errorlags} are uniquely identified.
\end{proposition}

The lower triangular structure has an important interpretive implication: the first row of $\bY_t$ responds
to the first row factor only, the second row of $\bY_t$ responds to the
first two row factors only, and so on; similarly for the columns. This implies that shocks to the first row factor are shocks to
the first row of the data matrix (or the first country), shocks to the second row factor are shocks to the
second row of the data matrix orthogonal to the first, and so forth. The leading
variables thus anchor the substantive content of each factor, and the
ordering of rows and columns of $\bY_t$ determines the labeling of the
rows and columns of $\bF_t$.

Whether this ordering is consequential depends on the use of the model.
Identification matters for the substantive labeling of factors and
loadings but not for forecasting, which depends on the
loadings and factors through the product $\bA \bF_t \bB^\top$ and
is invariant to rotations of the factor representation. Practitioners
concerned about ordering dependence can verify the robustness of their
forecasting results via permutation experiments across alternative
orderings; we report such an experiment for our application in Section~\ref{sec:permute}.

Three limitations of our specification are worth noting. First,
Assumption~\ref{assumption4} treats the elements of $\bF_t$ as having AR($q$) dynamics. Allowing for cross-factor dependencies via a
vector or matrix autoregression for $\bF_t$ would generalize the dynamics, but combined with the time-varying volatility, outlier adjustments, and Kronecker idiosyncratic covariance we adopt, it would complicate identification and posterior computation. Second, the common stochastic volatility specification in \eqref{eq:sv} imposes that all series share a single time-varying scale; allowing factor-specific or block-specific volatilities is a natural extension, which we briefly discuss in Section \ref{sec:hetsv}. Third, our triangular identification
gains interpretability at the cost of requiring the practitioner to choose an ordering of rows and columns. Adopting order-invariant identification strategies (see, e.g., \citet{chan2018invariant,kaufmann2019bayesian,leung2016order,fruhwirth2024gltcounts,fruhwirth2024sparse}) is a
promising direction for future research.


\subsection{Priors}\label{sec:priors}
We use natural conjugate priors for the transposed loadings $\bA'$ and $\bB'$ jointly with inverse-Wishart priors for $\vSigma_r$ and $\vSigma_c$:
\begin{equation}\label{eq:priors}
	\begin{aligned}
		\vSigma_r\sim\distn{IW}(\nu_r,\bS_r),\quad
		(\vecf(\bA')\given\vSigma_r)&\sim\distn{N}\left(\vecf(\bA_0'),\,\vSigma_r\otimes\bV_{\bA'}\right),\\
		\vSigma_c\sim\distn{IW}(\nu_c,\bS_c),\quad
		(\vecf(\bB')\given\vSigma_c)&\sim\distn{N}\left(\vecf(\bB_0'),\,\vSigma_c\otimes\bV_{\bB'}\right),
	\end{aligned}
\end{equation}
where $\bV_{\bA'}$ and $\bV_{\bB'}$ are $p_1\times p_1$ and $p_2\times p_2$ scale matrices, respectively. This conjugate structure has both a substantive and a computational advantage. Substantively, the prior implies $\cov(\bA_{i,\cdot}, \bA_{j,\cdot}\given\vSigma_r) = \vSigma_{r,i,j}\bV_{\bA'}$, so that the prior correlation between the $i$-th and $j$-th rows of $\bA$ along any direction in the factor space equals $\vSigma_{r,i,j}/\sqrt{\vSigma_{r,i,i}\vSigma_{r,j,j}}$. In other words, countries with more correlated idiosyncratic components are a priori expected to have similar factor-loading patterns; an analogous interpretation holds for the columns of $\bB$ through $\vSigma_c$. Computationally, the conjugate prior in \eqref{eq:priors} yields a matrix-normal-inverse-Wishart conditional posterior for $(\bA',\vSigma_r)$ and $(\bB',\vSigma_c)$, which we exploit for joint sampling of the loading matrices and the row/column covariances; see Section~\ref{sec:bayesian} below. In practice, $\bS_r$ and $\bS_c$ are taken to be diagonal, reflecting the prior belief that idiosyncratic risks are uncorrelated, while the data are allowed to inform any off-diagonal correlations.

For parsimony, we consider AR(1) dynamics for each factor entry; the extension to AR$(q)$ with $q > 1$ is straightforward, replacing the truncation $|\rho| < 1$ with truncation to the stationarity region $\mathcal{S}_q = \{\vrho \in \mathbb{R}^q : 1 - \sum_{s=1}^q \rho_s z^s \neq 0 \text{ for all } |z| \leq 1\}$ and the scalar variance in the prior for initial values below with the solution to the discrete Lyapunov equation. Under AR(1), the diagonal elements of $\bH_{\vrho}$ have truncated normal priors,
\begin{equation*}
	\rho_{j,k} \sim \distn{N}(\rho_{j,k,0}, V_{\rho_{j,k}})\, \bone(|\rho_{j,k}| < 1), \qquad j = 1, \ldots, p_1,\ k = 1, \ldots, p_2,
\end{equation*}
where $\bone(\cdot)$ denotes the indicator function. The factor innovation variances are $\lambda_{j,k}^2 \sim \distn{IG}(\nu_{\lambda_{j,k}}, S_{\lambda_{j,k}})$, and the initial value $f_{j,k,1}$ of each factor process is assigned the stationary prior
\begin{equation*}
	f_{j,k,1} \mid \rho_{j,k}, \lambda_{j,k}^2 \sim \distn{N}\!\left(0,\ \frac{\lambda_{j,k}^2}{1 - \rho_{j,k}^2}\right),
\end{equation*}
which is the unconditional variance of the stationary AR(1) process.

\subsection{Bayesian Estimation}\label{sec:bayesian}
Posterior draws are obtained by sequentially sampling from
(1) $p(\bA',\vSigma_r\given\bY,\bB,\bF,\vSigma_c)$;
(2) $p(\bB',\vSigma_c\given\bY,\bA,\bF,\vSigma_r)$;
(3) $p(\vecf(\bF_t)\given\bY_t,\bA,\bB,\vSigma_r,\vSigma_c,\vomega^2,\vrho)$ for $t=1,\ldots,T$; (4) $p(\lambda_{j,k}^2\given\bbf_{j,k},\rho_{j,k})$; (5) $p(\rho_{j,k,l}\given\bbf_{j,k},\lambda_{j,k}^2)$; and (6) $p(\omega_t\given\bA,\bF_t,\bB,\vSigma_c,\vSigma_r)$ for $t=1,\ldots,T$.

The conjugate prior in \eqref{eq:priors} delivers matrix-normal-inverse-Wishart conditional posteriors for $(\bA',\vSigma_r)$ and $(\bB',\vSigma_c)$, which permit joint sampling of the loading matrices and the corresponding row/column covariances rather than element-wise updates. Specifically,
\begin{equation}\label{eq:ASigreq}
	(\bA',\vSigma_r\given\blank)\sim\distn{NIW}\!\left(\widehat{\bA}',\,\bK_{\bA'}^{-1},\,\widehat{\nu}_r,\,\widehat{\bS}_r\right), \quad (\bB',\vSigma_c\given\blank)\sim\distn{NIW}\!\left(\widehat{\bB}',\,\bK_{\bB'}^{-1},\,\widehat{\nu}_c,\,\widehat{\bS}_c\right),
\end{equation}
with closed-form expressions for the moments given in \ref{app: estimation} of the Online Appendix. The triangular identification restrictions on $\bA$ and $\bB$ amount to linear constraints of the form $\bM_{\bA'}\vecf(\bA')=\ba_0$ and $\bM_{\bB'}\vecf(\bB')=\bb_0$ on the conditional posterior; we sample efficiently under these constraints using the algorithm of \citet{cong2017fast}, which generates unrestricted draws from the matrix-normal distribution and then projects them onto the constrained subspace. Sampling $\vSigma_c$ subject to the restriction $\sigma_{c,1,1}=1$ uses the algorithm of \citet{nobile2000comment}. Sampling the dynamic factors $\bF_t$ uses the precision sampler proposed by \cite{chan2009efficient}, sampling the autoregressive coefficients $\rho_{j,k}$ uses a Metropolis-Hastings sampler, sampling the stochastic volatility using the acceptance-rejection Metropolis-Hastings sampler proposed by \cite{chan2017stochastic}, sampling fat-tailed innovation states with the Inverse-Gamma posterior, and sampling the outlier components using a discrete grid sampling algorithm. Full algorithmic details, including closed-form expressions for $\widehat{\bA}'$, $\widehat{\bB}'$, $\bK_{\bA'}$, $\bK_{\bB'}$, $\widehat{\bS}_r$, $\widehat{\bS}_c$, the constrained sampling steps, and the full Markov chain are given in \ref{app: estimation} of the Online Supplemental Materials.

\subsection{An Extension: Heteroskedastic Time-Varying Volatility}\label{sec:hetsv}
A more flexible alternative is to give each idiosyncratic error its own stochastic volatility, in the spirit of \citet{cogley2005drifts} for VARs, replacing the Kronecker-structured covariance of $\vecf(\bE_t)$ with a diagonal one,
\begin{equation}\label{eq:errornew}
	\vecf(\bE_t)\sim\distn{N}\!\left(\bzero,\bD_t\right),\qquad
	\bD_t = \diag(\e^{h_{1,1,t}},\e^{h_{2,1,t}},\ldots,\e^{h_{n,k,t}}),
\end{equation}
where each log-volatility $h_{i,j,t}$ follows an independent stationary AR(1) as in \eqref{eq:sv}. This accommodates substantially richer volatility patterns at the cost of $nk$ latent processes in place of the single common one. The trade-off is twofold: being diagonal, \eqref{eq:errornew} removes all cross-row and cross-column correlation in the idiosyncratic component, which can be empirically restrictive, and it precludes the conjugate MNIW sampling of $(\bA',\vSigma_r)$ and $(\bB',\vSigma_c)$ that drives our baseline's efficiency.

\section{Model Comparison via Marginal Likelihood}\label{sec:selection}
With several models plausible, one main issue facing practitioners is the lack of tools to compare these models. We use marginal likelihoods to address three comparison questions jointly: the factor dimensions $(p_1, p_2)$, vector versus matrix dynamic factor model, and the idiosyncratic covariance structure (diagonal versus Kronecker, with or without time-varying volatility). The marginal likelihood is the natural Bayesian criterion here: it integrates over the parameter space rather than relying on point estimates, penalizes over-parameterization automatically, and delivers Bayes factors that quantify the relative evidence between any two models.

The tools most common in the factor-model literature answer a narrower question. Information criteria \citep{bai2002determining} and the eigenvalue-ratio criteria standard in the matrix factor literature \citep{wang2019factor, chen2020constrained, he2024matrix} select the factor dimension within a fixed specification, operating on the unconditional second moments of the data. They do not extend to the non-nested comparisons in this paper--vector versus matrix, exact versus approximate idiosyncratic covariance, homoskedastic versus stochastic-volatility errors--which alter the likelihood through dynamic and covariance structure rather than eigenvalue rank. The marginal likelihood places all of these on one scale.\footnote{The closest Bayesian treatment is \citet{qin2025bayesian}, who select factor dimensions and lag orders via fractional Bayes factors \citep{ohagan1995fractional} under diffuse priors; we instead use proper priors and compute the full-sample marginal likelihood directly. \citet{he2023one} develop a test for the one-way (vector) versus two-way (matrix) choice; our framework handles that choice alongside the covariance and volatility structure on the same scale.}

The obstacle to wider use of marginal likelihood is computational: the integral $p(\by) = \int p(\by\given\vtheta)\,p(\vtheta)\,\di\vtheta$ is high-dimensional and hard to evaluate in dynamic models. We compute it with the cross-entropy importance-sampling estimator of \citet{chan2015marginal}. With $p(\vtheta)$ the prior and $g(\vtheta)$ an importance density,
\begin{equation}\label{eq:ceestimate}
	\hat{p}_{IS}(\by) = \frac{1}{N}\sum_{n=1}^N
	\frac{p(\by\given\vtheta_n)\,p(\vtheta_n)}{g(\vtheta_n)},
	\qquad \vtheta_1,\ldots,\vtheta_N\stackrel{\text{iid}}{\sim}g(\vtheta).
\end{equation}
The estimator is unbiased and consistent whenever $g$ dominates $p(\by\given\blank)p(\blank)$, and its efficiency hinges on how closely $g$ approximates the posterior. If we use the posterior density as the importance density, then the associated importance sampling estimator has zero variance. However, in practice, the posterior density cannot be used as its normalizing constant is exactly the unknown quantity we aim to estimate. The cross-entropy method makes it convenient to find an importance density within a parametric family that is ``closest'' to the posterior density by minimizing the Kullback-Leibler divergence between them. \citet{chan2015marginal} show the minimizing hyperparameters in the parametric family coincide with the maximum likelihood estimator obtained by treating the posterior draws as data for these hyperparameters. Calibrating $g$ thus reduces to the following step: we run the Gibbs sampler of Section~\ref{sec:bayesian}, and then obtain optimal hyperparameters given the posterior draws by obtaining the maximum likelihood estimates. Because the importance draws are i.i.d. from $g$, numerical standard errors of the marginal-likelihood estimator follows directly; and because the calibrated $g$ closely approximates the posterior, that variance is small. We further reduce variance by integrating out the latent factors with the Kalman filter before applying~\eqref{eq:ceestimate}. The importance-sampling family for each parameter block, the integrated-likelihood expression, and the Kalman-filter recursions are given in \ref{app: ml} of the Online Supplemental Materials.

\section{Monte Carlo Studies}\label{sec:monte}
In this section, we first assess the accuracy of the factor estimates by comparing them to their true values across datasets of varying sizes. We then evaluate whether the marginal likelihood estimator can correctly identify the true model structure -- the dimension of the factor matrix, the vector-versus-matrix
specification, and the structure of idiosyncratic covariance.

\subsection{Performance of factor estimators}\label{sec:accuracy}

The data are generated from \eqref{eq:mdfm}--\eqref{eq:errorlags} with $q=1$. Free parameters in $\bA$ and $\bB$ are drawn from $\distn{U}(0,1)$, $\rho_{j,k}$ from $\distn{U}(0.8,0.9)$ for $j=1,\ldots,p_1$, $k=1,\ldots,p_2$, $\vSigma_c = 0.3\bI_k$,
$\vSigma_r = 0.5\bI_n$, and $\lambda_{j,k}^2 = 1$. We consider sample sizes $(n,k)\in\{(10,10),(20,15),(30,20)\}$, observation lengths
$T\in\{200,500,1000\}$, and factor matrix dimensions $(p_1,p_2)\in\{(3,2),(5,5)\}$.\footnote{For $(p_1,p_2)=(3,2)$, we use a Gibbs chain of 10{,}000 iterations after 5{,}000 burn-in. For $(p_1,p_2)=(5,5)$, we extend the
chain to 20{,}000 iterations after 10{,}000 burn-in. With larger factor matrices, proper initialization is important; estimates from a VDFM (1{,}000 posterior draws after 1{,}000 burn-in) work well as starting values. Geweke statistics confirm chain convergence in all designs.} We use the posterior mean as the point estimate and project the true factors onto the estimates, reporting adjusted $R^2$ values for each factor series.

\begin{figure}[H]
    \centering
    \includegraphics[width=0.9\linewidth]{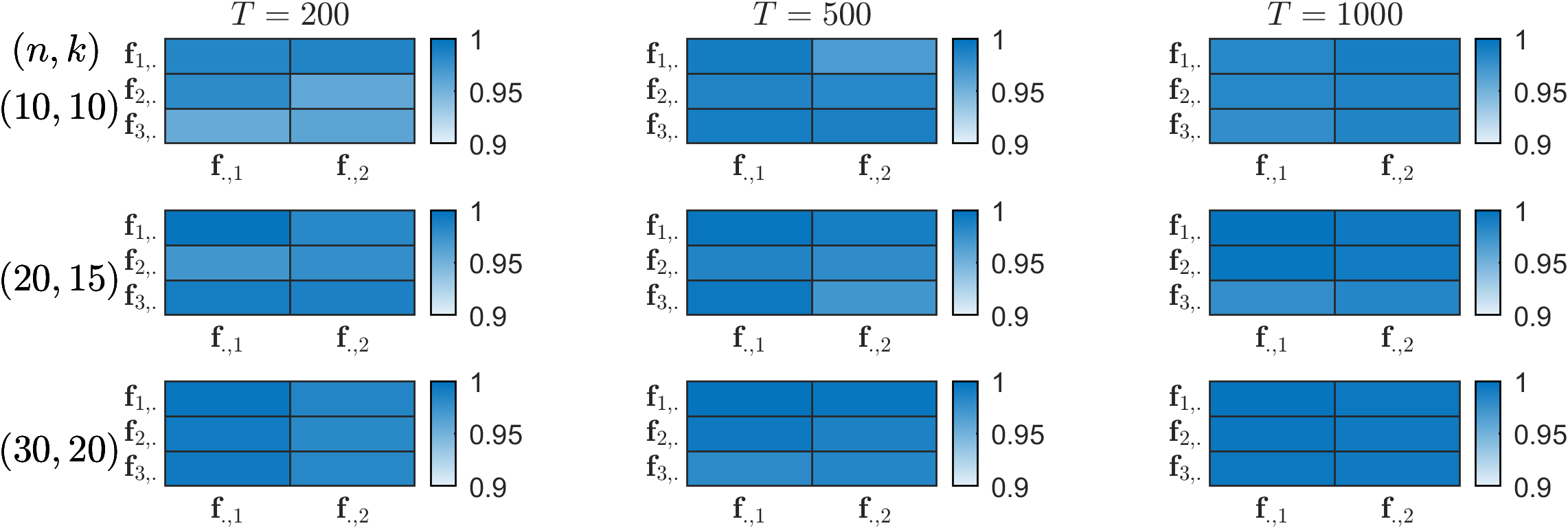}
    \caption{Adjusted $R^2$ from regressing the true factors on the estimates:
    $p_1 = 3$, $p_2 = 2$.}
    \label{fig:adr2_32}
\end{figure}

\begin{figure}[H]
    \centering
    \includegraphics[width=0.9\linewidth]{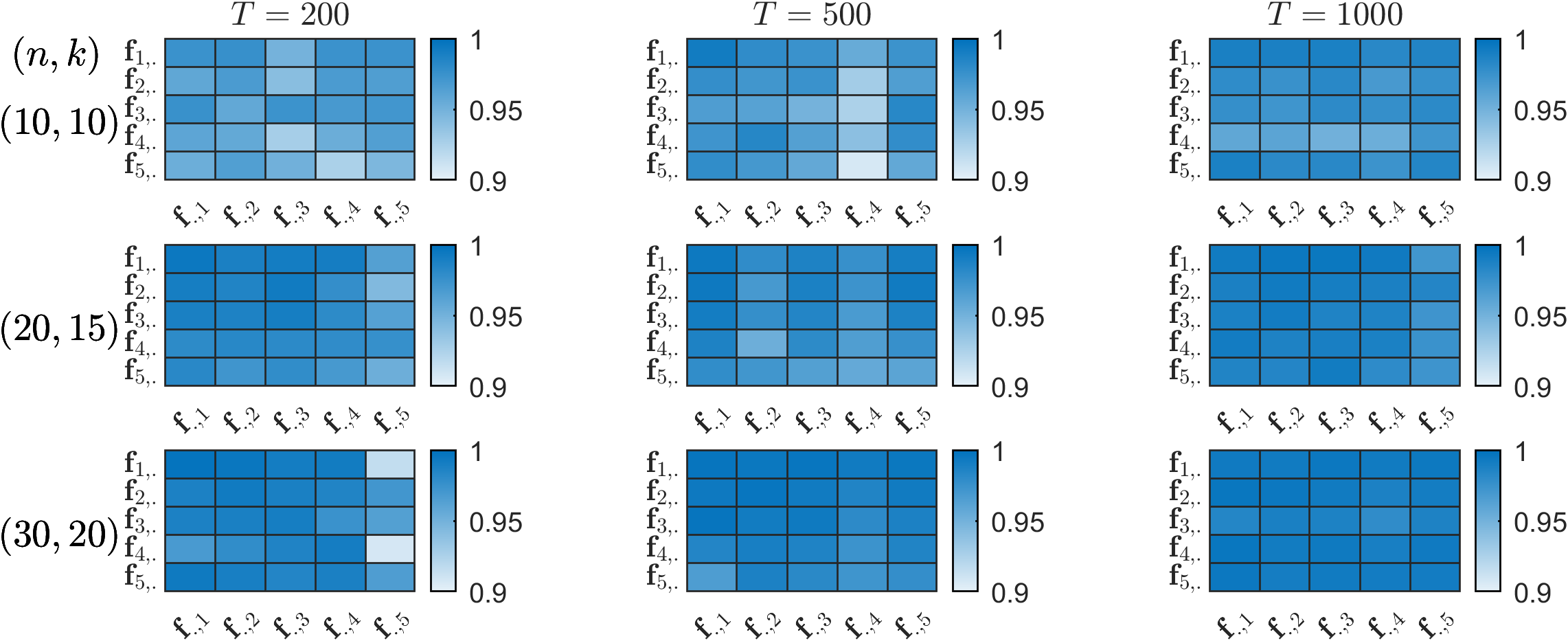}
    \caption{Adjusted $R^2$ from regressing the true factors on the estimates:
    $p_1 = 5$, $p_2 = 5$.}
    \label{fig:adr2_55}
\end{figure}

Figures~\ref{fig:adr2_32}--\ref{fig:adr2_55} display the adjusted $R^2$ values.
Each row corresponds to a different sample size and each column to a different
$T$. The minimum of the color axis is set to 0.9 (the smallest adjusted $R^2$ we
obtain across all designs is 0.91); detailed numerical values are reported in
\ref{app:simulation} of the Online Supplemental Materials. Two patterns are visible. First, factor
estimates closely track their true values across all designs. Second, accuracy
improves with $T$ and with $(n,k)$, in line with the inferential theory for
vectorized factor models \citep{bai2003inferential} and static matrix factor models
\citep{chen2023statistical}. Smaller factor matrix dimensions
$(p_1,p_2)=(3,2)$ deliver more accurate estimates than $(5,5)$, as expected.

\subsection{Identifying the dimension of the factor matrix}\label{sec:mc_dimension}

To evaluate the marginal likelihood estimator's ability to identify the correct factor matrix dimension, we estimate log marginal likelihoods over a grid of $(p_1,p_2)$ values using four datasets from Section~\ref{sec:accuracy}:

\begin{itemize}[font=\itshape, itemsep=1pt]
\setlength{\itemindent}{2.4em}
    \item [Dataset 1:]$n = 10$, $k = 10$, $T = 500$; true dimension: $p_1 = 3$, $p_2 = 2$.
    \item [Dataset 2:]$n = 20$, $k = 15$, $T = 500$; true dimension: $p_1 = 3$, $p_2 = 2$.
    \item [Dataset 3:]$n = 10$, $k = 10$, $T = 500$; true dimension: $p_1 = 5$, $p_2 = 5$.
    \item [Dataset 4:]$n = 20$, $k = 15$, $T = 500$; true dimension: $p_1 = 5$, $p_2 = 5$.
\end{itemize}

For datasets with $(p_1,p_2)=(3,2)$, we estimate models with $p_1$ and $p_2$
ranging from 1 to 5; for $(5,5)$, we estimate models with $p_1$ and $p_2$ ranging
from 3 to 7.

\begin{figure}[H]
    \centering
    \begin{minipage}{0.9\textwidth}
        \centering
        \includegraphics[width=0.72\linewidth]{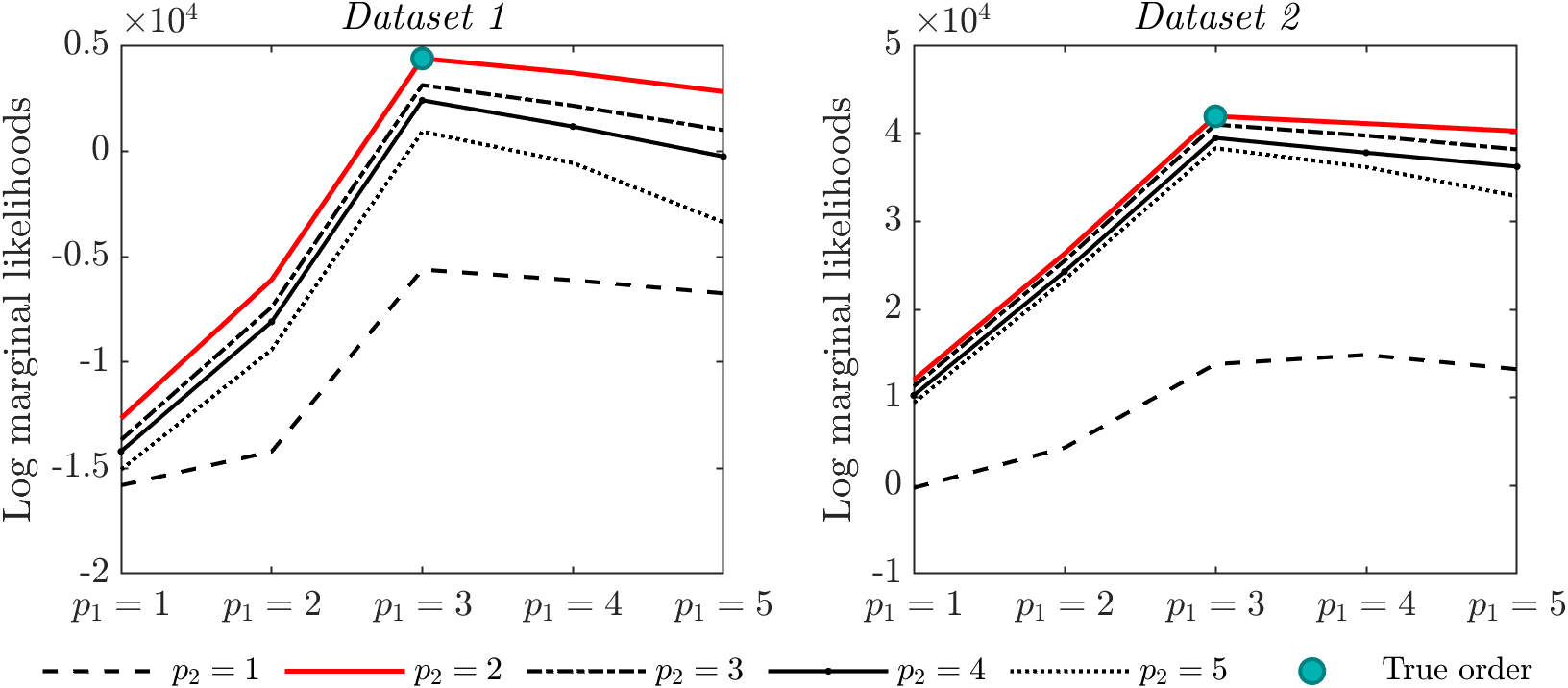}
    \end{minipage}

    \vspace{1em}

    \begin{minipage}{0.9\textwidth}
        \centering
        \includegraphics[width=0.72\linewidth]{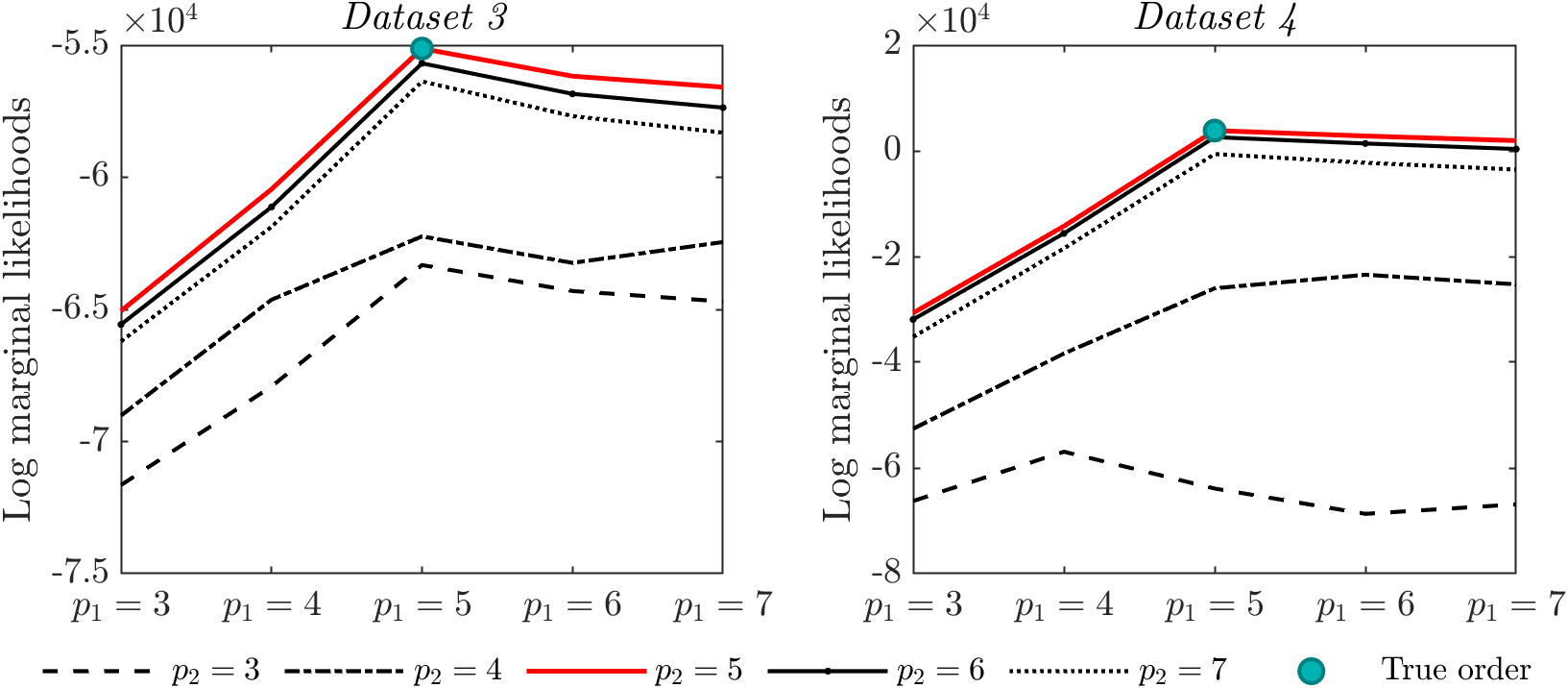}
    \end{minipage}

    \caption{Log marginal likelihood estimates for the four datasets. From top
    left to bottom right, the panels correspond to $(n,k,p_1,p_2) = (10,10,3,2)$,
    $(20,15,3,2)$, $(10,10,5,5)$, and $(20,15,5,5)$. Each line represents a
    different value of $p_2$, and the blue dot marks the true value for $p_1$ and
    $p_2$.}
    \label{fig:ml_dimension}
\end{figure}

Figure~\ref{fig:ml_dimension} reports the log marginal likelihood estimates. In all four datasets, the estimator correctly identifies
the true dimension --- the maximum log marginal likelihood is achieved at the true $(p_1,p_2)$. In addition, the estimates exhibit a consistent pattern: monotonically increasing as we approach the true dimension and monotonically decreasing afterwards. For example, when the true dimension is $(3,2)$, $\log\hat{p}(\by\given p_1=1)<\log\hat{p}(\by\given p_1=2)<\log\hat{p}(\by\given p_1=3)$
and $\log\hat{p}(\by\given p_1=3)>\log\hat{p}(\by\given p_1=4)>\log\hat{p}(\by\given p_1=5)$, with an analogous pattern for $p_2$.

\subsection{Distinguishing competing model structures}\label{sec:mc_structure}

We next examine whether the marginal likelihood estimator can distinguish between competing model \emph{structures}: vector versus matrix dynamic factor models (VDFM versus MDFM), and exact versus approximate factor models (diagonal versus Kronecker idiosyncratic covariance, with and without stochastic volatility).

\paragraph{VDFM versus MDFM}
Following Section~\ref{sec:accuracy}, we generate data from MDFMs with factor
dimensions $(p_1,p_2)\in\{(1,2),(2,1),(2,2)\}$ and from VDFMs with $k_f\in\{1,2,3\}$
factors. For each true model, we estimate the corresponding alternative
specifications and compare log marginal likelihoods. When the true model is an
MDFM, we estimate VDFMs with $k_f=1,\ldots,6$. When the true model is a VDFM, we
estimate MDFMs with $(p_1,p_2)\in\{(1,1),(1,2),(2,1),(2,2)\}$.

\begin{figure}[H]
    \centering
    \includegraphics[width=0.95\linewidth]{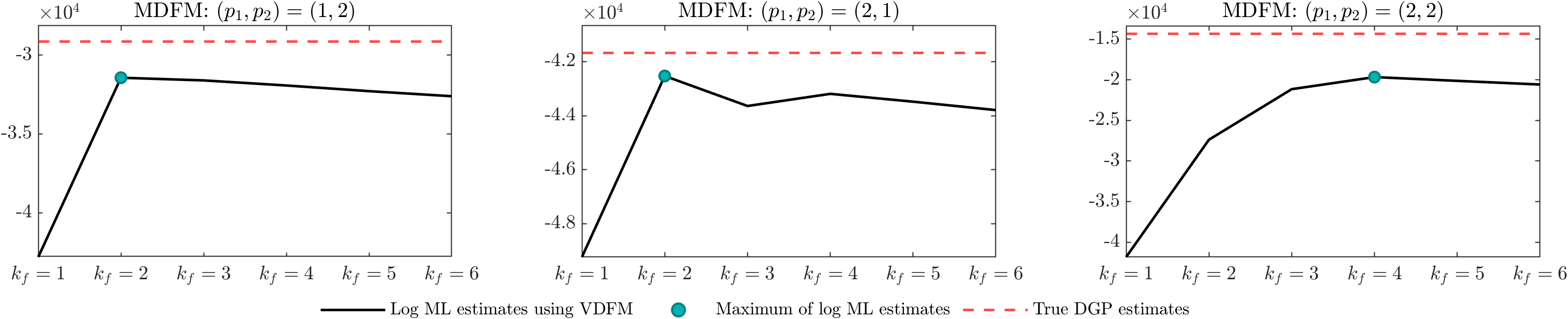}
    \caption{Log marginal likelihood estimates under VDFM specifications when the true
    model is an MDFM. The black line traces VDFM marginal likelihoods across
    $k_f=1,\ldots,6$ factors; the red dashed line marks the log marginal likelihood estimate of the true DGP: MDFM. The
    left, middle, and right panels correspond to MDFMs with
    $(p_1,p_2)=(1,2),(2,1),(2,2)$.}
    \label{fig:mdfm_ml}
\end{figure}

\begin{figure}[H]
    \centering
    \includegraphics[width=0.95\linewidth]{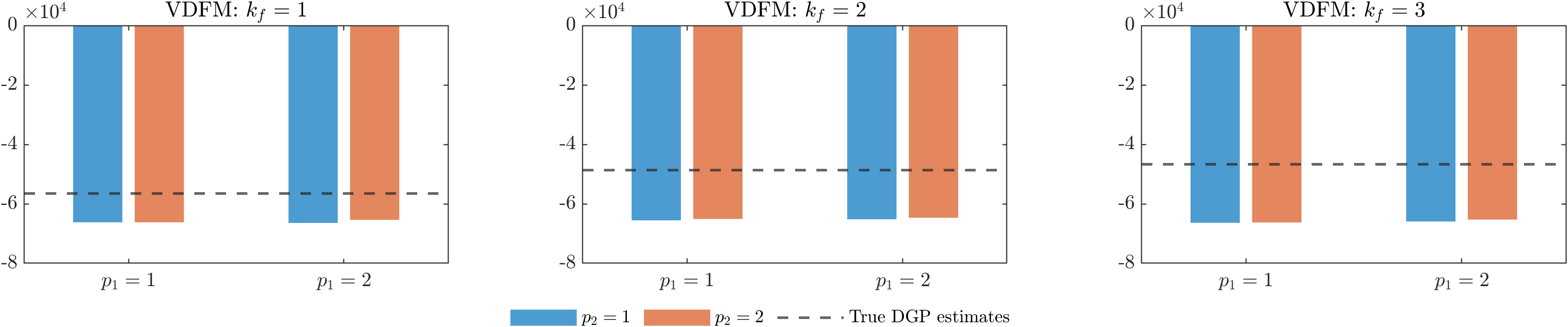}
    \caption{Log marginal likelihoods under MDFM specifications when the true
    model is a VDFM. The black dashed line marks the log marginal likelihood estimate of the true DGP: VDFM; bars
    correspond to competing MDFMs. The left, middle, and right panels correspond
    to VDFMs with $k_f=1,2,3$.}
    \label{fig:vdfm_ml}
\end{figure}

Figures~\ref{fig:mdfm_ml}--\ref{fig:vdfm_ml} report the results. In all six cases,
the true model achieves the highest marginal likelihood. A noteworthy pattern in
Figure~\ref{fig:mdfm_ml}: when the true model is MDFM, the best-performing VDFM is
the one whose number of factors equals the total number of MDFM factors --- e.g.,
$k_f=2$ for MDFMs with $(1,2)$ and $(2,1)$, and $k_f=4$ for $(2,2)$. The intuition
is that a VDFM with $p_1\times p_2$ factors can in principle span the same factor
space as an MDFM with dimensions $(p_1,p_2)$, but does so with $nk\times p_1p_2$
loadings rather than the $np_1+kp_2$ loadings of the MDFM, leading to weaker fit
when the matrix structure is true.

\paragraph{Diagonal versus Kronecker idiosyncratic covariance, with and without SV}
We next assess whether the estimator can distinguish three idiosyncratic-covariance
specifications:

\begin{itemize}
    \item \textbf{MDFM-diag:} diagonal idiosyncratic covariance, no stochastic
    volatility.
    \item \textbf{MDFM-cross:} Kronecker idiosyncratic covariance
    ($\vSigma_c\otimes\vSigma_r$), no stochastic volatility.
    \item \textbf{MDFM-sv:} diagonal idiosyncratic covariance with common
    stochastic volatility.
\end{itemize}

We generate 100 datasets of dimension $(n,k,T)=(20,20,100)$ with factor dimension
$(p_1,p_2)=(2,2)$. The free elements of $\bA$ and $\bB$ are drawn from
$\distn{N}(0,0.3^2)$. Row and column covariance matrices are drawn from
$\vSigma_r\sim\distn{IW}(n+2,\bI_n)$ and $\vSigma_c\sim\distn{IW}(k+2,\bI_k)$. The
log-volatility process has AR(1) coefficient 0.97 and innovation variance 0.1; the
factor innovation variance is $0.1\bI_{p_1p_2}$, with AR coefficients drawn from
$\distn{U}(0.8,0.9)$.

\begin{figure}[H]
    \centering
    \includegraphics[width=0.95\linewidth]{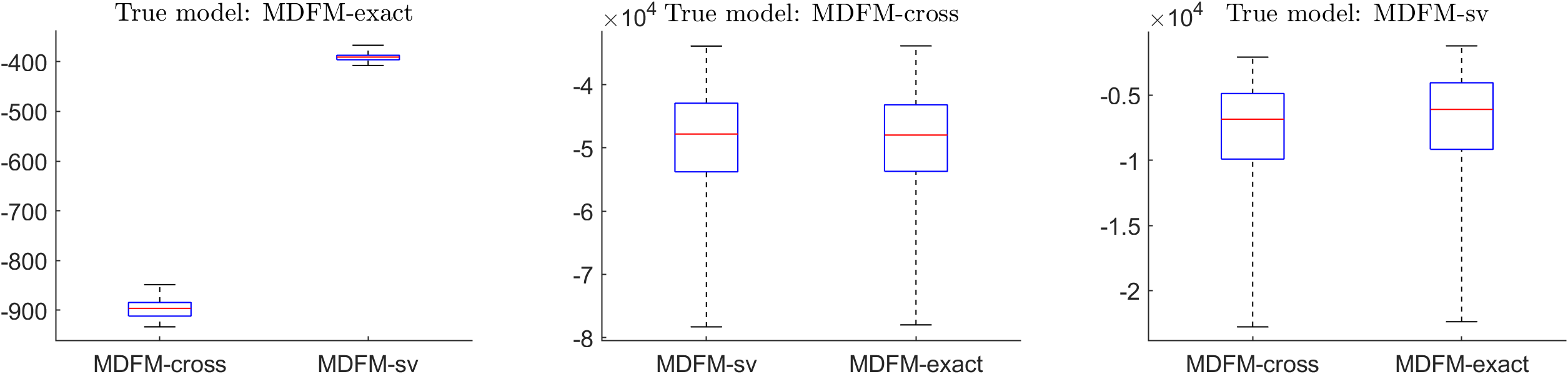}
    \caption{Boxplots of differences in log marginal likelihoods relative to the
    true model: MDFM-diag (left), MDFM-cross (middle), and MDFM-sv (right). A
    negative value indicates that the true model is favored.}
    \label{fig:ml_exactvsapproximate}
\end{figure}

For each dataset, we estimate the log marginal likelihood under each of the three
specifications and compute differences relative to the true model.
Figure~\ref{fig:ml_exactvsapproximate} reports boxplots of these differences. In
all three settings, the true model achieves the highest marginal likelihood,
indicating that the estimator reliably distinguishes the structure of the
idiosyncratic component. The differentiating power is largest when the true model
is MDFM-diag (the most parsimonious of the three), since the alternatives
correspond to over-specified models in that case.

\section{An Application to a Large OECD Macroeconomic Panel}\label{sec:application}
We illustrate the MDFM on a quarterly OECD macroeconomic panel covering 19
advanced economies and 10 indicators from 1995.Q1 to 2023.Q3. The
application is designed to showcase three features of the model
relative to existing alternatives: the economic interpretability of factors
under triangular identification; the empirical
relevance of the Kronecker idiosyncratic covariance structure for capturing
residual cross-cell dependence; and the predictive performance of iterated
forecasts from a coherent dynamic model relative to direct projections from
a static matrix factor benchmark.\footnote{A second application to the Fama--French
	$10 \times 10$ size-by-book-to-market portfolio panel, with comparable
	structure but a different empirical setting, is reported in \ref{app:fama} of
	the Online Supplemental Materials. In that application, we find that a vector DFM is favored by the data compared to a matrix DFM, based on the marginal likelihood estimates.}
\subsection{Data}
The dataset comprises 10 indicators for 19 developed economies from North America,
Europe, Asia, and Oceania, observed from 1995.Q1 to 2023.Q3 (115 quarters). The
indicators are real GDP, headline CPI, labor unit cost, exports, imports,
unemployment, energy CPI, household consumption, core CPI, and food CPI. Each
series is rendered stationary via first or logarithmic differencing and standardized
to zero mean and unit variance. Detailed variable definitions and transformation
codes are given in \ref{app:data_app1} of the Online Supplemental Materials.

For countries, we place the United States first as
the largest global economy, followed by the United Kingdom, Australia, Germany,
and Japan as anchors of their respective regional economies. For indicators, we
order real GDP first as the headline measure of real activity, followed by headline
CPI as the principal inflation measure, followed by labor unit cost for its
productivity interpretation. 

\subsection{Model Selection and Estimation}\label{subsec:est_selection}

Posterior inference is obtained via the Gibbs sampler described in Section \ref{sec:bayesian}. 10,000 posterior draws are used for inference after discarding 5,000 burn-in draws.

\paragraph{Model selection}
Tables~\ref{tab:ml_vdfm} and \ref{tab:ml_mdfm} report log marginal
likelihood estimates (with numerical standard errors in parentheses) for
VDFMs and MDFMs. We use the labels VDFM-diag and VDFM-sv for vector models
with diagonal idiosyncratic covariance, without and with common stochastic
volatility, and analogously MDFM-diag, MDFM-cross, and MDFM-cross-sv for
matrix models with diagonal, Kronecker, and Kronecker-plus-SV
specifications.

\begin{table}[H]
	\centering
	\caption{Log marginal likelihood estimates using VDFMs. Numerical standard
		errors in parentheses. The largest estimate in each panel is in bold.}
	\label{tab:ml_vdfm}
	\resizebox{0.7\textwidth}{!}{
		\begin{tabular}{ccccccccc}
			\toprule
			\toprule
			\multicolumn{4}{c}{VDFM-diag} &       & \multicolumn{4}{c}{VDFM-sv} \\
			\cmidrule{1-4}\cmidrule{6-9}    $k=1$   & $k=2$   & $k=3$   & $k=4$   &       & $k=1$   & $k=2$   & $k=3$   & $k=4$ \\
			$-26361$ & $-24786$ & $-23602$ & $-23017$ &       & $-22715$ & $-21642$ & $-21272$ & $-20910$ \\
			(0.1) & (0.3) & (0.9) & (1.3) &       & (0.2) & (0.6) & (0.7) & (1.0) \\
			$k=5$   & $k=6$   & $k=7$   & $k=8$   &       & $k=5$   & $k=6$   & $k=7$   & $k=8$ \\
			$\mathbf{-22944}$ & $-23001$ & $-23107$ & $-23280$ &       & $\mathbf{-20862}$ & $-21130$ & $-21202$ & $-21371$ \\
			(1.6) & (1.6) & (2.3) & (4.3) &       & (2.0) & (3.2) & (4.2) & (6.0) \\
			\bottomrule
			\bottomrule
	\end{tabular}}
\end{table}

\begin{table}[H]
	\centering
	\caption{Log marginal likelihood estimates using MDFMs. Numerical standard
		errors in parentheses. The largest estimate in each panel is in bold;
		the overall maximum is highlighted in red.}
	\label{tab:ml_mdfm}
	\resizebox{0.9\textwidth}{!}{
		\begin{tabular}{lccccccccccc}
			\toprule
			\toprule
			& \multicolumn{3}{c}{MDFM-diag} &       & \multicolumn{3}{c}{MDFM-cross} &       & \multicolumn{3}{c}{MDFM-cross-sv} \\
			\cmidrule{2-4}\cmidrule{6-8}\cmidrule{10-12}          & $p_2=1$ & $p_2=2$ & $p_2=3$ &       & $p_2=1$ & $p_2=2$ & $p_2=3$ &       & $p_2=1$ & $p_2=2$ & $p_2=3$ \\
			$p_1=1$ & $-26472$ & $-24796$ & $-23493$ &       & $-17968$ & $\mathbf{-17930}$ & $-17950$ &       & $-16144$ & \textcolor[rgb]{1,0,0}{$\mathbf{-16140}$} & $-16164$ \\
			& (0.2) & (0.3) & (0.7) &       & (0.3) & (0.4) & (0.8) &       & (0.6) & (0.4) & (0.5) \\
			$p_1=2$ & $-26310$ & $-24607$ & $-23221$ &       & $-18007$ & $-17979$ & $-18068$ &       & $-16260$ & $-16265$ & $-16337$ \\
			& (0.1) & (0.4) & (0.4) &       & (0.4) & (0.4) & (1.4) &       & (0.5) & (0.6) & (1.0) \\
			$p_1=3$ & $-26164$ & $-24469$ & $\mathbf{-23012}$ &       & $-18051$ & $-18058$ & $-18150$ &       & $-16336$ & $-16386$ & $-16464$ \\
			& (0.3) & (0.6) & (0.7) &       & (0.3) & (0.8) & (0.6) &       & (0.7) & (0.7) & (1.1) \\
			\bottomrule
			\bottomrule
	\end{tabular}}
\end{table}

There are three interesting findings. First, the data strongly favor the
inclusion of common stochastic volatility: in both vector and matrix models,
the SV variants dominate their constant-volatility analogues by thousands of
log points (right panels of Tables~\ref{tab:ml_vdfm} and \ref{tab:ml_mdfm}).
Second, the data favor Kronecker-structured idiosyncratic covariance over
the diagonal alternative within the matrix framework. Third, the matrix
structure dominates the vector representation: the best MDFM achieves a log
marginal likelihood of $-16{,}140$, more than $4{,}700$ log points above the
best VDFM (at $-20{,}862$). The optimal MDFM has a $1 \times 2$ factor
matrix, indicating one country dimension and two indicator dimensions,
whereas the best VDFM requires five factors. The $1 \times 2$ shape of
$\bF_t$ implies that variation along the indicator dimension is richer than
variation along the country dimension in this panel. We hereafter
estimate the MDFM with $(p_1, p_2) = (1, 2)$ and refer to it simply as the
MDFM.

\subsection{In-Sample Results}\label{sec:insample}
\begin{figure}[H]
	\centering
	\includegraphics[width=\linewidth]{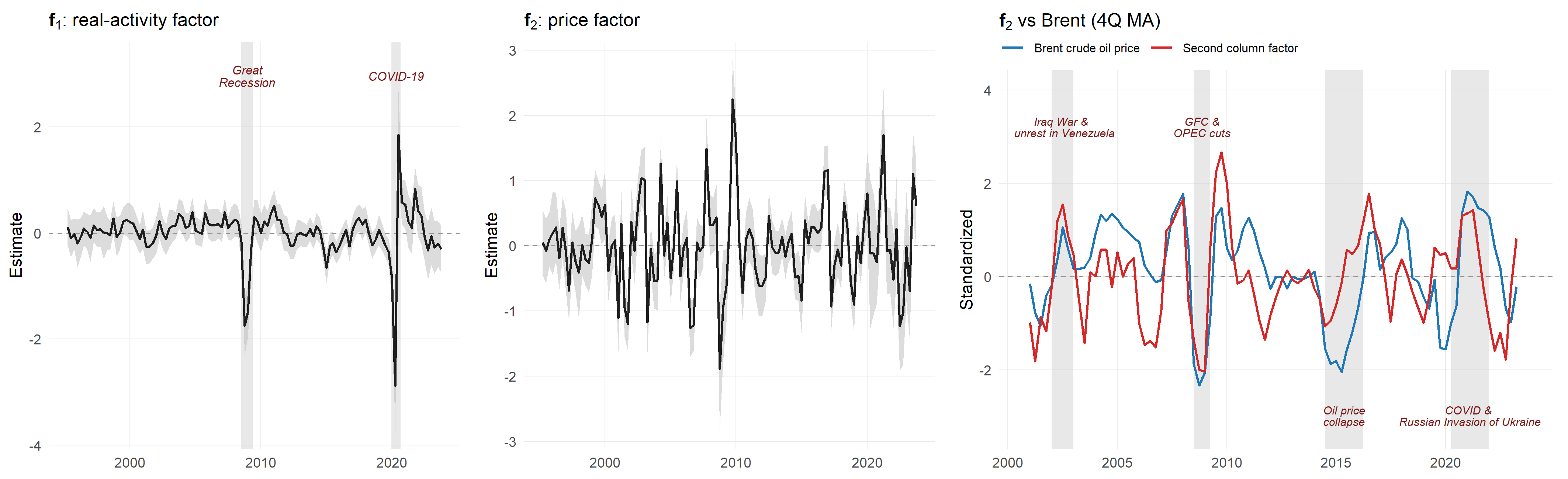}
	\caption{Left: real-activity factor $\bbf_1$ with shaded recession
		episodes. Middle: price factor $\bbf_2$ with $90\%$ credible band.
		Right: four-quarter moving averages of standardized Brent crude oil
		price growth (blue) and the second column factor (red); shaded bands
		mark major oil-market episodes (war-premium spike during the 2003
		Iraq War and unrest in Venezuela; the 2008--2009 oil-price collapse
		and OPEC production cuts; the 2014--2016 oil-price collapse driven
		by U.S.\ shale supply and weak global demand; and the COVID-19
		demand shock and 2022 Russian invasion of Ukraine).}
	\label{fig:factors_oil}
\end{figure}

\paragraph{Factor series and the price--oil link}
Figure~\ref{fig:factors_oil} displays posterior estimates of the two
factor series with $90\%$ credible bands, alongside a comparison of the
price factor to Brent crude oil price growth.

The real-activity factor $\bbf_1$ traces the global business cycle. Two
episodes dominate the series: a sharp contraction during the 2008--2009
Great Recession and a much larger spike in 2020 corresponding to the
COVID-19 pandemic. The early-2000s slowdown is visible but quantitatively
modest, consistent with the recession being concentrated in the United
States and the United Kingdom rather than synchronized across the panel.

The price factor $\bbf_2$ has its strongest external corroboration in the
right-hand panel of Figure~\ref{fig:factors_oil}: a four-quarter moving
average of $\bbf_2$ tracks four-quarter Brent crude oil price growth
throughout the sample, with aligned peaks and troughs at the 2002–2003 Iraq war and civil
unrest in Venezuela, the 2008--2009 Great Recession and OPEC’s production cuts, the
2014–2016 oil price collapse and the 2022 Russian invasion of Ukraine. The posterior correlation between the two smoothed series is
approximately $0.66$. 

\begin{figure}[H]
	\centering
	\includegraphics[width=0.65\linewidth]{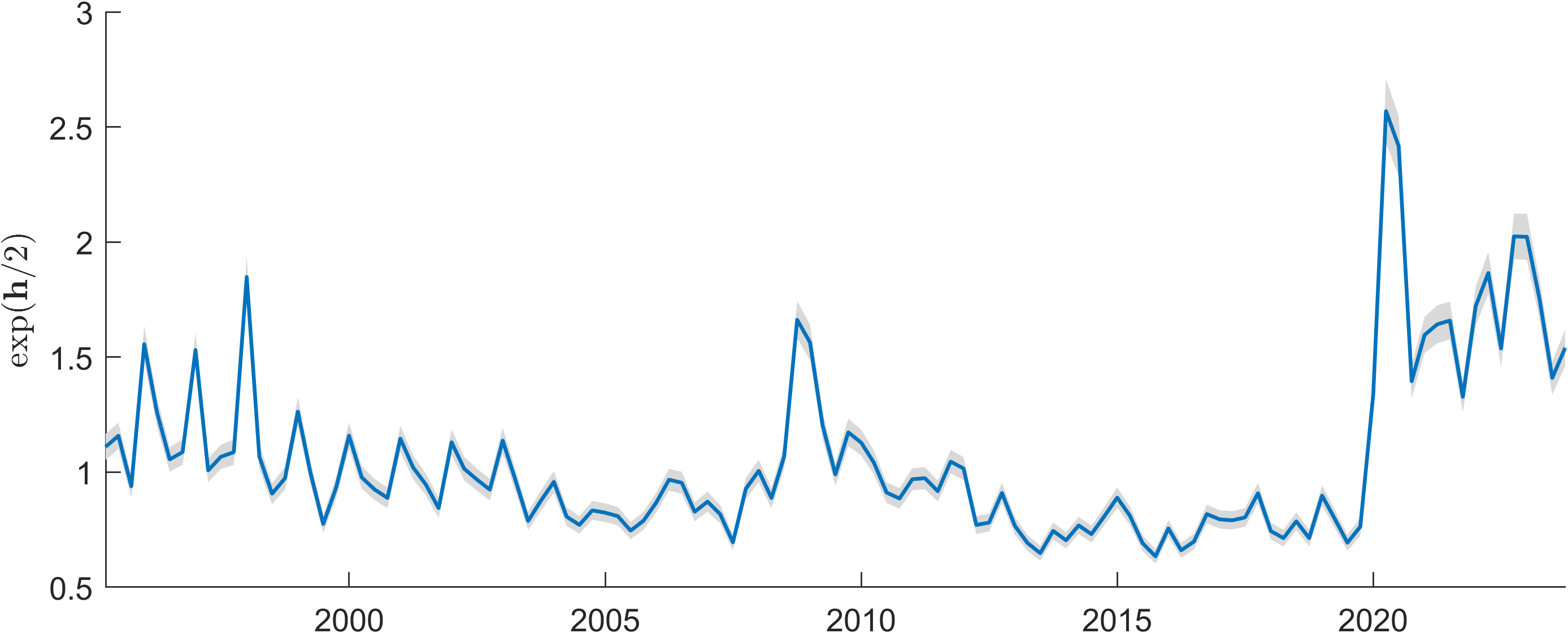}
	\caption{Posterior mean of common stochastic volatility,
		$\widehat{\omega}_t = \exp(\widehat{\bh}_t / 2)$, with $90\%$
		credible band.}
	\label{fig:volatility}
\end{figure}

\begin{figure}[H]
	\centering
	\includegraphics[width=\linewidth]{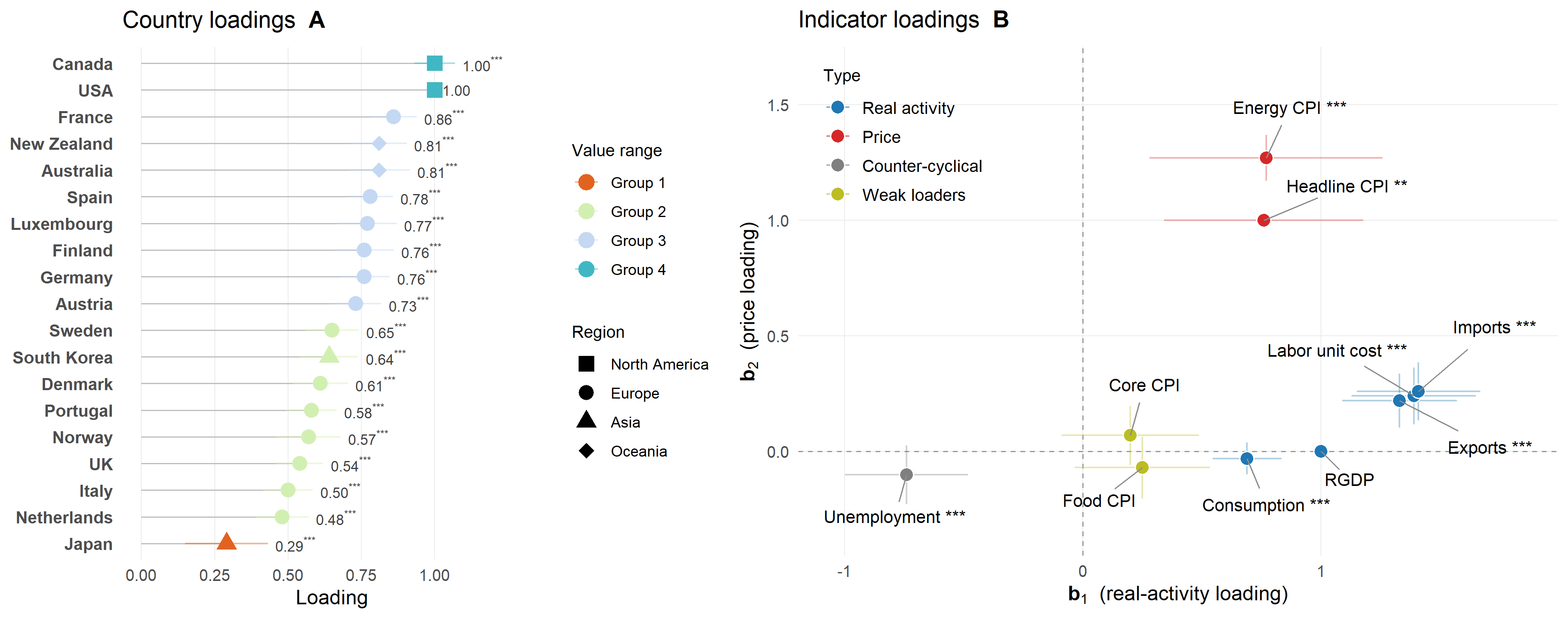}
	\caption{Loading estimates with $90\%$ credible intervals. Left: country
		loadings $\widehat{\bA}$, sorted ascending, colored by value-range
		group and shaped by region. The numeric value of each loading appears
		to the right of its credible interval, with significance stars as
		superscripts. The USA loading is fixed to one for identification.
		Right: indicator loadings $\widehat{\bB}$ in the $(\bb_1, \bb_2)$
		plane, colored by economic role.}\label{fig:loadings}
\end{figure}

\paragraph{Loadings}
The latent structure of the global macroeconomy can be read off the row and
column loading matrices. Figure~\ref{fig:loadings} reports posterior
estimates side by side: the country loadings $\widehat{\bA}$ on the left and
the indicator loadings $\widehat{\bB}$ on the right, with $90\%$ credible
intervals from $1.645 \times$ posterior standard deviations on each
axis.\footnote{Since the factor matrix has only one row, $\bA$ is
	effectively a $19 \times 1$ vector; we retain the matrix notation for
	consistency.}

The country loadings reveal a striking heterogeneity in exposure to the factors that does not map cleanly onto geography.\footnote{We sort these loading estimates and compute the
	posterior probabilities that the differences between neighboring values are greater than 0.
	We then group countries and indicators by comparing these posterior probabilities against
	a 0.9 threshold: when the probability exceeds 0.9, it indicates that the neighboring values
	are significantly different, so they are placed in separate groups.} Japan
stands alone at the bottom of the distribution with a loading of $0.29$. A mid-tier of European
economies plus Korea (NLD, ITA, GBR, NOR, PRT, DNK, KOR, SWE) loads between
$0.48$ and $0.65$. A higher-tier group containing the continental European
core and Oceania (AUT, DEU, FIN, LUX, ESP, AUS, NZL, FRA) loads between
$0.73$ and $0.86$. Canada and the United States sit at $1.00$. It is clear that geographic
proximity matters, as the two North American economies cluster together at
the top, and continental Europe is concentrated in the upper-mid range, although it is not the only factor. Japan and Korea fall in different groups
despite their proximity, and Australia and New Zealand are grouped together with
continental European economies rather than with Asia.

The indicator loadings sort the ten variables into four economic roles.
Imports, labor unit cost, and exports load most strongly on the real-activity
factor $\bb_1$ (loadings between $1.3$ and $1.4$) with small but positive
loadings on the price factor $\bb_2$. RGDP and consumption load positively
on $\bb_1$ alone. Unemployment is the only counter-cyclical indicator, with
a large negative $\bb_1$ loading of approximately $-0.6$ and a $\bb_2$
loading indistinguishable from zero. Energy CPI and headline CPI load
heavily on $\bb_2$ (loadings around $1.3$ and $1.0$ respectively), with
smaller but positive $\bb_1$ components. Core CPI and food CPI are weak
loaders on both factors: their $90\%$ credible intervals overlap the origin
on at least one axis, indicating that most of their variation is left to
the idiosyncratic component. The two-dimensional view in the right panel of Figure~\ref{fig:loadings} shows that
the real-activity and price factors are approximately orthogonal in their
effect on indicators. Indicators with large $\bb_1$ have small $\bb_2$, and
vice versa. 

\paragraph{The stochastic volatility and idiosyncratic correlations}
The common stochastic volatility component $\widehat{\omega}_t$
(Figure~\ref{fig:volatility}) isolates three episodes of elevated common
uncertainty: the 1997--1998 Asian financial crisis (driven particularly by
Japan and Korea), the 2008--2009 Great Recession, and the COVID-19
pandemic starting in 2020. The COVID-19 spike is by far the largest, consistent with the recent
literature documenting the importance of stochastic volatility and outlier
modeling for post-2020 macroeconomic data. Importantly, volatility
remains elevated through 2023 rather than returning to its pre-2020
baseline.

\begin{figure}[H]
	\centering
	\includegraphics[width=0.85\linewidth]{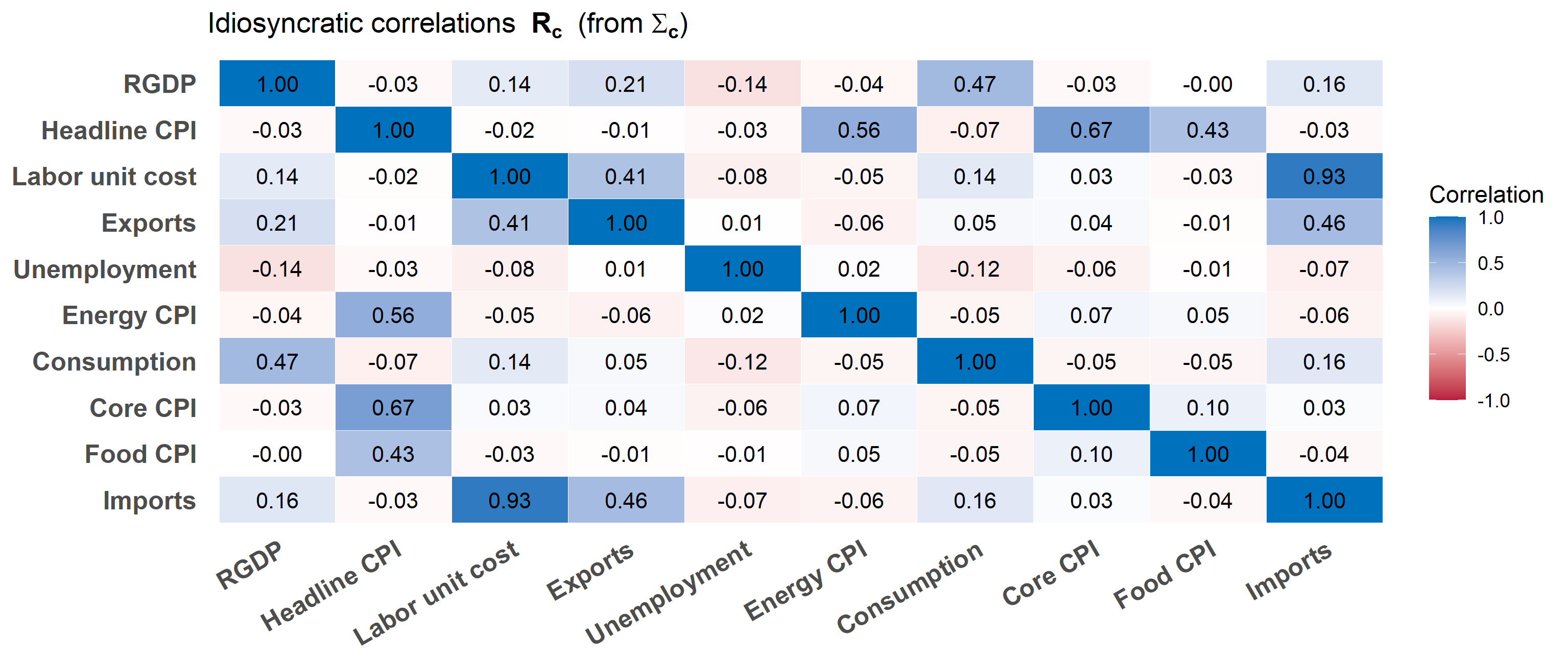}
	\caption{Idiosyncratic correlation matrix $\bR_c$ across indicators,
		standardized from $\widehat{\vSigma}_c$.}
	\label{fig:Rc}
\end{figure}

\begin{figure}[H]
	\centering
	\includegraphics[width=0.85\linewidth]{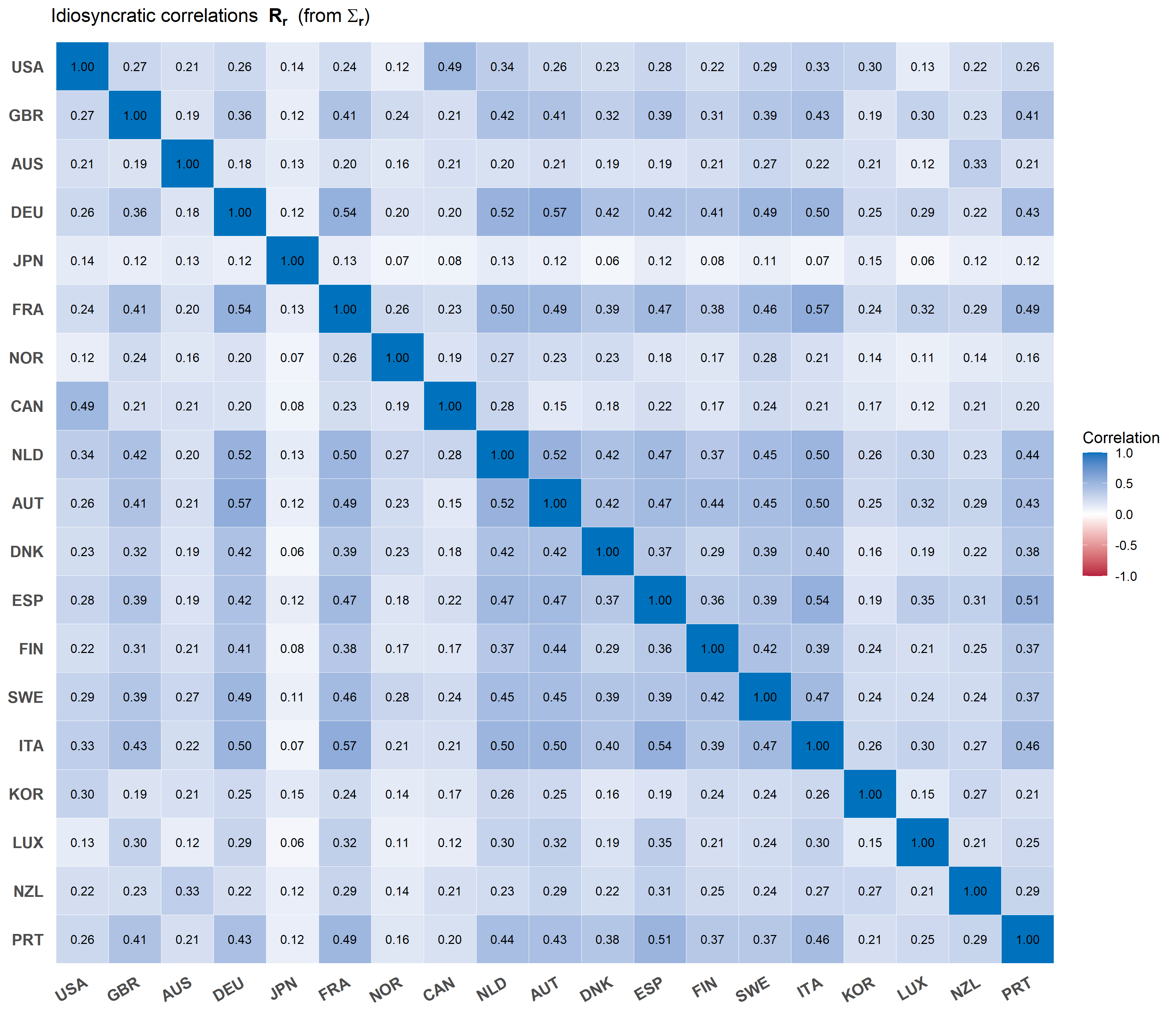}
	\caption{Idiosyncratic correlation matrix $\bR_r$ across countries,
		standardized from $\widehat{\vSigma}_r$.}
	\label{fig:Rr}
\end{figure}

In addition, there exist significant correlations among countries and indicators in idiosyncratic components. Figures~\ref{fig:Rc} and
\ref{fig:Rr} display the corresponding correlation matrices $\bR_c$ and
$\bR_r$, obtained by standardizing $\vSigma_c$ and $\vSigma_r$ to unit
diagonal. Correlations are a more interpretable summary than raw
covariances because they are not inflated by the very different residual
variances across rows and columns.

Consider first the indicator correlations in $\bR_c$. The idiosyncratic correlation is not diffuse but concentrated in a few interpretable blocks. First, real activity retains a clear correlation, with RGDP and consumption correlated at 0.47. In addition, the four price indices form a tight block among themselves (headline-core 0.67, headline-energy 0.56, headline-food 0.43). A third block links unit labor costs (ULC), imports, and exports. Notably, this block is unrelated to the price block, indicating that this idiosyncratic correlation  is driven by real-side forces rather than by prices. A possible explanation is a terms-of-trade gain that at the same time raises real wages relative to productivity, strengthens the currency and draws in imports. 

The idiosyncratic country correlations in $\bR_r$ line up along regional lines. The clearest grouping is an European block, i.e., France, Germany, Italy, the Netherlands, Austria, Spain, Portugal, and Finland, with correlations of about 0.4--0.57. The tightest pairs are neighbors and major trading partners: France--Italy (0.57), Germany--Austria (0.57), France--Germany (0.54), and Spain--Italy (0.54). Sweden, although it keeps its own currency, tracks this group closely as well, reflecting its trade ties with the European Union. The United Kingdom and Luxembourg attach to the group more loosely -- the UK at 0.36--0.43 and Luxembourg at 0.29--0.35. The US and Canada form their own tight pair (0.49), yet the transatlantic link is somewhat weaker -- the US correlates with European economies at around 0.13--0.34. Japan is almost an island: its correlation with every other OECD economy in the dataset stays below 0.15. Korea and Australia are among the most weakly connected economies after Japan, peaking at 0.30 and 0.33 respectively; Australia's only sizeable tie is to New Zealand (0.33), a small Australasian pair that mirrors the US--Canada link.

\subsection{Out-of-Sample Forecasting}\label{sec:sequential_forecasting}

We assess predictive performance of the MDFM via a recursive out-of-sample exercise where at each period $t$ of the evaluation sample, we use only data up to that period to estimate the models and compute the $h$-step-ahead forecasts for period $t+h$. The forecast evaluation period starts from 2007Q4 till the end of the sample. The benchmark throughout is the
static matrix factor model of \citet{wang2019factor}, hereafter MFM. 

\paragraph{MFM dimension selection and forecast construction}
The MFM selects factor dimensions $(\widehat{p}_1, \widehat{p}_2)$ via an
eigenvalue-ratio rule applied to the lag-$h_0$ autocovariance matrices
$\widehat{\bM}_1$ and $\widehat{\bM}_2$.\footnote{
		$\widehat{\bM}_1 \;=\; \sum_{h=1}^{h_0} \widehat{\vSigma}_h^{(1)}\,
		\big[\widehat{\vSigma}_h^{(1)}\big]^\top, \qquad
		\widehat{\bM}_2 \;=\; \sum_{h=1}^{h_0} \widehat{\vSigma}_h^{(2)}\,
		\big[\widehat{\vSigma}_h^{(2)}\big]^\top
		$, where $\widehat{\vSigma}_h^{(1)}$ and $\widehat{\vSigma}_h^{(2)}$ are
	row- and column-wise sample autocovariance matrices at lag $h$, and
	$h_0$ is a user-specified upper lag. Details of a pre-forecasting analysis of MFM is provided in  \ref{app:wang}.} Applied to our OECD panel, the
rule selects $\widehat{p}_1 = 1$ and $\widehat{p}_2 = 1$, agreeing with our
marginal likelihood selection on the row dimension but selecting one fewer
column factor. The selection is robust to the lag parameter $h_0$: across
$h_0 \in \{1, 2, 3, 4\}$ the rule uniformly returns $(1, 1)$, in-sample
RMSE under the auto-selected dimensions varies by less than $0.2\%$, and
the loading subspaces drift by less than $0.08$
(Table~\ref{tab:h0_sensitivity} in the Online Supplemental Materials). The
eigenvalue-ratio rule is known to under-select the number of rows and columns in
moderate-$T$ panels when the true value is small
\citep[Table~4]{wang2019factor}, and forcing $(p_1, p_2) = (1, 2)$ in the
MFM uniformly improves its in-sample RMSE by 6--7 percentage points across
all $h_0$ values, providing evidence independent of our Bayesian framework
that a second column factor may carry genuine signal. We nonetheless use the
auto-selected MFM as the out-of-sample benchmark to maintain a fair
comparison that respects the MFM's native model-selection rule.

The two methods differ in the construction of multi-step forecasts. The
MDFM produces \emph{iterated} forecasts as a direct byproduct of the specified
dynamics of latent states: each MCMC draw of the factor matrix
$\bF_t$ and the volatility path is propagated forward
through the model's state and observation equations given the parameter draw, yielding draws of
$\bY_{t+h}$ that integrate over parameter, factor, and volatility
uncertainty. The MFM forecasts are direct projections in the spirit of
\citet{stock2002macroeconomic}. At each origin $t$ we extract the scalar
factor, then estimate horizon-specific OLS projections of $Y_{i,j,t+h}$ on a constant and factor estimate $f_t$ for each (country, indicator) pair $(i,j)$ and horizon $h$.\footnote{Direct projection is the natural choice because the MFM does not specify a dynamic law of motion for the latent factor. Although density forecasts can be constructed for direct projections using bootstrap or simulation-based methods, such forecasts require additional assumptions and implementation choices. In this paper we focus on comparing the forecasting performance of MDFM and MFM as specified. }

We report results at three horizons: $h=1$, $h=4$, and $h=8$ (one-quarter, one-year, and two-year horizons).
Per-cell RMSFE ratios and
Diebold--Mariano test results are reported in
Figure~\ref{fig:oos_rmsfe_appendix} in the Online Supplemental Materials.
Table~\ref{tab:pooled_by_indicator_country} summarizes the pooled
performance by indicator (pooled across countries) and by country
(pooled across indicators).

\begin{table}[H]
	\centering
	\caption{Out-of-sample RMSFE ratios MDFM / MFM, pooled across 
		cells. The left panel pools across the 19 OECD countries within each 
		indicator. The middle and right panels pool across the 10 indicators 
		within each country. Pooled RMSFEs are computed from squared errors 
		summed over all cells and all expanding-window forecast origins before 
		taking the ratio. Values below 1 indicate that MDFM outperforms 
		MFM and are shown in \textbf{bold}. Diebold--Mariano significance 
		based on the pooled loss differential with Bartlett HAC variance 
		estimation: $^{*}\,p<0.10$, $^{**}\,p<0.05$, $^{***}\,p<0.01$. The 
		bottom row reports the fully pooled ratio across all $19 \times 10 = 190$ 
		cells.}
	\label{tab:pooled_by_indicator_country}
	\resizebox{\textwidth}{!}{
		\begin{tabular}{lccc@{\hspace{1.5em}}lccc@{\hspace{1.5em}}lccc}
			\toprule\toprule
			Indicator & $h{=}1$ & $h{=}4$ & $h{=}8$ &
			Country & $h{=}1$ & $h{=}4$ & $h{=}8$ &
			Country & $h{=}1$ & $h{=}4$ & $h{=}8$ \\
			\midrule
			RGDP                  & \textbf{0.911}$^{***}$ & \textbf{0.926}$^{***}$ & 1.060       &
			USA                   & \textbf{0.974}$^{***}$ & \textbf{0.970}$^{**}$  & 1.028       &
			DNK                   & \textbf{0.976}$^{***}$ & \textbf{0.933}$^{***}$ & \textbf{0.999}       \\
			Headline CPI          & \textbf{0.997}         & \textbf{0.996}         & \textbf{0.989}       &
			GBR                   & \textbf{0.971}$^{***}$ & \textbf{0.973}$^{**}$  & \textbf{0.976}       &
			ESP                   & \textbf{0.954}$^{***}$ & \textbf{0.955}$^{***}$ & \textbf{0.990}       \\
			Labor unit cost       & \textbf{0.985}$^{***}$ & \textbf{0.958}$^{***}$ & 1.018$^{*}$ &
			AUS                   & \textbf{0.983}$^{*}$   & \textbf{0.965}$^{*}$   & \textbf{0.986}       &
			FIN                   & \textbf{0.956}$^{***}$ & \textbf{0.945}$^{***}$ & 1.039$^{***}$ \\
			Exports               & \textbf{0.982}$^{***}$ & \textbf{0.955}$^{***}$ & 1.057$^{***}$ &
			DEU                   & \textbf{0.965}$^{***}$ & \textbf{0.943}$^{***}$ & 1.008       &
			SWE                   & \textbf{0.966}$^{***}$ & \textbf{0.961}$^{**}$  & 1.057$^{*}$ \\
			Unemployment          & \textbf{0.942}$^{***}$ & \textbf{0.941}$^{***}$ & \textbf{0.947}$^{*}$ &
			JPN                   & 1.000                  & \textbf{0.975}$^{*}$   & 1.016       &
			ITA                   & \textbf{0.957}$^{***}$ & \textbf{0.928}$^{***}$ & 1.028       \\
			Energy CPI            & 1.026$^{***}$          & 1.008                  & 1.027$^{***}$ &
			FRA                   & \textbf{0.982}$^{*}$   & \textbf{0.953}$^{***}$ & 1.046$^{***}$ &
			KOR                   & \textbf{0.996}         & \textbf{0.978}$^{*}$   & 1.023       \\
			Household consumption & \textbf{0.898}$^{***}$ & \textbf{0.929}$^{***}$ & \textbf{0.964}       &
			NOR                   & \textbf{0.973}$^{**}$  & \textbf{0.951}$^{***}$ & \textbf{0.981}       &
			LUX                   & \textbf{0.975}$^{*}$   & \textbf{0.967}$^{**}$  & 1.006       \\
			Core CPI              & \textbf{0.982}         & \textbf{0.982}         & \textbf{0.981}       &
			CAN                   & \textbf{0.985}         & \textbf{0.988}         & 1.063$^{***}$ &
			NZL                   & \textbf{0.991}         & \textbf{0.981}         & 1.020       \\
			Food CPI              & \textbf{0.967}$^{***}$ & \textbf{0.905}$^{***}$ & 1.008       &
			NLD                   & \textbf{0.967}$^{***}$ & \textbf{0.918}$^{***}$ & \textbf{0.986}       &
			PRT                   & \textbf{0.983}$^{**}$  & \textbf{0.960}$^{***}$ & \textbf{0.987}       \\
			Imports               & \textbf{0.972}$^{***}$ & \textbf{0.953}$^{***}$ & 1.071$^{***}$ &
			AUT                   & \textbf{0.976}$^{**}$  & \textbf{0.937}$^{***}$ & 1.034$^{**}$ &
			&                        &                        &             \\
			\midrule
			\multicolumn{12}{l}{\textit{All cells pooled:} 
				$h{=}1$: \textbf{0.975}$^{***}$, \quad
				$h{=}4$: \textbf{0.957}$^{***}$, \quad
				$h{=}8$: 1.016$^{***}$} \\
			\bottomrule\bottomrule
	\end{tabular}}
\end{table}

The MDFM substantially outperforms MFM at the 1-quarter
and 1-year horizons in pooled RMSFE, with statistically significant
improvements of $2.5\%$ at $h = 1$ and $4.3\%$ at $h = 4$. In particular, at $h=1$, 27 out of the 29 pooled indicator-country ratios fall below one. At $h=4$, MDFM outperforms MFM even more broadly and the only exception is energy CPI for which MFM outperforms MDFM insignificantly. The MDFM's advantage attenuates and partially reverses at the 2-year horizon. The fully-pooled $h=8$ ratio rises to 1.016, indicating that MDFM is marginally less accurate than MFM in pooled terms at the 2-year horizon, where the MFM's gains are significant at the $1\%$ level for exports, imports and energy CPI among indicators, and for France, Canada, and Finland among countries.

\begin{figure}[H]
	\centering
	\includegraphics[width=0.8\linewidth]{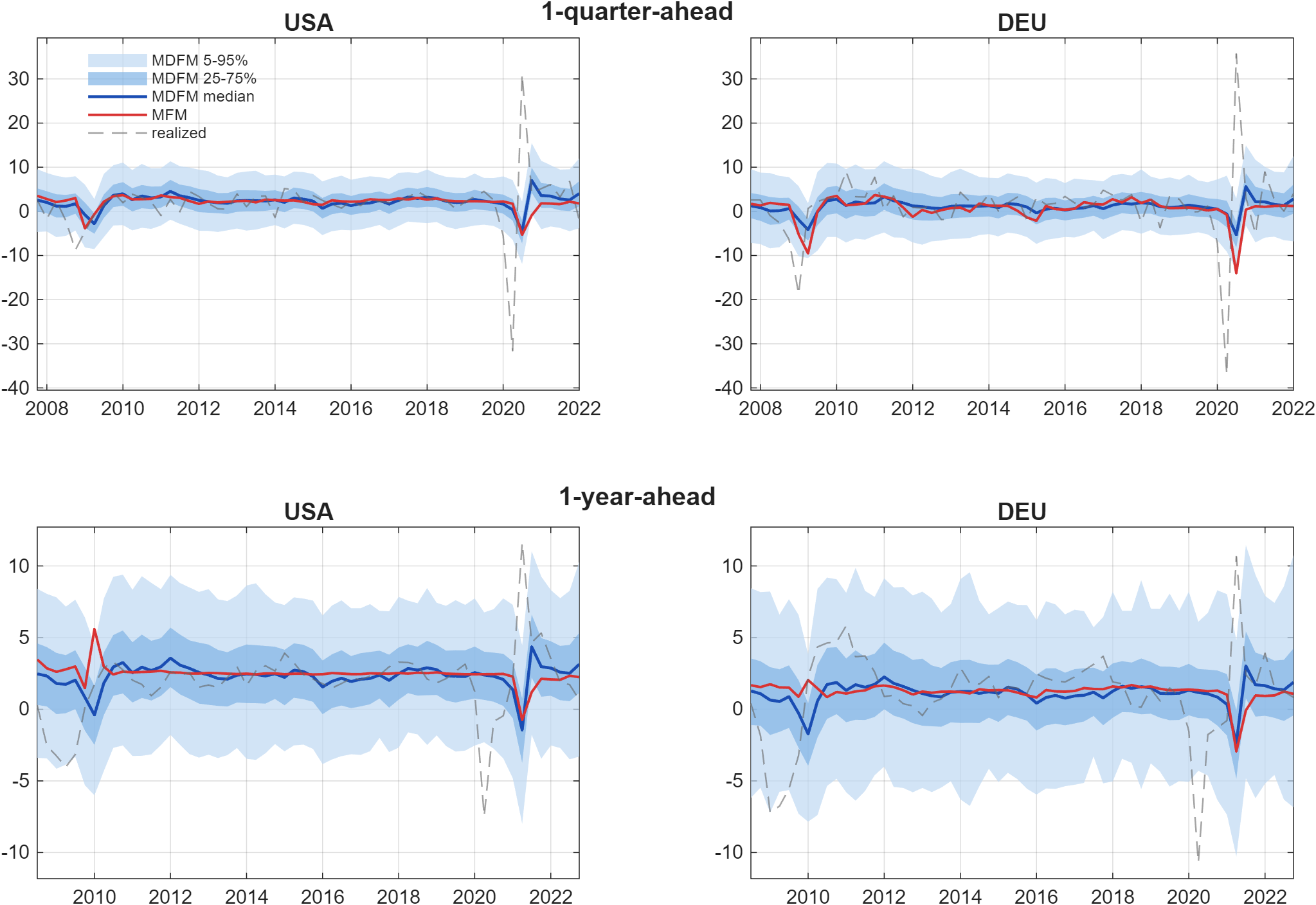}
	\caption{Posterior predictive fan charts for real GDP growth
		(annualized percent change) in the United States and Germany. The
		top row shows 1-quarter-ahead forecasts; the bottom row shows
		4-quarter (1-year) average forecasts, with both MDFM posterior
		predictive draws and realized growth averaged across
		$h = 1, \ldots, 4$. Shaded bands show $25\%$--$75\%$ (dark blue)
		and $5\%$--$95\%$ (light blue) MDFM posterior predictive intervals;
		the solid blue line is the MDFM posterior median; the solid red
		line is the MFM direct-projection point forecast; the dashed
		grey line shows realized values.}
	\label{fig:fancharts}
\end{figure}

In Figure \ref{fig:fancharts}, we show the predictive density for real GDP growth in the US and Germany at two horizons. The top panels show 1-quarter-ahead forecasts and realizations; the bottom panels show 1-year-ahead forecasts. The light blue shade represent the 5\%-95\% interval of the predictive density, while the darker blue represent the 25\%-75\% interval. The full G7 comparison is reported in Figure \ref{fig:fancharts_g7} of the Online Supplement Materials.

The incorporation of time-varying volatility lets the bands respond to changing risk: they widen during volatile episodes---modestly at the 1-quarter horizon and markedly in the 1-year panels, where they fan out around the 2008–09 crisis and the 2020–22 period. In addition, it is clear that  the 1-year panels expose a finite-sample weakness of the MFM direct projection: after the 2008–09 contraction the MFM forecast for the US jumps to about 5\% in early 2010, which is well ahead of the realized recovery, with a subtler version for Germany. The MDFM's iterated forecasts instead propagate the factor through the AR dynamics, yielding forecasts that better match the slow realized recovery.

\subsection{Robustness to data ordering}\label{sec:permute}
The lower-triangular identification scheme adopted in
Section~\ref{sec:identification} fixes the location of the latent factor
matrix at the cost of conditioning the loadings on the chosen ordering of
rows and columns. To assess whether this choice materially affects the
out-of-sample evidence, we re-run the entire expanding-window pipeline
under seven alternative orderings of the OECD panel: full reversal of
rows and columns, alphabetical, three random permutations with fixed
seeds, a rows-only random permutation (columns held at baseline), and a
columns-only random permutation (rows held at baseline). For each
ordering we permute the predictive draws back to baseline coordinates
before computing forecast losses.
Table~\ref{tab:permutation_pooled_rmsfe} reports the pooled RMSFE across
the panel for each ordering. It is clear that the pooled RMSFE varies by less than six percent across the eight orderings at
every horizon. The pooled forecasting performance of the MDFM is
therefore practically invariant to the ordering of the OECD panel.

\begin{table}[H]
	\centering
	\caption{Pooled out-of-sample RMSFE across the OECD panel for the
		MDFM under eight orderings of rows and columns. RMSFE is pooled over
		all $19 \times 10$ cells and all expanding-window forecast origins.}
	\label{tab:permutation_pooled_rmsfe}
	\begin{tabular}{lccc}
		\toprule\toprule
		Ordering & $h = 1$ & $h = 4$ & $h = 8$ \\
		\midrule
		Baseline (economic ordering)            & 1.131 & 0.656 & 0.431 \\
		Reverse rows and columns                & 1.087 & 0.646 & 0.428 \\
		Alphabetical                            & 1.110 & 0.641 & 0.419 \\
		Random \#1 (seed 1001)                  & 1.123 & 0.638 & 0.412 \\
		Random \#2 (seed 1002)                  & 1.117 & 0.654 & 0.430 \\
		Random \#3 (seed 1003)                  & 1.081 & 0.645 & 0.424 \\
		Rows random, columns baseline           & 1.137 & 0.673 & 0.446 \\
		Rows baseline, columns random           & 1.124 & 0.649 & 0.424 \\
		\midrule
		Range across orderings (\%)             & 5.2   & 5.5   & 8.3   \\
		\bottomrule\bottomrule
	\end{tabular}
\end{table}


\section{Concluding Remarks and Future Research}\label{sec:conclusion}
We have proposed a class of Bayesian dynamic factor models for high-dimensional matrix-valued time series, with autoregressive factor dynamics, stochastic volatility and outlier components, and a Kronecker-structured idiosyncratic covariance that captures cross-row and cross-column correlation. An efficient Gibbs sampler exploits the Kronecker structure to sample the loading matrices and row/column covariances jointly. For model comparison, we adapt the cross-entropy importance-sampling estimator of \citet{chan2015marginal} to select the factor-matrix dimension, distinguish vector from matrix specifications, and identify the structure of the idiosyncratic component. Applied to a 19-country, 10-indicator OECD panel, the model recovers a two-factor structure---a real-activity factor and a price factor that comoves with global oil prices---with idiosyncratic risks clustering within the European Union and volatility spiking during the Asian financial crisis, the Great Recession, and the COVID-19 pandemic. Out of sample, it delivers statistically significant forecast gains over a static matrix factor benchmark.

Several extensions are natural directions for future work. First, replacing the AR(1) dynamics for each cell of $\bF_t$ with a matrix autoregressive
process would allow cross-cell dependence in factor evolution, at the cost of
additional identification and computational challenges. Second, a structural
matrix factor model that imposes economic restrictions on $\bA$, $\bB$, or
$\bF_t$ would enable analysis of cross-country shock transmission. Third, sparse
priors on the loadings could automate the selection of relevant factor cells. We
leave these to future research.





\bibliographystyle{elsarticle-harv}
{\setstretch{0.9}
\bibliography{bibliography}}

@article{aguilar2000bayesian,
  title   = {{Bayesian dynamic factor models and portfolio allocation}},
  author  = {Aguilar, Omar and West, Mike},
  journal = {Journal of Business \& Economic Statistics},
  volume  = {18},
  number  = {3},
  pages   = {338--357},
  year    = {2000},
  publisher = {Taylor \& Francis}
}

@article{bai2003inferential,
  title   = {{Inferential theory for factor models of large dimensions}},
  author  = {Bai, Jushan},
  journal = {Econometrica},
  volume  = {71},
  number  = {1},
  pages   = {135--171},
  year    = {2003},
  publisher = {Wiley Online Library}
}

@article{bai2015identification,
  title   = {{Identification and Bayesian estimation of dynamic factor models}},
  author  = {Bai, Jushan and Wang, Peng},
  journal = {Journal of Business \& Economic Statistics},
  volume  = {33},
  number  = {2},
  pages   = {221--240},
  year    = {2015},
  publisher = {Taylor \& Francis}
}

@article{barigozzi2024dynamic,
  title   = {{The dynamic, the static, and the weak: factor models and the analysis of high-dimensional time series}},
  author  = {Barigozzi, Matteo and Hallin, Marc},
  journal = {Journal of Time Series Analysis},
  year    = {2024},
  publisher = {Wiley Online Library}
}

@article{carriero2015bayesian,
  title   = {{Bayesian VARs: specification choices and forecast accuracy}},
  author  = {Carriero, Andrea and Clark, Todd E. and Marcellino, Massimiliano},
  journal = {Journal of Applied Econometrics},
  volume  = {30},
  number  = {1},
  pages   = {46--73},
  year    = {2015},
  publisher = {Wiley Online Library}
}

@article{carriero2016common,
  title   = {{Common drifting volatility in large Bayesian VARs}},
  author  = {Carriero, Andrea and Clark, Todd E. and Marcellino, Massimiliano},
  journal = {Journal of Business \& Economic Statistics},
  volume  = {34},
  number  = {3},
  pages   = {375--390},
  year    = {2016},
  publisher = {Taylor \& Francis}
}

@article{carriero2019large,
  title   = {{Large Bayesian vector autoregressions with stochastic volatility and non-conjugate priors}},
  author  = {Carriero, Andrea and Clark, Todd E. and Marcellino, Massimiliano},
  journal = {Journal of Econometrics},
  volume  = {212},
  number  = {1},
  pages   = {137--154},
  year    = {2019},
  publisher = {Elsevier}
}

@article{carriero2024capturing,
  title   = {{Capturing macro-economic tail risks with Bayesian vector autoregressions}},
  author  = {Carriero, Andrea and Clark, Todd E. and Marcellino, Massimiliano},
  journal = {Journal of Money, Credit and Banking},
  volume  = {56},
  number  = {5},
  pages   = {1099--1127},
  year    = {2024},
  publisher = {Wiley Online Library}
}

@article{chamberlain1983arbitrage,
  title   = {{Arbitrage, factor structure, and mean-variance analysis on large asset markets}},
  author  = {Chamberlain, Gary and Rothschild, Michael},
  journal = {Econometrica},
  pages   = {1281--1304},
  year    = {1983},
  publisher = {JSTOR}
}

@article{chan2009efficient,
  title   = {{Efficient simulation and integrated likelihood estimation in state space models}},
  author  = {Chan, Joshua C.~C. and Jeliazkov, Ivan},
  journal = {International Journal of Mathematical Modelling and Numerical Optimisation},
  volume  = {1},
  number  = {1-2},
  pages   = {101--120},
  year    = {2009},
  publisher = {Inderscience Publishers}
}

@article{chan2015marginal,
  title   = {{Marginal likelihood estimation with the cross-entropy method}},
  author  = {Chan, Joshua C.~C. and Eisenstat, Eric},
  journal = {Econometric Reviews},
  volume  = {34},
  number  = {3},
  pages   = {256--285},
  year    = {2015},
  publisher = {Taylor \& Francis}
}

@article{chan2018bayesian,
  title   = {{Bayesian model comparison for time-varying parameter VARs with stochastic volatility}},
  author  = {Chan, Joshua C.~C. and Eisenstat, Eric},
  journal = {Journal of Applied Econometrics},
  volume  = {33},
  number  = {4},
  pages   = {509--532},
  year    = {2018},
  publisher = {Wiley Online Library}
}

@article{chan2023comparing,
  title   = {{Comparing stochastic volatility specifications for large Bayesian VARs}},
  author  = {Chan, Joshua C.~C.},
  journal = {Journal of Econometrics},
  volume  = {235},
  number  = {2},
  pages   = {1419--1446},
  year    = {2023},
  publisher = {Elsevier}
}

@article{chan2023large,
  title   = {{Large Bayesian matrix autoregressions}},
  author  = {Chan, Joshua C.~C. and Qi, Yaling},
  journal = {Journal of Econometrics},
  note    = {forthcoming},
  year    = {2025}
}

@article{chang2023modelling,
  title   = {{Modelling matrix-variate time series via dynamic tensor factor models}},
  author  = {Chang, Jinyuan and Han, Yuefeng and Yu, Rong and Zhang, Bo},
  journal = {Working paper},
  year    = {2023}
}

@article{chen2020constrained,
  title   = {{Constrained factor models for high-dimensional matrix-variate time series}},
  author  = {Chen, Elynn Y. and Tsay, Ruey S. and Chen, Rong},
  journal = {Journal of the American Statistical Association},
  year    = {2020},
  publisher = {Taylor \& Francis}
}

@article{chen2021autoregressive,
  title   = {{Autoregressive models for matrix-valued time series}},
  author  = {Chen, Rong and Xiao, Han and Yang, Dan},
  journal = {Journal of Econometrics},
  volume  = {222},
  number  = {1},
  pages   = {539--560},
  year    = {2021},
  publisher = {Elsevier}
}

@article{chen2022factor,
  title   = {{Factor models for high-dimensional tensor time series}},
  author  = {Chen, Rong and Yang, Dan and Zhang, Cun-Hui},
  journal = {Journal of the American Statistical Association},
  volume  = {117},
  number  = {537},
  pages   = {94--116},
  year    = {2022},
  publisher = {Taylor \& Francis}
}

@article{chen2023statistical,
  title   = {{Statistical inference for high-dimensional matrix-variate factor models}},
  author  = {Chen, Elynn Y. and Fan, Jianqing},
  journal = {Journal of the American Statistical Association},
  volume  = {118},
  number  = {542},
  pages   = {1038--1055},
  year    = {2023},
  publisher = {Taylor \& Francis}
}

@article{chen2023time,
  title   = {{Time-varying matrix factor models}},
  author  = {Chen, Bin and Chen, Elynn Y. and Bolivar, Sebastian and Chen, Rong},
  journal = {arXiv preprint arXiv:2404.01546},
  year    = {2024}
}

@article{chib1994bayes,
  title   = {{Bayes inference in regression models with ARMA (p, q) errors}},
  author  = {Chib, Siddhartha and Greenberg, Edward},
  journal = {Journal of Econometrics},
  volume  = {64},
  number  = {1-2},
  pages   = {183--206},
  year    = {1994},
  publisher = {Elsevier}
}

@article{chib2006analysis,
  title   = {{Analysis of high dimensional multivariate stochastic volatility models}},
  author  = {Chib, Siddhartha and Nardari, Federico and Shephard, Neil},
  journal = {Journal of Econometrics},
  volume  = {134},
  number  = {2},
  pages   = {341--371},
  year    = {2006},
  publisher = {Elsevier}
}

@article{cogley2005drifts,
  title   = {{Drifts and volatilities: monetary policies and outcomes in the post WWII US}},
  author  = {Cogley, Timothy and Sargent, Thomas J.},
  journal = {Review of Economic Dynamics},
  volume  = {8},
  number  = {2},
  pages   = {262--302},
  year    = {2005},
  publisher = {Elsevier}
}

@article{cong2017fast,
  title   = {{Fast simulation of hyperplane-truncated multivariate normal distributions}},
  author  = {Cong, Yulai and Chen, Bo and Zhou, Mingyuan},
  journal = {Bayesian Analysis},
  volume  = {12},
  number  = {4},
  pages   = {1017--1037},
  year    = {2017}
}

@article{cross2016forecasting,
  title   = {{Forecasting structural change and fat-tailed events in Australian macroeconomic variables}},
  author  = {Cross, Jamie and Poon, Aubrey},
  journal = {Economic Modelling},
  volume  = {58},
  pages   = {34--51},
  year    = {2016},
  publisher = {Elsevier}
}

@article{fama1992cross,
  title   = {{The cross-section of expected stock returns}},
  author  = {Fama, Eugene F. and French, Kenneth R.},
  journal = {Journal of Finance},
  volume  = {47},
  number  = {2},
  pages   = {427--465},
  year    = {1992},
  publisher = {Wiley Online Library}
}

@article{giglio2025test,
  title   = {{Test assets and weak factors}},
  author  = {Giglio, Stefano and Xiu, Dacheng and Zhang, Dake},
  journal = {Journal of Finance},
  volume  = {80},
  number  = {1},
  pages   = {259--319},
  year    = {2025},
  publisher = {Wiley Online Library}
}

@article{he2023one,
  title   = {{One-way or two-way factor model for matrix sequences?}},
  author  = {He, Yong and Kong, Xinbing and Trapani, Lorenzo and Yu, Long},
  journal = {Journal of Econometrics},
  volume  = {235},
  number  = {2},
  pages   = {1981--2004},
  year    = {2023},
  publisher = {Elsevier}
}

@article{he2024matrix,
  title   = {{Matrix factor analysis: from least squares to iterative projection}},
  author  = {He, Yong and Kong, Xinbing and Yu, Long and Zhang, Xinsheng and Zhao, Changwei},
  journal = {Journal of Business \& Economic Statistics},
  volume  = {42},
  number  = {1},
  pages   = {322--334},
  year    = {2024},
  publisher = {Taylor \& Francis}
}

@article{jacquier2004bayesian,
  title   = {{Bayesian analysis of stochastic volatility models with fat-tails and correlated errors}},
  author  = {Jacquier, Eric and Polson, Nicholas G. and Rossi, Peter E.},
  journal = {Journal of Econometrics},
  volume  = {122},
  number  = {1},
  pages   = {185--212},
  year    = {2004},
  publisher = {Elsevier}
}

@article{kastner2017efficient,
  title   = {{Efficient Bayesian inference for multivariate factor stochastic volatility models}},
  author  = {Kastner, Gregor and Fr\"uhwirth-Schnatter, Sylvia and Lopes, Hedibert F.},
  journal = {Journal of Computational and Graphical Statistics},
  volume  = {26},
  number  = {4},
  pages   = {905--917},
  year    = {2017},
  publisher = {Taylor \& Francis}
}

@article{koop2013forecasting,
  title   = {{Forecasting with medium and large Bayesian VARs}},
  author  = {Koop, Gary M.},
  journal = {Journal of Applied Econometrics},
  volume  = {28},
  number  = {2},
  pages   = {177--203},
  year    = {2013},
  publisher = {Wiley Online Library}
}

@article{kose2003international,
  title   = {{International business cycles: world, region, and country-specific factors}},
  author  = {Kose, M. Ayhan and Otrok, Christopher and Whiteman, Charles H.},
  journal = {American Economic Review},
  volume  = {93},
  number  = {4},
  pages   = {1216--1239},
  year    = {2003}
}

@article{li2022leverage,
  title   = {{Leverage, asymmetry, and heavy tails in the high-dimensional factor stochastic volatility model}},
  author  = {Li, Mengheng and Scharth, Marcel},
  journal = {Journal of Business \& Economic Statistics},
  volume  = {40},
  number  = {1},
  pages   = {285--301},
  year    = {2022},
  publisher = {Taylor \& Francis}
}

@article{liu2019helping,
  title   = {{Helping effects against curse of dimensionality in threshold factor models for matrix time series}},
  author  = {Liu, Xialu and Chen, Elynn Y.},
  journal = {arXiv preprint arXiv:1904.07383},
  year    = {2019}
}

@article{lopes2004bayesian,
  title   = {{Bayesian model assessment in factor analysis}},
  author  = {Lopes, Hedibert F. and West, Mike},
  journal = {Statistica Sinica},
  pages   = {41--67},
  year    = {2004},
  publisher = {JSTOR}
}

@article{nobile2000comment,
  title   = {{Comment: Bayesian multinomial probit models with a normalization constraint}},
  author  = {Nobile, Agostino},
  journal = {Journal of Econometrics},
  volume  = {99},
  number  = {2},
  pages   = {335--345},
  year    = {2000},
  publisher = {Elsevier}
}

@article{poncela2021factor,
  title   = {{Factor extraction using Kalman filter and smoothing: this is not just another survey}},
  author  = {Poncela, Pilar and Ruiz, Esther and Miranda, Karen},
  journal = {International Journal of Forecasting},
  volume  = {37},
  number  = {4},
  pages   = {1399--1425},
  year    = {2021},
  publisher = {Elsevier}
}

@article{qin2025bayesian,
  title   = {{Bayesian dynamic matrix factor models}},
  author  = {Qin, Lei and Wang, Yan and Zhu, Yan and Shia, Ben-Chang},
  journal = {Journal of Business \& Economic Statistics},
  pages   = {1--13},
  year    = {2025},
  publisher = {Taylor \& Francis}
}

@incollection{sargent1977business,
  title   = {{Business cycle modeling without pretending to have too much a priori economic theory}},
  author  = {Sargent, Thomas J. and Sims, Christopher A. and others},
  booktitle = {New Methods in Business Cycle Research},
  volume  = {1},
  pages   = {145--168},
  year    = {1977}
}

@article{stock2005implications,
  title   = {{Implications of dynamic factor models for VAR analysis}},
  author  = {Stock, James H. and Watson, Mark W.},
  journal = {NBER Working Paper No. 11467},
  year    = {2005}
}

@incollection{stock2012dynamic,
  title   = {{Dynamic factor models}},
  author  = {Stock, James H. and Watson, Mark W.},
  booktitle = {The Oxford Handbook of Economic Forecasting},
  pages   = {35--59},
  year    = {2012},
  publisher = {Oxford University Press}
}

@article{stock2016core,
  title   = {{Core inflation and trend inflation}},
  author  = {Stock, James H. and Watson, Mark W.},
  journal = {Review of Economics and Statistics},
  volume  = {98},
  number  = {4},
  pages   = {770--784},
  year    = {2016},
  publisher = {MIT Press}
}

@article{wang2019factor,
  title   = {{Factor models for matrix-valued high-dimensional time series}},
  author  = {Wang, Dong and Liu, Xialu and Chen, Rong},
  journal = {Journal of Econometrics},
  volume  = {208},
  number  = {1},
  pages   = {231--248},
  year    = {2019},
  publisher = {Elsevier}
}

@article{yu2022projected,
  title   = {{Projected estimation for large-dimensional matrix factor models}},
  author  = {Yu, Long and He, Yong and Kong, Xinbing and Zhang, Xinsheng},
  journal = {Journal of Econometrics},
  volume  = {229},
  number  = {1},
  pages   = {201--217},
  year    = {2022},
  publisher = {Elsevier}
}

@article{yu2024dynamic,
  title   = {{Dynamic matrix factor models for high dimensional time series}},
  author  = {Yu, Ruofan and Chen, Rong and Xiao, Han and Han, Yuefeng},
  journal = {arXiv preprint arXiv:2407.05624},
  year    = {2024}
}

@article{chan2018large,
	title   = {{Large Bayesian VARs: A flexible Kronecker error covariance structure}},
	author  = {Chan, Joshua C.~C.},
	journal = {Journal of Business \& Economic Statistics},
	volume  = {38},
	number  = {1},
	pages   = {68--79},
	year    = {2020},
	publisher = {Taylor \& Francis}
}

@article{banbura2010large,
	title   = {{Large Bayesian vector auto regressions}},
	author  = {Ba{\'n}bura, Marta and Giannone, Domenico and Reichlin, Lucrezia},
	journal = {Journal of Applied Econometrics},
	volume  = {25},
	number  = {1},
	pages   = {71--92},
	year    = {2010},
	publisher = {Wiley Online Library}
}

@article{stock2002macroeconomic,
	author = {Stock, James H. and Watson, Mark W.},
	title = {Macroeconomic Forecasting Using Diffusion Indexes},
	journal = {Journal of Business and Economic Statistics},
	volume = {20},
	number = {2},
	pages = {147--162},
	year = {2002}
}

@article{marcellino2006comparison,
	author = {Marcellino, Massimiliano and Stock, James H. and Watson, Mark W.},
	title = {A Comparison of Direct and Iterated Multistep {AR} Methods for Forecasting Macroeconomic Time Series},
	journal = {Journal of Econometrics},
	volume = {135},
	number = {1--2},
	pages = {499--526},
	year = {2006}
}

@incollection{west2003bayesian,
	author    = {West, Mike},
	title     = {{Bayesian Factor Regression Models in the ``Large $p$, Small $n$'' Paradigm}},
	booktitle = {Bayesian Statistics 7},
	editor    = {Bernardo, J. M. and Bayarri, M. J. and Berger, J. O. and Dawid, A. P. and Heckerman, D. and Smith, A. F. M. and West, M.},
	publisher = {Oxford University Press},
	pages     = {733--742},
	year      = {2003}
}

@article{bernanke2005measuring,
	author  = {Bernanke, Ben S. and Boivin, Jean and Eliasz, Piotr},
	title   = {{Measuring the Effects of Monetary Policy: A Factor-Augmented Vector Autoregressive (FAVAR) Approach}},
	journal = {Quarterly Journal of Economics},
	volume  = {120},
	number  = {1},
	pages   = {387--422},
	year    = {2005},
	doi     = {10.1162/0033553053327452}
}

@article{bai2013principal,
	author  = {Bai, Jushan and Ng, Serena},
	title   = {{Principal Components Estimation and Identification of Static Factors}},
	journal = {Journal of Econometrics},
	volume  = {176},
	number  = {1},
	pages   = {18--29},
	year    = {2013},
	doi     = {10.1016/j.jeconom.2013.03.007}
}

@article{chan2018invariant,
	author  = {Chan, Joshua and Le{\'o}n-Gonz{\'a}lez, Roberto and Strachan, Rodney W.},
	title   = {{Invariant Inference and Efficient Computation in the Static Factor Model}},
	journal = {Journal of the American Statistical Association},
	volume  = {113},
	number  = {522},
	pages   = {819--828},
	year    = {2018},
	doi     = {10.1080/01621459.2017.1287080}
}

@article{fruhwirth2024sparse,
	author  = {Fr{\"u}hwirth-Schnatter, Sylvia and Hosszejni, Darjus and Lopes, Hedibert F.},
	title   = {{Sparse Bayesian Factor Analysis When the Number of Factors is Unknown}},
	journal = {Bayesian Analysis},
	year    = {2024},
	note    = {Advance publication},
	doi     = {10.1214/24-BA1423}
}

@article{fruhwirth2024gltcounts,
	author  = {Fr{\"u}hwirth-Schnatter, Sylvia and Hosszejni, Darjus and Lopes, Hedibert Freitas},
	title   = {{When It Counts --- Econometric Identification of the Basic Factor Model Based on {GLT} Structures}},
	journal = {Econometrics},
	volume  = {12},
	number  = {4},
	pages   = {1--29},
	year    = {2024},
	doi     = {10.3390/econometrics12040026}
}

@article{del2008dynamic,
	author  = {Del Negro, Marco and Otrok, Christopher},
	title   = {{Dynamic Factor Models with Time-Varying Parameters: Measuring Changes in International Business Cycles}},
	journal = {Federal Reserve Bank of New York Staff Reports},
	number  = {326},
	year    = {2008}
}

@article{huber2020multi,
	author  = {Huber, Florian and Pfarrhofer, Michael and Piribauer, Philipp},
	title   = {{A Multi-Country Dynamic Factor Model with Stochastic Volatility for Euro Area Business Cycle Analysis}},
	journal = {Journal of Forecasting},
	volume  = {39},
	number  = {6},
	pages   = {911--926},
	year    = {2020},
	doi     = {10.1002/for.2667}
}

@article{kaufmann2019bayesian,
	author  = {Kaufmann, Sylvia and Schumacher, Christian},
	title   = {{Bayesian Estimation of Sparse Dynamic Factor Models with Order-Independent and Ex-Post Mode Identification}},
	journal = {Journal of Econometrics},
	volume  = {210},
	number  = {1},
	pages   = {116--134},
	year    = {2019},
	doi     = {10.1016/j.jeconom.2018.11.008}
}

@article{leung2016order,
	author  = {Leung, Dennis and Drton, Mathias},
	title   = {{Order-Invariant Prior Specification in {B}ayesian Factor Analysis}},
	journal = {Statistics and Probability Letters},
	volume  = {111},
	pages   = {60--66},
	year    = {2016},
	doi     = {10.1016/j.spl.2016.01.006}
}

@article{geweke1996measuring,
	author  = {Geweke, John and Zhou, Guofu},
	title   = {{Measuring the Pricing Error of the Arbitrage Pricing Theory}},
	journal = {Review of Financial Studies},
	volume  = {9},
	number  = {2},
	pages   = {557--587},
	year    = {1996},
	doi     = {10.1093/rfs/9.2.557}
}

@article{ohagan1995fractional,
	author  = {O'Hagan, Anthony},
	title   = {{Fractional Bayes Factors for Model Comparison}},
	journal = {Journal of the Royal Statistical Society, Series B},
	volume  = {57},
	number  = {1},
	pages   = {99--118},
	year    = {1995}
}

@article{fong2020marginal,
	title={On the marginal likelihood and cross-validation},
	author={Fong, Edwin and Holmes, Chris C},
	journal={Biometrika},
	volume={107},
	number={2},
	pages={489--496},
	year={2020},
	publisher={Oxford University Press}
}

@article{adrian2019vulnerable,
	author  = {Adrian, Tobias and Boyarchenko, Nina and Giannone, Domenico},
	title   = {{Vulnerable Growth}},
	journal = {American Economic Review},
	volume  = {109},
	number  = {4},
	pages   = {1263--1289},
	year    = {2019}
}

@article{chan2025bayesian,
	title={Bayesian model comparison for large Bayesian VARs after the COVID-19 pandemic},
	author={Chan, Joshua CC and Yu, Xuewen and Zhang, Wei},
	journal={Journal of Econometrics},
	pages={106072},
	year={2025},
	publisher={Elsevier}
}

@article{justiniano2008time,
	title={The time-varying volatility of macroeconomic fluctuations},
	author={Justiniano, Alejandro and Primiceri, Giorgio E},
	journal={American Economic Review},
	volume={98},
	number={3},
	pages={604--641},
	year={2008},
	publisher={American Economic Association}
}

@article{doz2012quasi,
	title={A quasi-maximum likelihood approach for large, approximate dynamic factor models},
	author={Doz, Catherine and Giannone, Domenico and Reichlin, Lucrezia},
	journal={Review of Economics and Statistics},
	volume={94},
	number={4},
	pages={1014--1024},
	year={2012},
	publisher={MIT Press}
}

@article{bai2002determining,
	title={Determining the number of factors in approximate factor models},
	author={Bai, Jushan and Ng, Serena},
	journal={Econometrica},
	volume={70},
	number={1},
	pages={191--221},
	year={2002},
	publisher={Wiley}
}

@article{stock2002forecasting,
	title={Forecasting using principal components from a large number of predictors},
	author={Stock, James H. and Watson, Mark W.},
	journal={Journal of the American Statistical Association},
	volume={97},
	number={460},
	pages={1167--1179},
	year={2002},
	publisher={Taylor \& Francis}
}

@article{forni2000generalized,
	title={The generalized dynamic-factor model: identification and estimation},
	author={Forni, Mario and Hallin, Marc and Lippi, Marco and Reichlin, Lucrezia},
	journal={Review of Economics and Statistics},
	volume={82},
	number={4},
	pages={540--554},
	year={2000},
	publisher={MIT Press}
}

@article{chan2017stochastic,
	title={The stochastic volatility in mean model with time-varying parameters: An application to inflation modeling},
	author={Chan, Joshua CC},
	journal={Journal of Business \& Economic Statistics},
	volume={35},
	number={1},
	pages={17--28},
	year={2017},
	publisher={Taylor \& Francis}
}

\clearpage
\appendix
\numberwithin{equation}{section}
\numberwithin{table}{section}
\numberwithin{figure}{section}

\begin{center}
{\LARGE\bfseries Online Supplemental Materials}\\[0.75em]
{\large Bayesian Dynamic Factor Models for High-Dimensional Matrix-Valued Time Series}
\end{center}
\bigskip

\section{Proof of Propositions}\label{app:proofs}\label{app: identification}
\subsection{Proof of \textit{Proposition 1}}

\textit{Proof of Proposition 1}: Without the loss of generality, we prove one of the two cases in Proposition \ref{prop1}. That is, we assume $\var(\bu_t) = \bI_{p_1p_2}$ and $\bA$ is a lower-triangular matrix with ones on the diagonal, while $\bB$ is a lower-triangular matrix with strictly positive diagonal elements.

As shown in \eqref{eq:mdfmrotate}, we may identify a rotation of $\bF_t$, given by $\bC\bF_t\bD'$. 
\begin{equation}\label{eq:mdfmrotate}
    \bY_t = \bA\bC^{-1}\bC\bF_t\bD'(\bD')^{-1}\bB' + \bE_t,
\end{equation}
where $\bC$ and $\bD$ are $p_1\times p_1$ and $p_2\times p_2$ invertible matrices.

We use $\tilde{\bF}_t$ to denote the rotated factor matrix: $\tilde{\bF}_t \equiv\bC\bF_t\bD'$, and we use $\tilde{\bbf}_t$ to denote the vectorized $\bF_t$: $\tilde{\bbf}_t \equiv (\bD\otimes\bC)\bbf_t$.

Let
\begin{equation*}   
 \bA =
\begin{bmatrix}
    1&0&...&0\\
    a_{21}&1&...&0\\
    \vdots&\vdots&\ddots&\vdots\\
    a_{p_11}&a_{p_12}&...&1\\
    \vdots&\vdots&\ddots&\vdots\\
    a_{n1}&a_{n2}&...&a_{np_1}
\end{bmatrix},
\bC^{-1} = 
\begin{bmatrix}
    c_{11}&...&c_{1p_1}\\
    \vdots&\ddots&\vdots\\
    c_{p_11}&...&c_{p_1p_1}
\end{bmatrix}
\end{equation*}

Then the rotated factor loadings $\bA\bC^{-1}$ needs to be a lower triangular matrix with ones on the diagonal as well, that is,
\begin{equation}\label{eq:rotatedloadings}
    \begin{bmatrix}
    1&0&...&0\\
    a_{21}&1&...&0\\
    \vdots&\vdots&\ddots&\vdots\\
    a_{p_11}&a_{p_12}&...&1\\
    \vdots&\vdots&\ddots&\vdots\\
    a_{n1}&a_{n2}&...&a_{np_1}
\end{bmatrix}
\begin{bmatrix}
    c_{11}&...&c_{1p_1}\\
    \vdots&\ddots&\vdots\\
    c_{p_11}&...&c_{p_1p_1}
\end{bmatrix}
=
\begin{bmatrix}
    1&0&...&0\\
    a^*_{21}&1&...&0\\
    \vdots&\vdots&\ddots&\vdots\\
    a^*_{p_11}&a^*_{p_12}&...&1\\
    \vdots&\vdots&\ddots&\vdots\\
    a^*_{n1}&a^*_{n2}&...&a^*_{np_1}
\end{bmatrix}
\end{equation}

For \eqref{eq:rotatedloadings} to hold, we must have $c_{i,j} = 0$ for any $i,j$ such that $i<j$ and $c_{i,i} = 1$, or $\bC^{-1}$ is lower triangular with ones on the diagonal. 

Similarly,
\begin{equation}\label{eq:rotatedloadingsB}
    \begin{bmatrix}
    b_{11}&0&...&0\\
    b_{21}&b_{22}&...&0\\
    \vdots&\vdots&\ddots&\vdots\\
    b_{p_21}&b_{p_22}&...&b_{p_2p_2}\\
    \vdots&\vdots&\ddots&\vdots\\
    b_{n1}&b_{n2}&...&b_{np_2}
\end{bmatrix}
\begin{bmatrix}
    d_{11}&...&d_{1p_2}\\
    \vdots&\ddots&\vdots\\
    d_{p_21}&...&d_{p_2p_2}
\end{bmatrix}
=
\begin{bmatrix}
    b^*_{11}&0&...&0\\
    b^*_{21}&b^*_{22}&...&0\\
    \vdots&\vdots&\ddots&\vdots\\
    b^*_{p_11}&b^*_{p_12}&...&b^*_{p_2p_2}\\
    \vdots&\vdots&\ddots&\vdots\\
    b^*_{n1}&b^*_{n2}&...&b^*_{np_2}
\end{bmatrix}
\end{equation}

For \eqref{eq:rotatedloadingsB} to hold, we must have $d_{ij} = 0$ for any $i,j$ such that $i<j$, or $\bD^{-1}$ is lower triangular given the assumption that $b_{ii}\neq 0$, $b^*_{ii}\neq 0$, for $i=1,...,p_2$.

Define $\bbf_t \equiv \vecf(\bF_t)$. Consider the case $q = 1$, we rewrite \eqref{eq:errorlags} as follows
\begin{equation}\label{eq:vecfactor}
    \bbf_t = \bH_{\vrho}\bbf_{t-1}+\bu_t,
\end{equation}
where $\bH_{\vrho}$ is a diagonal matrix with $\vrho = (\rho_{1,1,t},...,\rho_{p_1,p_2,t})'$ on the diagonal. $\bu_t = (u_{1,1,t},...,u_{p_1,p_2,t})'$, $\bu_t \sim \distn{N}(\bzero, \Lambda_t)$, where $\Lambda_1 = \diag(\lambda^2_{1,1}/(1-\rho_{1,1}^2),...,\lambda^2_{p_1,p_2}/(1-\rho_{p_1,p_2}^2))$ for $t = 1$, and $\Lambda_t = \diag(\lambda^2_{1,1},...,\lambda^2_{p_1,p_2})$ for $t=2,...,T$.

Define $\bM \equiv \bD\otimes\bC$, multiply \eqref{eq:vecfactor} by $\bM$ on both side, we have
\begin{equation}
    \bM\bbf_t = \bM\bH_{\vrho}\bbf_{t-1} + \bM\bu_t.
\end{equation}

Therefore
\begin{equation}
    \tilde{\bbf}_t = \bM\bH_{\vrho}\bM^{-1}\tilde{\bbf}_{t-1}+\bM\bu_t.
\end{equation}

The observation equation after the rotation becomes
\begin{equation}\label{eq:rotation}
    \bY_t = \bA\bC^{-1}\tilde{\bF}_t(\bD')^{-1}\bB'+\bE_t.
\end{equation}


Given the condition that $\var(\bu_t) = \bI_{p_1p_2}$, $\var(\bM\bu_t)$ should be an identity matrix as well. That is, $\bM\var(\bu_t)\bM' = \bI_{p_1p_2}$. Therefore, we have $\bM\bM' = \bI_{p_1p_2}$. Or $\bM$ is an orthogonal matrix. Therefore, we have
\begin{equation*}
    \begin{aligned}
        \bM\bM' = \bI \Leftrightarrow (\bD\otimes\bC)(\bD\otimes\bC)' = \bI \Leftrightarrow (\bD\bD')\otimes(\bC\bC') = \bI,
    \end{aligned}
\end{equation*}
which holds if and only if $\bD\bD' = \bI_{p_2}$ and $\bC\bC' = \bI_{p_1}$, given that C and D are lower triangular matrices and the diagonal elements of $\bC$ is ones.

This means that $\bC$ and $\bD$ are orthogonal matrices. An orthogonal matrix that is lower triangular must be diagonal. Therefore, the rotation matrix $\bC$ is an identity matrix. Given that $b_{ii}>0$ for $i = 1,...,p_2$, we must have that the rotation matrix $\bD$ is also an identity matrix. 

This proves that the proposed assumptions in \textit{MDFM1} fully identify the factor matrix and the factor loading matrices.


\subsection{Proof of \textit{Proposition 2}}

\textit{Proof of Proposition 2}: Similar to the proof of proposition 1, the rotated factor loadings $\bC^{-1}$ needs to be a lower triangular matrix, as shown in \eqref{eq:rotatedloadings}. Additionally, given we have ones on the diagonal of $\bA$, $\bC^{-1}$ needs to have ones on its diagonal as well. Similarly, $\bD^{-1}$ needs to be a lower triangular matrix with ones on its diagonal. Therefore, the matrix $\bM$ is a lower triangular matrix with ones on its diagonal.

Again, we need $\cov(\bM\bu_t) = \cov(\bu_t)$, i.e., $\bM\vLambda_t\bM' = \vLambda_t$, where $\vLambda_t$ is a diagonal matrix. Given that the diagonal elements in $\vLambda_t$ must be larger than 0, this requires that $m_{i,j}$ for all $i>j$ must be zero, for $\bM\vLambda_t\bM'$ to only have non-zero terms on its diagonal and match $\vLambda_t$. Therefore, $\bM$ must be identity matrix.

This proves that assumptions \ref{assumption1}, \ref{assumption4} and \ref{assumption5} fully identifies the factor matrix and the factor loading matrices.

\clearpage

\section{Bayesian Estimation for MDFM with Stochastic Volatility}\label{app: estimation}
Recall the dynamic factor model for matrix-valued time series with stochastic volatility
\begin{align}
    \bY_t &= \bA\bF_t\bB' + \bE_t, \quad \vecf(\bE_t)\sim\distn{N}(\bzero,\omega_t\vSigma_c\otimes\vSigma_r),\label{eq:mdfm1} \\ 
    \vecf(\bF_t) &= \bH_{\vrho_1}\vecf(\bF_{t-1})+\ldots+\bH_{\vrho_q}\vecf(\bF_{t-q})+\bu_t, \quad \bu_t\sim\distn{N} (\bzero,\vLambda_t),\label{eq:errorlags1}
\end{align}
where $\bA$ is a $n\times p_1$ matrix of factor loadings, $\bB$ is a $k \times p_2$ matrix of factor loadings, $\bF_t$ is a $p_1\times p_2$ latent matrix-valued time series of common factors, $\bE_t$ is a $n\times k$ idiosyncratic component, $\vecf(\blank)$ is a vectorizing function, $\bH_{\vrho_l}$ is a diagonal matrix of autoregressive coefficients $(\rho_{1,l},\ldots,\rho_{p_1p_2,l})'$, $l = 1,\ldots,q$, and $\vLambda_t$ is a covariance matrix for the error in factor evolution process. 

We use a natural conjugate prior for the transpose of factor loadings: $\bA'$ and $\bB'$. In addition, we use inverse-Wishart prior for $\vSigma_r$ and $\vSigma_c$:
\begin{equation}
    \begin{aligned}
        \vSigma_r\sim\distn{IW}(\nu_r,\bS_r),\quad (\vecf(\bA')|\vSigma_r)&\sim\distn{N}(\vecf(\bA_0'),\vSigma_r\otimes\bV_{\bA'}),\\
        \vSigma_c\sim\distn{IW}(\nu_c,\bS_c),\quad (\vecf(\bB')|\vSigma_c)&\sim\distn{N}(\vecf(\bB_0'),\vSigma_c\otimes\bV_{\bB'}).
    \end{aligned}
\end{equation}

The autoregressive coefficient $\rho_{j,k,l}$ is assumed to have a truncated normal prior on the interval $(-1,1)$:
\begin{equation*}
    \rho_{j,k,l}\sim\distn{TN}(\rho_{j,k,l,0},V_{\rho_{j,k,l}}), \quad j = 1,...,p_1, \quad k=1,...,p_2,\quad l = 1,\ldots,q.
\end{equation*}

The prior variance $\lambda_{j,k}^2$ is assumed to have a inverse-gamma prior: $\distn{IG}(\nu_{\lambda_{j,k}},S_{\lambda_{j,k}})$. We also treat the first $q$ factors as unknown, and use the following prior
\begin{equation*}   f_{j,k,l}\sim\distn{N}\left(0,\frac{\lambda^2_{j,k}}{1-\sum_{m=1}^q\rho_{j,k,m}^2}\right),\quad l= 1,\ldots,q.
\end{equation*}


For identification, we use assumptions \ref{assumption1}, \ref{assumption4} and \ref{assumption5}. We employ Markov Chain Monte Carlo (MCMC) methods to obtain a draw from the joint posterior of the latent factors and parameters of the model. Specifically, the following steps are carried out:

\textbf{1. Sampling from $(\bA',\vSigma_r|\bY,\bB,\bF,\vSigma_c)$}\\
We sample  $(\bA',\vSigma_r)$ conditional on the latent factors and other parameters from a normal-inverse-Wishart distribution:
\begin{equation*}
    (\bA',\vSigma_r|\blank)\sim\distn{NIW}(\widehat{\bA}', \bK_{\bA'}^{-1},\widehat{\nu}_r,\widehat{\bS}_r),
\end{equation*}
where
\begin{equation*}
\begin{aligned}
    &\bK_{\bA'} = \bV_{\bA'}^{-1}+\sum_{t=1}^T\omega_t^{-1}\bF_t\bB'\vSigma_c^{-1}\bB\bF_t',\quad \widehat{\bA}' = \bK_{\bA'}^{-1}\left(\bV_{\bA'}^{-1}\bA_0'+\sum_{t=1}^T\omega_t^{-1}\bF_t\bB'\vSigma_c^{-1}\bY_t'\right)\\
    &\widehat{\nu}_r = \nu_r+Tk,\quad \widehat{\bS}_r = \bS_r + \bA_0\bV_{\bA'}^{-1}\bA_0'+\sum_{t=1}^T\omega_t^{-1}\bY_t\vSigma_c^{-1}\bY_t'-\widehat{\bA}\bK_{\bA'}\widehat{\bA}'.
\end{aligned}
\end{equation*}

With the constraints for identification, we cannot directly sample from the above normal-inverse-Wishart distribution. Here we outline the sampling scheme for $\bA'$ with the structure constraints. To that end, we first represent the restrictions as a system of linear restrictions. For example, for $\bA'$, we represent the restrictions that $\bA$ is a lower triangular matrix with ones on the diagonal using $\bM_{\bA'}\vecf(\bA')=\ba_0$. Assuming $n>p_1$, $\bM_{\bA'} = (m_{i,j})$ is a $p_1(p_1+1)/2\times np_1$ selection matrix, and $\ba_0$ is a $p_1(p_1+1)/2\times 1$ vector consisting of ones and zeros. Then we apply Algorithm 2 in \cite{cong2017fast} or Algorithm 1 in \cite{chan2023large} to efficiently sample ($\vecf(\bA')|\blank) \sim \distn{N}(\vecf(\widehat{\bA}'),\vSigma_r\otimes\bK_{\bA'}^{-1})$ such that $\bM_{\bA'}\vecf(\bA') = \ba_0$. In particular, one can first sample $\vecf(\bA_u')$ from the unconstrained conditional posterior distribution in Step 1, and then return
\begin{equation*}
    \vecf(\bA') = \vecf(\bA_u')+(\vSigma_r\otimes\bK_{\bA'}^{-1})\bM_{\bA'}'\left(\bM_{\bA'}(\vSigma_r\otimes\bK_{\bA'}^{-1})\bM_{\bA'}'\right)^{-1}(\ba_0-\bM_{\bA'}\vecf(\bA_u')),
\end{equation*}
which can be realized by the following four steps:
\begin{enumerate}[label={(\arabic*)}]
    \item Compute $\bC = \bC_{\vSigma_r^{-1}}\otimes\bC_{\bK_{\bA'}}$, where $\bC_{\vSigma_r^{-1}}$ is the lower Cholesky factor of $\vSigma_r^{-1}$, and $\bC_{\bK_{\bA'}}$ is the lower Cholesky factor of $\bK_{\bA'}$;
    \item Solve $\bC\bC'\bU = \bM_{\bA'}'$ for $\bU$;
    \item Solve $\bM_{\bA'}\bU\bV = \bU'$ for $\bV$;
    \item Return $\vecf(\bA') = \vecf(\bA_u') + \bV'(\ba_0-\bM_{\bA'}\vecf(\bA_u'))$.
\end{enumerate}

\textbf{2. Sampling from $(\bB',\vSigma_c|\bY,\bA,\bF,\vSigma_r)$}\\
Similar to step 1, $(\bB, \vSigma_c)$ are drawn from a normal-inverse-Wishart distribution:
\begin{equation*}
    (\bB, \vSigma_c|\blank)\distn{NIW}(\widehat{\bB}',\bK_{\bB'}^{-1},\widehat{\nu}_c,\widehat{\bS}_{c}),
\end{equation*}
where
\begin{equation*}
    \begin{aligned}
        &\bK_{\bB'} = \bV_{\bB'}^{-1}+\sum_{t=1}^T\omega_t^{-1}\bF_t'\bA'\vSigma_r^{-1}\bA\bF_t,\quad \widehat{\bB}' = \bK_{\bB'}^{-1}\left(\bV_{\bB'}^{-1}\bB_0'+\sum_{t=1}^T\omega_t^{-1}\bF_t'\bA'\vSigma_r^{-1}\bY_t\right)\\
        &\widehat{\nu}_c = \nu_c+Tn,\quad\widehat{\bS}_{c} = \bS_c+\bB_0\bV_{\bB'}^{-1}\bB_0'+\sum_{t=1}^T\omega_t^{-1}\bY_t'\vSigma_r^{-1}\bY_t-\widehat{\bB}\bK_{\bB'}\widehat{\bB}'.
    \end{aligned}
\end{equation*}

We sample $(\bB, \vSigma_c|\blank)$ in two steps. First, we sample $\vSigma_c$ marginally from $(\vSigma_c\given \bY, \bA,\bF, \vSigma_r)\sim\distn{IW}(\widehat{\bS}_c,\nu_c+Tn)$ with the normalization restriction that $\sigma_{c,1,1} = 1$. This can be done using the algorithm in \cite{nobile2000comment} described below. Then we simulate $(\vecf(\bB')\given\bY,\bA,\bF,\vSigma_r,\vSigma_c)\sim\distn{N}(\vecf(\widehat{\bB}),\vSigma_c\otimes\bK^{-1}_{\bB'})$, which can be done using the algorithm described in step 1.

The algorithm in \cite{nobile2000comment} can be realized by the following steps:
\begin{enumerate}[label={(\arabic*)}]
    \item Exchange row/column 1 and $n$ in the matrix $\widehat{\bS}_c$. Denote this matrix as $\widehat{\bS}_c^{Trans}$.
    \item Construct a lower triangular matrix $\vDelta$ such that
    \begin{itemize}
        \item $\delta_{ii}$ equal to the square root of $\chi^2_{\widehat{\nu}_c+1-i}$ for $i=1,\ldots,n-1$;
        \item $\delta_{nn} = (l_{nn})^{-1}$, where $l_{nn}$ is the $(n,n)$-th element in the Cholesky decomposition of $(\widehat{\bS}_c^{Trans})^{-1}$, denoted as $\bL$
        \item $\delta_{ij}$ equal to $\distn{N}(0,1)$ random variates, $i>j$.
    \end{itemize}
    \item Set $\vSigma_c = (\bL^{-1})'(\vDelta^{-1})'\vDelta^{-1}\bL^{-1}$.
    \item Exchange the row/column 1 and $n$ of $\vSigma_c$ back.
\end{enumerate}

\textbf{3. Sampling from $(\vecf(\bF_t)|\bY_t,\bA,\bB,\vSigma_r, \vSigma_c,\vomega^2,\vrho)$}, $t=1,\ldots,T$\\
We sample the factors by $t$. Specifically, conditional on parameters, $\vecf(\bF_t)$ from a normal distribution:
\begin{equation*}
    (\vecf(\bF_t)|\blank)\sim\distn{N}(\widehat{\bbf}_t,\bK_{\bbf_t}^{-1}),
\end{equation*}
where
\begin{equation*} \bK_{\bbf_t}=\omega_t^{-1}\bB'\vSigma_c^{-1}\bB\otimes\bA'\vSigma_r^{-1}\bA+\vLambda^{-1}, \quad \widehat{\bbf}_t = \bK_{\bbf_t}^{-1}\left[\omega_t^{-1}(\bB'\vSigma_c^{-1}\otimes\bA'\vSigma_r^{-1})\vecf(\bY_t)+\vLambda^{-1}\bH_{\vrho}\bbf_{t-1}\right].
\end{equation*}

\textbf{Step 4. Sampling from $(\lambda_{j,k}^2|\bbf_{j,k},\vrho_{j,k}),\quad j=1,...,p_1, k=1,...,p_2$}\\
It is clear that $(\lambda_{j,k}^2|\bbf_{j,k},\vrho_{j,k})\sim\distn{IG}(\widehat{\nu}_{\lambda_{j,k}},\widehat{S}_{\lambda_{j,k}})$, \sloppy where $\widehat{\nu}_{\lambda_{j,k}} = \nu_{\lambda_{j,k}}+\frac{T}{2}$, and $\widehat{S}_{\lambda_{j,k}} = S_{\lambda_{j,k}}+\frac{1}{2}\left[\sum_{t=1}^qf_{j,k,t}^2(1-\sum_m\rho_{j,k,m}^2)+\sum_{t=q+1}^T(f_{j,k,t} - \rho_{j,k,1}f_{j,k,t-1}-...-\rho_{j,k,q}f_{j,k,q})^2\right]$.

\textbf{Step 5. Sampling from $(\vrho_{j,k}|\bbf_{j,k},\lambda^2_{j,k}),\quad j=1,...,p_1, k=1,...,p_2$}\\
Note that $\vrho_{j,k}$ is a $q\times 1$ vector: $\vrho_{j,k} = (\rho_{j,k,1},\ldots,\rho_{j,k,q})'$. We rewrite \eqref{eq:errorlags} as follows:
\begin{equation}
    \tilde{\bbf}_{j,k} = \tilde{\bF}_{j,k}\vrho_{j,k}+\bu_{j,k},\quad \bu_{j,k}\sim\distn{N}(\bzero,\lambda_{j,k}\bI_{T-q}),
\end{equation}
where $\tilde{\bbf}_{j,k} = (f_{j,k,q+1},\ldots,f_{j,k,T})'$, and
\begin{equation*}
    \tilde{\bF}_{j,k} =
    \begin{bmatrix}
        f_{j,k,1}&f_{j,k,2}&\cdots&f_{j,k,q}\\
        f_{j,k,2}&f_{j,k,3}&\cdots&f_{j,k,q+1}\\
        \vdots&\cdots&\cdots&\vdots\\
        f_{j,k,T-q}&f_{j,k,T-q+1}&\cdots&f_{j,k,T}
    \end{bmatrix}.
\end{equation*}

Following \cite{chib1994bayes} and \cite{chan2009efficient}, we design an Metropolis-Hastings algorithm with proposal $\vrho_{j,k}^*\sim\distn{N}(\hat{\vrho}_{j,k},\bK_{\vrho_{j,k}}^{-1})$, where $\bK_{\vrho_{j,k}} = \bV_{\vrho_{j,k}}^{-1}+\tilde{\bF}'_{j,k}\tilde{\bF}_{j,k}/\lambda_{j,k}^2$, $\hat{\vrho}_{j,k}=\bK_{\vrho_{j,k}}^{-1}(\bV_{\vrho_{j,k}}^{-1}\vrho_{j,k,0}+\tilde{\bF}_{j,k}'\tilde{\bbf}_{j,k}/\lambda_{j,k}^2)$. The proposed value $\vrho_{j,k}^*$ is accepted with probablity
\begin{equation*}
    \alpha_{MH}(\vrho_{j,k},\vrho_{j,k}^*) = \min\left\{1,\frac{f_{\distn{N}}(\bbf_{j,k,1:q}|\bzero,\lambda_{j,k}^2/(1-\sum_m\rho_{j,k,m}^{*2})\bI_q)}{f_{\distn{N}}(\bbf_{j,k,1:q}|\bzero,\lambda_{j,k}^2/(1-\sum_m\rho_{j,k,m}^{2})\bI_q)}\right\}.
\end{equation*}
\textbf{5. Sampling the time-varying volatility}\\
For clearer illustration, assume that we have only one type of time-varying volatility. The following three steps correspond to each type.

\textbf{5.1 Common stochastic volatility: sampling from $(\bh|\bY,\bA,\bF,\bB,\vSigma_c,\vSigma_r)$}\\
The conditional posterior for $\bh$ is not a standard distribution. In this paper, we follow \cite{chan2017stochastic} for this purpose. In particular, we first obtain the mode of the log density of $(\bh|\blank)$ as well as the negative Hessian evaluated at the mode, denoted as $\widehat{\bh}$ and $\bK_{\bh}$, respectively. Then we use $\distn{N}(\widehat{\bh},\bK_{\bh}^{-1})$ as the proposal distribution, and sample $\bh$ using an acceptance-rejection Metropolis-Hasting step. Samplers for $\phi$ and $\sigma_h^2$ are standard and we omit the details in this paper.

\textbf{5.2 The explicit outlier components: sampling from $(\bo,p_{\bo}|\bY,\bA,\bF,\bB,\vSigma_c,\vSigma_r)$}\\
We follow \cite{stock2016core} to discretize the support of $o_{t}$ to simplify estimation. Specifically, we use a grid with points at $1,2,3,...,20$. The likelihood can be easily evaluated at these grid points. Finally, a draw from the full conditional posterior distribution of $o_{t}$ can be obtained using the inverse transform method. 

The conditional distribution of $p_{o_i}$ is a Beta distribution:
\begin{equation*}
    (p_{o_i}|\bo_i)\sim\distn{B}(a_{p_{o_i}}+n_2, b_{p_{o_i}}+n_1),
\end{equation*}
where $n_1 = \sum_{t=1}^T\bI(o_{i,t}=1)$ is the number of ``regular" periods, and $n_2 = T-\sum_{t=1}^T\bI(o_{i,t}=1)$ is the number of ``outlier" periods.

\textbf{5.3 Fat-tailed innovations: sampling from $(q_t^2|\bY,\bA,\bF,\bB,\vSigma_c,\vSigma_r)$, $t= 1,\ldots,T$}
Conditional on the factors and parameters, the posterior for $q^2_t$ has an inverse-gamma distribution:
\begin{equation*}   (q^2_t|\blank)\sim\distn{IG}((nk+l)/2,(s_t^2+l)/2),
\end{equation*}
where $s_t^2 = \tr\left[\vSigma_c^{-1}(\bY_t-\bA\bF_t\bB)'\vSigma_r^{-1}(\bY_t-\bA\bF_t\bB)\right]$.

\clearpage

\section{Estimating Marginal Likelihoods}\label{app: ml}
This section describes the method we use to obtain integrated likelihood and the importance-sampling densities and the choice of the importance sampling density. For illustration, we consider $q = 1$.
\subsection{Integrated Likelihood}
Model \eqref{eq:mdfm}--\eqref{eq:errorlags} can be rewritten as follows
\begin{equation}\label{eq:yt}
    \begin{aligned}
        &\by_t = (\bB\otimes\bA)\bbf_t+\vepsilon_t,\quad \vepsilon_t\sim\distn{N}(\bzero,\vSigma_c\otimes\vSigma_r),\\
        &\bbf\given\vrho,\vOmega \sim\distn{N}\left(\bzero,\left[\bH_{\vrho}'(\bI_T\otimes\vOmega)^{-1}\bH_{\vrho}\right]^{-1}\right),
    \end{aligned}
\end{equation}

System \eqref{eq:yt} can be rewritten as follows
\begin{equation}
\begin{aligned}
&\by = (\bI_T\otimes\bA\otimes\bB)\bbf + \vepsilon,\quad\vepsilon\sim\distn{N}(\bzero,\bI_T\otimes(\vSigma_c\otimes\vSigma_r)),\\
&\bbf\given\vrho,\vOmega \sim\distn{N}\left(\bzero,\left[\bH_{\vrho}'(\bI_T\otimes\vOmega)^{-1}\bH_{\vrho}\right]^{-1}\right). 
\end{aligned}   
\end{equation}

It is easy to integrate out $\bbf$ and we can get the following likelihood
\begin{equation}
    \by\given\bA,\bB, \vSigma_c,\vSigma_r,\vOmega,\vrho\sim\distn{N}(\widebar{\by},\widebar{\bD}_{\by}),
\end{equation}
where 
\begin{equation*}
    \begin{aligned}
        \widebar{\by} &= \Em\left[\Em\left(\by\given\bbf,\bA,\bB, \vSigma_c,\vSigma_r,\vOmega,\vrho\right)\given\bA,\bB, \vSigma_c,\vSigma_r,\vOmega,\vrho\right]\\
        &=\Em\left[(\bI_T\otimes\bB\otimes\bA)\bbf\given\bA,\bB, \vSigma_c,\vSigma_r,\vOmega,\vrho\right]\\
        &=(\bI_T\otimes\bB\otimes\bA)\Em[\bbf\given\bA,\bB, \vSigma_c,\vSigma_r,\vOmega,\vrho]\\
        &=\bzero,
    \end{aligned}
\end{equation*}

and
\begin{equation*}
    \begin{aligned}
        \widebar{\bD}_{\by} &= \Em\left\{\left[\Var\left(\by\given\bbf,\bA,\bB, \vSigma_c,\vSigma_r,\vOmega,\vrho\right)\given\blank\right\}+\Var\left(\Em[\by\given\bbf,\bA,\bB, \vSigma_c,\vSigma_r,\vOmega]\given\blank\right)\right]\\
        &= \bI_T\otimes\vSigma_c\otimes\vSigma_r+(\bI_T\otimes\bB\otimes\bA)[\bH_{\rho}'(\bI_T\otimes\vOmega)^{-1}\bH_{\rho}]^{-1}(\bI_T\otimes\bB'\otimes\bA').
    \end{aligned}
\end{equation*}

It can be very costly to compute the inverse of the covariance matrix $\widebar{\bD}_{\by}$. Therefore, here we use Kalman filter. In particular, it is not difficult to show that the marginal distribution for $\bbf_t\equiv \vecf(\bF_t)$ is as follows:
\begin{equation*}
    \begin{aligned}
        (\bbf_1 \given \vrho,\vlambda) &\sim\distn{N}(\bzero,\vLambda_1)\\
        (\bbf_t\given \vrho,\vlambda)&\sim\distn{N}(\bzero,\vLambda_t+\bH_{\vrho}\vLambda_{t-1}\bH_{\vrho}'), \quad t =2,\ldots,T, \\
    \end{aligned}
\end{equation*}
where for $t = 2,\ldots,T$, $\vLambda_t = \diag(\lambda_{1,1}^2,\lambda_{2,1}^2,...,\lambda_{p_1,p_2}^2)$, and for $t=1$, $\vLambda_1 = \diag(\lambda_{1,1}^2/(1-\rho_{1,1}^2),\lambda_{2,1}^2/(1-\rho_{2,1}^2),...,\lambda_{p_1,p_2}^2/(1-\rho_{p_1,p_2}^2))$. $\bH_{\vrho} = \diag(\rho_{1,1},\rho_{2,1},...,\rho_{p_1,p_2})$

Therefore, the integrated likelihood at time $t$ is:
\begin{equation*}
    (\by_t\given\bA,\bB,\vSigma_c,\vSigma_r) \sim\distn{N}(\bzero,\widebar{\bD}_{\by_t}),
\end{equation*}
where 
\begin{equation*}
    \begin{aligned}
        &\widebar{\bD}_{\by_1} = \vSigma_c\otimes\vSigma_r+(\bB\otimes\bA)\vLambda_1(\bB'\otimes\bA')\\
        &\widebar{\bD}_{\by_t} = \vSigma_c\otimes\vSigma_r+(\bB\otimes\bA)(\vLambda_t +\bH_{\vrho}\vLambda_{t-1}\bH_{\vrho}')(\bB'\otimes\bA'),\quad t = 2,\ldots,T.
    \end{aligned}
\end{equation*}

\subsection{Finding the Optimal Importance-sampling Densities}
The next step is to find the maximum likelihood estimators for the hyperparameters in the importance-sampling density. The importance-sampling density is denoted as
\begin{equation}
\begin{aligned}
        f(\vtheta;\bv) &= f(\bA,\vSigma,\vOmega,\vrho;\bv) \\
        &= f(\bA;\widebar{\bA},\widebar{\bD}_{\bA})\cdot f(\vSigma_c;\Psi_c, \nu_c)\cdot    f(\vSigma_r;\Psi_r, \nu_r)\cdot f(\vlambda;\nu_{\lambda},S_{\lambda})\cdot f(\vrho;\widebar{\vrho},\widebar{\bD}_{\vrho}).
\end{aligned}
\end{equation}

In terms of the parameteric family, we use Gaussian density for $f(\bA;\widebar{\bA},\widebar{\bD}_{\bA})$, where $\widebar{\bA}$ and $\widebar{\bD}$ are the corresponding mean and covariance matrix. We use inverse Wishart densities for $f(\vSigma_c;\nu_c,\Psi_c)$ as well as $f(\vSigma_r;\nu_r,\Psi_r)$. We use inverse gamma density for We use the truncated normal density on the interval $(-1, 1)$ for $f(\vrho;\widebar{\vrho},\widebar{\bD}_{\vrho})$, where $\widebar{\vrho}$ and $\widebar{\bD}_{\vrho}$ are the corresponding mean and covariance matrix. we use inverse-gamma distribution for $f(\vlambda;\nu_{\lambda},S_{\lambda})$. 

In order to obtain the maximum likelihood estimators for the parameters in inverse Wishart distribution, we first use maximum likelihood estimation on the Wishart distribution given the posterior samples, and then compute the degree of freedom and scale matrix of the inverse Wishart distribution using 
\textit{Lemma 1}.

\textit{Lemma 1}: $\vSigma$ follows an inverse Wishart distribution if $\bK \equiv \vSigma^{-1}$ follows a Wishart distribution, formally expressed as
\begin{equation}  \vSigma\sim\distn{IW}_d(\delta-d+1,\vPsi^{-1}) \Leftrightarrow \bK = \vSigma^{-1}\sim\distn{W}_d(\delta,\vPsi),
\end{equation}
where $d$ is the dimension of the matrix $\vSigma$, $\delta$ is the degree of freedom of the Wishart distribution, and $\vPsi$ is the scale matrix. 

A Wishart distribution is defined as:
\begin{equation*}
    f(\bK\given\vPsi,\delta) = \frac{\lvert\bK\rvert^{\frac{\delta-d-1}{2}}}{2^{\frac{\delta d}{2}}\lvert\vPsi\rvert^{\frac{\delta}{2}}\Gamma_d\left(\frac{\delta}{2}\right)}\exp\left\{-\frac{1}{2}\tr(\bK\vPsi^{-1})\right\}.
\end{equation*}

We assume that each matrix is drawn independently from the same Wishart distribution $\distn{W}(\vPsi,\delta)$. Therefore, we can model the joint distribution as:
\begin{equation*}
    f(\bK_1,...,\bK_M\given\vPsi,\delta)=\prod_{m=1}^M\frac{\lvert\bK_m\rvert^{\frac{\delta-d-1}{2}}}{2^{\frac{\delta d}{2}}\lvert\vPsi\rvert^{\frac{\delta}{2}}\Gamma_d\left(\frac{\delta}{2}\right)}\exp\left\{-\frac{1}{2}\tr(\bK_m\vPsi^{-1})\right\}.
\end{equation*}

The log-likelihood function is therefore
\begin{equation*}
\begin{aligned}
    \log f(\bK_1,...,\bK_M\given\vPsi,\delta) = &-\frac{\delta d M}{2}\log 2 - \frac{\delta M}{2}\log\lvert\vPsi\rvert-M\log\Gamma_d\left(\frac{\delta}{2}\right)+\\
    &\frac{\delta-d-1}{2}\sum_{m=1}^M\log\lvert\bK_m\rvert-\frac{1}{2}\tr\left(\sum_{m=1}^M\bK_m\vPsi^{-1}\right).
\end{aligned}
\end{equation*}

The first derivative of the log-likelihood function with respect to the scale matrix $\vPsi$ is equal to
\begin{equation} \label{eq: PsiMLE}
\frac{\di\log(f(\bK_1,...,\bK_M\given\vPsi,\delta))}{\di\vPsi} = -\frac{M\delta}{2}\vPsi^{-1}+\frac{1}{2}\vPsi^{-1}\sum_{m=1}^M\bK_m\vPsi^{-1},
\end{equation}
where two results are used
\begin{enumerate}
    \item $\frac{\partial \lvert\bX\rvert}{\partial \bX} = \lvert\bX\rvert\bX^{-1}$;
    \item $\frac{\partial \tr(\bA\bX^{-1})}{\partial \bX}=-\bX^{-1}\bA\bX^{-1}$.
\end{enumerate}

From equation \eqref{eq: PsiMLE} we obtain a function of the MLE of $\vPsi$ with respect to the degree of freedom $\delta$
\begin{equation}\label{eq:Psi}
    \widehat{\vPsi}^{mle} = \frac{1}{M\delta}\sum_{m=1}^M\bK_m.
\end{equation}

In order to obtain the MLE for the degree of freedom, a straightforward way is to find the first order condition and second order condition to maximize the log-likelihood function with respect to $\delta$. We then use the Newton-type methods to find the estimate for $\widehat{\delta}$. 

In particular, the first derivative of the log-likelihood function after we plug in \eqref{eq:Psi} is
\begin{equation}
\begin{aligned}
    \frac{\partial \log f(\bK_1,...,\bK_M\given\delta)}{\partial \delta} = &-\frac{dM}{2}(\log 2 + 1)+\frac{Md}{2}\log\delta-\frac{M}{2}\log\left|M^{-1}\sum_m\bK_m\right|\\
    &-\frac{M}{2}\psi_d\left(\frac{\delta}{2}\right)+\frac{1}{2}\sum_m\log\lvert\bK_m\rvert.
\end{aligned}   
\end{equation}

The second derivative is
\begin{equation*}
    \frac{\partial^2 \log f(\bK_1,...,\bK_M\given\delta)}{\partial \delta^2} =-\frac{Md}{2\delta}-\frac{M}{4}\psi_d^{(2)}\left(\frac{1}{2}\delta\right).
\end{equation*}
Maximum likelihood estimators for parameters for normal distributions and inverse gamma distributions are straightforward to obtain so that we omit the details here. 

\clearpage

\section{Additional Simulation Results}\label{app:simulation}
\begin{table}[H]
  \centering
    \caption{Adjusted $R^2$ from regressing the true factors on the estimates: $p_1 = 3$, $p_2 = 2$}
    \begin{tabular}{lcccccccc}
    \hline
    \hline
    \rowcolor[rgb]{ .514,  .8,  .922} (n, k) & \multicolumn{2}{c}{T = 200} &       & \multicolumn{2}{c}{T = 500} &       & \multicolumn{2}{c}{T = 1000} \\
    \hline
    \rowcolor[rgb]{ .922,  .992,  1}  & 0.98  & 0.98  &       & 0.99  & 0.97  &       & 0.98  & 0.99 \\
    \rowcolor[rgb]{ .922,  .992,  1}       & 0.98  & 0.96  &       & 0.98  & 0.98  &       & 0.98  & 0.99 \\
    \rowcolor[rgb]{ .922,  .992,  1}       \multirow{-3}[1]{*}{(10, 10)}& 0.96  & 0.96  &       & 0.99  & 0.99  &       & 0.98  & 0.99 \\
    \rowcolor[rgb]{ .514,  .8,  .922} Average & \multicolumn{2}{c}{0.97} &       & \multicolumn{2}{c}{0.98} &       & \multicolumn{2}{c}{0.98} \\
    \hline
    \rowcolor[rgb]{ .922,  .992,  1}  & 1.00  & 0.98  &       & 1.00  & 0.99  &       & 1.00  & 0.99 \\
    \rowcolor[rgb]{ .922,  .992,  1}       & 0.97  & 0.98  &       & 0.99  & 0.98  &       & 1.00  & 0.99 \\
    \rowcolor[rgb]{ .922,  .992,  1}      \multirow{-3}[1]{*}{(20, 15)} & 0.99  & 0.99  &       & 0.99  & 0.97  &       & 0.98  & 0.98 \\
    \rowcolor[rgb]{ .514,  .8,  .922} Average & \multicolumn{2}{c}{0.98} &       & \multicolumn{2}{c}{0.99} &       & \multicolumn{2}{c}{0.99} \\
    \hline
    \rowcolor[rgb]{ .922,  .992,  1}  & 1.00  & 0.98  &       & 1.00  & 1.00  &       & 1.00  & 0.99 \\
    \rowcolor[rgb]{ .922,  .992,  1}       & 0.99  & 0.98  &       & 0.99  & 0.99  &       & 0.99  & 0.99 \\
    \rowcolor[rgb]{ .922,  .992,  1}      \multirow{-3}[1]{*}{(30, 20)} & 0.99  & 0.98  &       & 0.98  & 0.98  &       & 0.99  & 0.99 \\
    \rowcolor[rgb]{ .514,  .8,  .922} Average & \multicolumn{2}{c}{0.99} &       & \multicolumn{2}{c}{0.99} &       & \multicolumn{2}{c}{0.99} \\
    \hline
    \hline
    \end{tabular}%
\end{table}%

\begin{table}[H]
  \centering
  \caption{Adjusted $R^2$ from regressing the true factors on the estimates: $p_1 = 5$, $p_2 = 5$}
  \resizebox{\textwidth}{!}{
    \begin{tabular}{cccccccccccccccccc}
    \hline
    \hline
    \rowcolor[rgb]{ .514,  .8,  .922} $(n, k)$& \multicolumn{5}{c}{$T=200$}             &       & \multicolumn{5}{c}{$T=500$}             &       & \multicolumn{5}{c}{$T=1000$} \\
    \hline
    \rowcolor[rgb]{ .922,  .992,  1}  & 0.97  & 0.98  & 0.95  & 0.97  & 0.97  &       & 0.99  & 0.98  & 0.97  & 0.95  & 0.97  &       & 0.99  & 0.99  & 0.99  & 0.98  & 0.98 \\
     \rowcolor[rgb]{ .922,  .992,  1}      & 0.96  & 0.97  & 0.94  & 0.97  & 0.97  &       & 0.98  & 0.97  & 0.97  & 0.93  & 0.97  &       & 0.98  & 0.97  & 0.98  & 0.97  & 0.98 \\
      \rowcolor[rgb]{ .922,  .992,  1}   & 0.97  & 0.96  & 0.97  & 0.97  & 0.97  &       & 0.97  & 0.96  & 0.95  & 0.92  & 0.98  &       & 0.98  & 0.97  & 0.98  & 0.98  & 0.98 \\
    \rowcolor[rgb]{ .922,  .992,  1}     & 0.96  & 0.96  & 0.93  & 0.95  & 0.96  &       & 0.97  & 0.98  & 0.96  & 0.94  & 0.98  &       & 0.96  & 0.96  & 0.95  & 0.95  & 0.97 \\
\rowcolor[rgb]{ .922,  .992,  1} \multirow{-5}[1]{*}{(10, 10)}         & 0.95  & 0.96  & 0.95  & 0.93  & 0.95  &       & 0.98  & 0.97  & 0.96  & 0.91  & 0.96  &       & 0.99  & 0.98  & 0.98  & 0.97  & 0.98 \\
    \rowcolor[rgb]{ .514,  .8,  .922} Average & \multicolumn{5}{c}{0.96}              &       & \multicolumn{5}{c}{0.96}              &       & \multicolumn{5}{c}{0.98} \\
    \hline
    \rowcolor[rgb]{ .922,  .992,  1} & 0.99  & 0.99  & 0.99  & 0.99  & 0.96  &       & 0.99  & 0.98  & 0.99  & 0.98  & 0.99  &       & 0.99  & 0.99  & 0.99  & 0.99  & 0.97 \\
        \rowcolor[rgb]{ .922,  .992,  1}  & 0.99  & 0.99  & 0.99  & 0.98  & 0.94  &       & 0.99  & 0.97  & 0.99  & 0.97  & 0.99  &       & 0.99  & 0.99  & 0.99  & 0.99  & 0.98 \\
      \rowcolor[rgb]{ .922,  .992,  1}    & 0.99  & 0.99  & 0.99  & 0.98  & 0.96  &       & 0.99  & 0.98  & 0.98  & 0.97  & 0.99  &       & 0.99  & 0.99  & 0.99  & 0.99  & 0.97 \\
      \rowcolor[rgb]{ .922,  .992,  1}    & 0.98  & 0.98  & 0.98  & 0.98  & 0.98  &       & 0.99  & 0.95  & 0.98  & 0.97  & 0.98  &       & 0.99  & 0.99  & 0.99  & 0.99  & 0.97 \\
    \rowcolor[rgb]{ .922,  .992,  1}    \multirow{-5}[1]{*}{(20, 15)} & 0.98  & 0.97  & 0.98  & 0.97  & 0.95  &       & 0.98  & 0.97  & 0.96  & 0.96  & 0.96  &       & 0.99  & 0.98  & 0.99  & 0.98  & 0.97 \\
    \rowcolor[rgb]{ .514,  .8,  .922} Average & \multicolumn{5}{c}{0.98}              &       & \multicolumn{5}{c}{0.98}              &       & \multicolumn{5}{c}{0.99} \\
    \hline
    \rowcolor[rgb]{ .922,  .992,  1}  & 1.00  & 1.00  & 0.99  & 0.99  & 0.92  &       & 1.00  & 0.99  & 1.00  & 0.99  & 0.99  &       & 0.99  & 0.99  & 0.99  & 0.99  & 0.99 \\
        \rowcolor[rgb]{ .922,  .992,  1}   & 0.99  & 0.99  & 0.99  & 0.99  & 0.97  &       & 0.99  & 1.00  & 0.99  & 0.99  & 0.99  &       & 1.00  & 0.99  & 0.99  & 0.99  & 0.99 \\
       \rowcolor[rgb]{ .922,  .992,  1}     & 0.99  & 0.99  & 0.99  & 0.97  & 0.96  &       & 1.00  & 0.99  & 0.99  & 0.98  & 0.99  &       & 0.98  & 0.99  & 0.99  & 0.98  & 0.99 \\
        \rowcolor[rgb]{ .922,  .992,  1}  & 0.97  & 0.98  & 0.99  & 0.99  & 0.91  &       & 0.98  & 0.99  & 0.99  & 0.97  & 0.99  &       & 1.00  & 0.99  & 0.99  & 0.99  & 0.99 \\
        \rowcolor[rgb]{ .922,  .992,  1} \multirow{-5}[1]{*}{(30, 20)} & 0.99  & 0.99  & 0.98  & 0.99  & 0.97  &       & 0.97  & 0.99  & 0.98  & 0.97  & 0.98  &       & 0.99  & 0.99  & 0.99  & 0.99  & 0.99 \\
    \rowcolor[rgb]{ .514,  .8,  .922} Average & \multicolumn{5}{c}{0.98}              &       & \multicolumn{5}{c}{0.99}              &       & \multicolumn{5}{c}{0.99} \\
    \hline
    \hline
    \end{tabular}}
\end{table}%

\clearpage

\section{Data: Multinational Macroeconomic Panel}\label{app:data_app1}
Table \ref{tab:list} describes the list of variables we use for the first application. We attach the link of the website we downloaded the specific variable to the variable name in the table. The second column of \ref{tab:list} is the stationarity transformation for each variable.

\begin{table}[H]
  \centering
  \caption{List of variables}
  \label{tab:list}
    \begin{tabular}{lc}
    \toprule
    \toprule
    Variable & Transformation \\
    \midrule
    \href{https://data-explorer.oecd.org/vis?df[ds]=dsDisseminateFinalDMZ&df[id]=DSD_NAMAIN1%40DF_QNA_EXPENDITURE_GROWTH_OECD&df[ag]=OECD.SDD.NAD&df[vs]=1.0&pd=1990-Q1%2C2024-Q1&dq=Q..OECD%2BTUR%2BG7%2BUSA%2BGBR%2BCHE%2BSWE%2BESP%2BPRT%2BNOR%2BNZL%2BNLD%2BMEX%2BLUX%2BKOR%2BJPN%2BITA%2BISR%2BDEU%2BFRA%2BFIN%2BDNK%2BCAN%2BAUT%2BAUS...B1GQ......G1.&to[TIME_PERIOD]=false&vw=ov}{Real GDP} & No transformation \\
    \href{https://data-explorer.oecd.org/vis?df[ds]=dsDisseminateFinalDMZ&df[id]=DSD_NAMAIN1%40DF_QNA_EXPENDITURE_USD&df[ag]=OECD.SDD.NAD&df[vs]=1.1&dq=Q..OECD%2BG7%2BUSA%2BGBR%2BCHE%2BSWE%2BESP%2BPRT%2BNOR%2BNLD%2BNZL%2BMEX%2BLUX%2BKOR%2BJPN%2BITA%2BIRL%2BDEU%2BFRA%2BFIN%2BDNK%2BCAN%2BAUT%2BAUS.S1M..P3.....LR..&pd=1990-Q1%2C2024-Q2&to[TIME_PERIOD]=false}{Consumption} & $\Delta\log(x)$ \\
    \href{https://data-explorer.oecd.org/vis?fs[0]=Frequency%20of%20observation%2C0%7CQuarterly%23Q%23&fs[1]=Topic%2C1%7CEconomy%23ECO%23%7CProductivity%23ECO_PRO%23&pg=0&fc=Topic&snb=1&df[ds]=dsDisseminateFinalDMZ&df[id]=DSD_PDB%40DF_PDB_ULC_Q&df[ag]=OECD.SDD.TPS&df[vs]=1.0&dq=OECD%2BUSA%2BGBR%2BCHE%2BSWE%2BESP%2BPRT%2BNOR%2BNZL%2BNLD%2BLUX%2BKOR%2BJPN%2BITA%2BIRL%2BDEU%2BFRA%2BFIN%2BDNK%2BCAN%2BAUT%2BAUS.Q.ULCE..IX...S.&pd=1990-Q1%2C2024-Q1&to[TIME_PERIOD]=false}{Labor unit costs} & $\Delta x$ \\
    \href{https://data-explorer.oecd.org/vis?fs[0]=Topic%2C1%7CEmployment%23JOB%23%7CUnemployment%20indicators%23JOB_UNEMP%23&fs[1]=Reference%20area%2C0%7CAustralia%23AUS%23&fs[2]=Reference%20area%2C0%7CAustria%23AUT%23&fs[3]=Reference%20area%2C0%7CCanada%23CAN%23&fs[4]=Reference%20area%2C0%7CSwitzerland%23CHE%23&fs[5]=Reference%20area%2C0%7CGermany%23DEU%23&fs[6]=Reference%20area%2C0%7CDenmark%23DNK%23&fs[7]=Reference%20area%2C0%7CSpain%23ESP%23&fs[8]=Reference%20area%2C0%7CFinland%23FIN%23&fs[9]=Reference%20area%2C0%7CFrance%23FRA%23&fs[10]=Reference%20area%2C0%7CG7%23G7%23&fs[11]=Reference%20area%2C0%7CUnited%20Kingdom%23GBR%23&fs[12]=Reference%20area%2C0%7CGreece%23GRC%23&fs[13]=Reference%20area%2C0%7CIreland%23IRL%23&fs[14]=Reference%20area%2C0%7CIsrael%23ISR%23&fs[15]=Reference%20area%2C0%7CItaly%23ITA%23&fs[16]=Reference%20area%2C0%7CJapan%23JPN%23&fs[17]=Reference%20area%2C0%7CKorea%23KOR%23&fs[18]=Reference%20area%2C0%7CLuxembourg%23LUX%23&fs[19]=Reference%20area%2C0%7CMexico%23MEX%23&fs[20]=Reference%20area%2C0%7CNetherlands%23NLD%23&fs[21]=Reference%20area%2C0%7CNorway%23NOR%23&fs[22]=Reference%20area%2C0%7CNew%20Zealand%23NZL%23&fs[23]=Reference%20area%2C0%7COECD%23OECD%23&fs[24]=Reference%20area%2C0%7CPortugal%23PRT%23&fs[25]=Reference%20area%2C0%7CSweden%23SWE%23&fs[26]=Reference%20area%2C0%7CUnited%20States%23USA%23&fs[27]=Frequency%20of%20observation%2C0%7CQuarterly%23Q%23&pg=0&fc=Frequency%20of%20observation&snb=6&vw=tb&df[ds]=dsDisseminateFinalDMZ&df[id]=DSD_LFS%40DF_IALFS_UNE_M&df[ag]=OECD.SDD.TPS&df[vs]=1.0&dq=AUS%2BAUT%2BCAN%2BCHE%2BDEU%2BDNK%2BESP%2BFIN%2BFRA%2BG7%2BGBR%2BGRC%2BIRL%2BISR%2BITA%2BJPN%2BKOR%2BLUX%2BMEX%2BNLD%2BNOR%2BNZL%2BOECD%2BPRT%2BSWE%2BUSA..._Z.Y._T.Y_GE15..Q&pd=1990-Q1%2C2024-Q2&to[TIME_PERIOD]=false&ly[cl]=TIME_PERIOD&ly[rw]=REF_AREA}{Unemployment} & $\Delta x$ \\
    \href{https://data-explorer.oecd.org/vis?fs[0]=Topic%2C1%7CEconomy%23ECO%23%7CPrices%23ECO_PRI%23&pg=0&fc=Topic&bp=true&snb=16&df[ds]=dsDisseminateFinalDMZ&df[id]=DSD_PRICES%40DF_PRICES_ALL&df[ag]=OECD.SDD.TPS&df[vs]=1.0&pd=1950-Q1%2C2024-Q2&dq=OECD%2BG7%2BUSA%2BGBR%2BTUR%2BCHE%2BSWE%2BESP%2BSVN%2BSVK%2BPRT%2BPOL%2BNOR%2BNZL%2BNLD%2BMEX%2BLUX%2BLTU%2BLVA%2BKOR%2BJPN%2BITA%2BISR%2BIRL%2BHUN%2BISL%2BGRC%2BFRA%2BDEU%2BFIN%2BEST%2BDNK%2BCZE%2BCOL%2BCAN%2BBEL%2BAUT%2BAUS.Q.N%2BHICP.CPI.PA.CP01%2B_TXCP01_NRG%2BCP045_0722%2BSERV%2B_T.N.GY&to[TIME_PERIOD]=false&vw=tb&ly[cl]=TIME_PERIOD&ly[rs]=METHODOLOGY%2CEXPENDITURE&ly[rw]=REF_AREA}{Headline CPI} & $\Delta x$ \\
    \href{https://data-explorer.oecd.org/vis?fs[0]=Topic%2C1%7CEconomy%23ECO%23%7CPrices%23ECO_PRI%23&pg=0&fc=Topic&bp=true&snb=16&df[ds]=dsDisseminateFinalDMZ&df[id]=DSD_PRICES%40DF_PRICES_ALL&df[ag]=OECD.SDD.TPS&df[vs]=1.0&pd=1950-Q1%2C2024-Q2&dq=OECD%2BG7%2BUSA%2BGBR%2BTUR%2BCHE%2BSWE%2BESP%2BSVN%2BSVK%2BPRT%2BPOL%2BNOR%2BNZL%2BNLD%2BMEX%2BLUX%2BLTU%2BLVA%2BKOR%2BJPN%2BITA%2BISR%2BIRL%2BHUN%2BISL%2BGRC%2BFRA%2BDEU%2BFIN%2BEST%2BDNK%2BCZE%2BCOL%2BCAN%2BBEL%2BAUT%2BAUS.Q.N%2BHICP.CPI.PA.CP01%2B_TXCP01_NRG%2BCP045_0722%2BSERV%2B_T.N.GY&to[TIME_PERIOD]=false&vw=tb&ly[cl]=TIME_PERIOD&ly[rs]=METHODOLOGY%2CEXPENDITURE&ly[rw]=REF_AREA}{Energy CPI} & $\Delta x$ \\
    \href{https://data-explorer.oecd.org/vis?fs[0]=Topic%2C1%7CEconomy%23ECO%23%7CPrices%23ECO_PRI%23&pg=0&fc=Topic&bp=true&snb=16&df[ds]=dsDisseminateFinalDMZ&df[id]=DSD_PRICES%40DF_PRICES_ALL&df[ag]=OECD.SDD.TPS&df[vs]=1.0&pd=1950-Q1%2C2024-Q2&dq=OECD%2BG7%2BUSA%2BGBR%2BTUR%2BCHE%2BSWE%2BESP%2BSVN%2BSVK%2BPRT%2BPOL%2BNOR%2BNZL%2BNLD%2BMEX%2BLUX%2BLTU%2BLVA%2BKOR%2BJPN%2BITA%2BISR%2BIRL%2BHUN%2BISL%2BGRC%2BFRA%2BDEU%2BFIN%2BEST%2BDNK%2BCZE%2BCOL%2BCAN%2BBEL%2BAUT%2BAUS.Q.N%2BHICP.CPI.PA.CP01%2B_TXCP01_NRG%2BCP045_0722%2BSERV%2B_T.N.GY&to[TIME_PERIOD]=false&vw=tb&ly[cl]=TIME_PERIOD&ly[rs]=METHODOLOGY%2CEXPENDITURE&ly[rw]=REF_AREA}{Food CPI} & $\Delta x$ \\
    \href{https://data-explorer.oecd.org/vis?fs[0]=Topic%2C1%7CEconomy%23ECO%23%7CPrices%23ECO_PRI%23&pg=0&fc=Topic&bp=true&snb=16&df[ds]=dsDisseminateFinalDMZ&df[id]=DSD_PRICES%40DF_PRICES_ALL&df[ag]=OECD.SDD.TPS&df[vs]=1.0&pd=1950-Q1%2C2024-Q2&dq=OECD%2BG7%2BUSA%2BGBR%2BTUR%2BCHE%2BSWE%2BESP%2BSVN%2BSVK%2BPRT%2BPOL%2BNOR%2BNZL%2BNLD%2BMEX%2BLUX%2BLTU%2BLVA%2BKOR%2BJPN%2BITA%2BISR%2BIRL%2BHUN%2BISL%2BGRC%2BFRA%2BDEU%2BFIN%2BEST%2BDNK%2BCZE%2BCOL%2BCAN%2BBEL%2BAUT%2BAUS.Q.N%2BHICP.CPI.PA.CP01%2B_TXCP01_NRG%2BCP045_0722%2BSERV%2B_T.N.GY&to[TIME_PERIOD]=false&vw=tb&ly[cl]=TIME_PERIOD&ly[rs]=METHODOLOGY%2CEXPENDITURE&ly[rw]=REF_AREA}{Core CPI} & $\Delta x$ \\
    \href{https://data-explorer.oecd.org/vis?fs[0]=Topic%2C1%7CEconomy%23ECO%23%7CPrices%23ECO_PRI%23&pg=0&fc=Topic&bp=true&snb=16&df[ds]=dsDisseminateFinalDMZ&df[id]=DSD_PRICES%40DF_PRICES_ALL&df[ag]=OECD.SDD.TPS&df[vs]=1.0&pd=1950-Q1%2C2024-Q2&dq=OECD%2BG7%2BUSA%2BGBR%2BTUR%2BCHE%2BSWE%2BESP%2BSVN%2BSVK%2BPRT%2BPOL%2BNOR%2BNZL%2BNLD%2BMEX%2BLUX%2BLTU%2BLVA%2BKOR%2BJPN%2BITA%2BISR%2BIRL%2BHUN%2BISL%2BGRC%2BFRA%2BDEU%2BFIN%2BEST%2BDNK%2BCZE%2BCOL%2BCAN%2BBEL%2BAUT%2BAUS.Q.N%2BHICP.CPI.PA.CP01%2B_TXCP01_NRG%2BCP045_0722%2BSERV%2B_T.N.GY&to[TIME_PERIOD]=false&vw=tb&ly[cl]=TIME_PERIOD&ly[rs]=METHODOLOGY%2CEXPENDITURE&ly[rw]=REF_AREA}{Imports} & $\Delta\log(x)$ \\
    \href{https://data-explorer.oecd.org/vis?fs[0]=Topic%2C1%7CEconomy%23ECO%23%7CPrices%23ECO_PRI%23&pg=0&fc=Topic&bp=true&snb=16&df[ds]=dsDisseminateFinalDMZ&df[id]=DSD_PRICES%40DF_PRICES_ALL&df[ag]=OECD.SDD.TPS&df[vs]=1.0&pd=1950-Q1%2C2024-Q2&dq=OECD%2BG7%2BUSA%2BGBR%2BTUR%2BCHE%2BSWE%2BESP%2BSVN%2BSVK%2BPRT%2BPOL%2BNOR%2BNZL%2BNLD%2BMEX%2BLUX%2BLTU%2BLVA%2BKOR%2BJPN%2BITA%2BISR%2BIRL%2BHUN%2BISL%2BGRC%2BFRA%2BDEU%2BFIN%2BEST%2BDNK%2BCZE%2BCOL%2BCAN%2BBEL%2BAUT%2BAUS.Q.N%2BHICP.CPI.PA.CP01%2B_TXCP01_NRG%2BCP045_0722%2BSERV%2B_T.N.GY&to[TIME_PERIOD]=false&vw=tb&ly[cl]=TIME_PERIOD&ly[rs]=METHODOLOGY%2CEXPENDITURE&ly[rw]=REF_AREA}{Exports} & $\Delta\log(x)$ \\
    \bottomrule
    \bottomrule
    \end{tabular}%

\end{table}%

\begin{figure}[H]
	\centering
	\includegraphics[width=\linewidth]{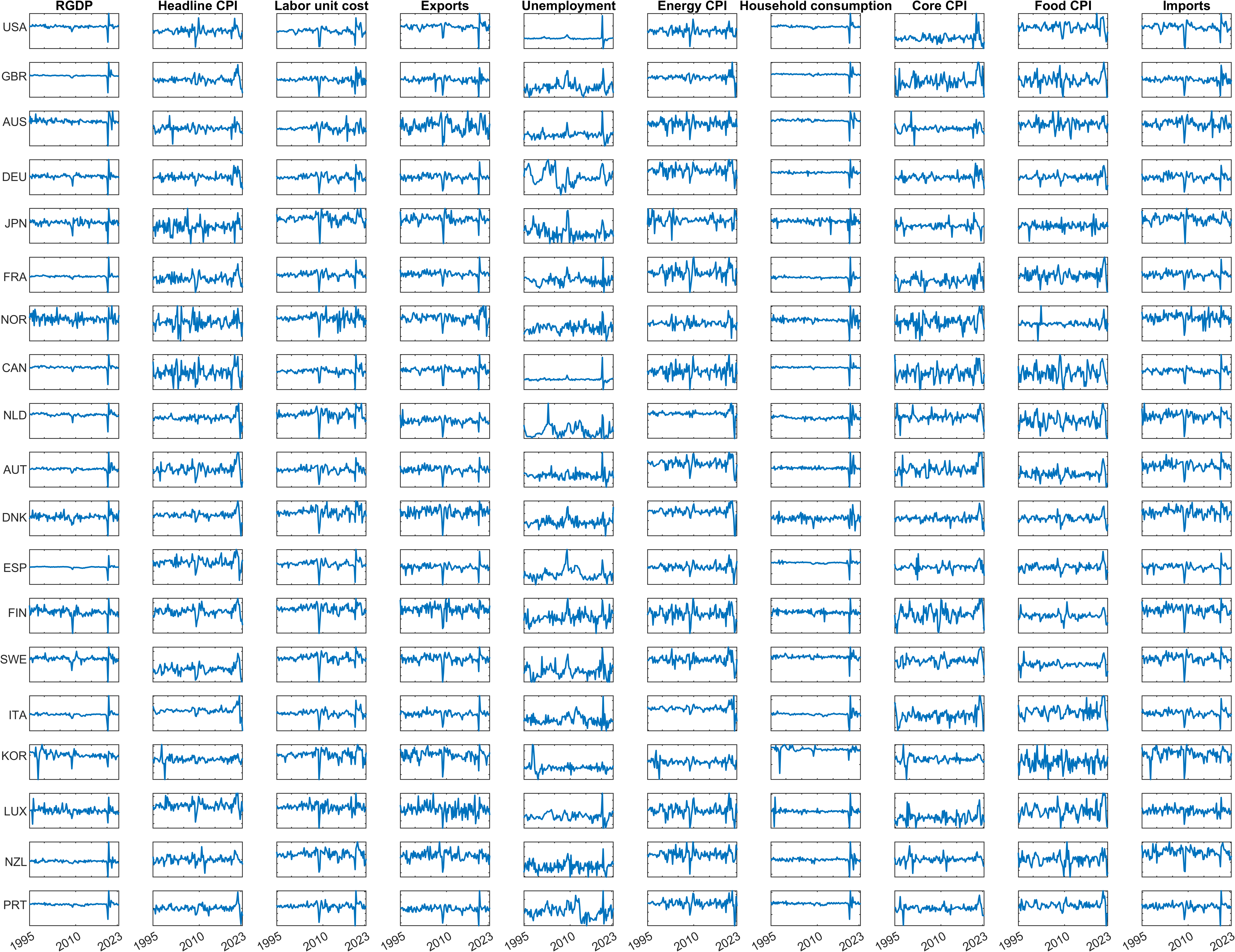}
	
	\caption{The ten macroeconomic indicators (by columns) for 19 countries (by rows). The horizontal axis represents time and the vertical axis represents the standardized growth rates. The ranges of the vertical values are not the same. The order of countries and indicators is the order adopted in the estimation.}
	\label{fig:data}
\end{figure}

\section{The \texorpdfstring{\cite{wang2019factor}}{Wang et al.\ (2019)} Benchmark: Implementation and Comparison}\label{app:wang}

This appendix documents the implementation of the
\citet{wang2019factor} estimator used as the frequentist benchmark
throughout the paper, and reports the auxiliary diagnostics that informed
the comparison reported in Section~\ref{sec:application} of the main text.
The main text emphasizes the out-of-sample forecasting comparison; this
appendix collects the methodological details and the in-sample diagnostics.

\subsection{Estimator and rank-selection rule}

\citet{wang2019factor} estimate the matrix factor model
$\bY_t = \bA \bF_t \bB^\top + \bE_t$ in a frequentist framework. The
loading spaces $\mathrm{span}(\bA)$ and $\mathrm{span}(\bB)$ are recovered
as the leading eigenspaces of two symmetric positive semidefinite
matrices constructed from lagged sample autocovariances:
\begin{equation}
	\widehat{\bM}_1 \;=\; \sum_{h=1}^{h_0} \widehat{\vSigma}_h^{(1)}\,
	\big[\widehat{\vSigma}_h^{(1)}\big]^\top, \qquad
	\widehat{\bM}_2 \;=\; \sum_{h=1}^{h_0} \widehat{\vSigma}_h^{(2)}\,
	\big[\widehat{\vSigma}_h^{(2)}\big]^\top,
	\label{eq:wang_M}
\end{equation}
where $\widehat{\vSigma}_h^{(1)}$ and $\widehat{\vSigma}_h^{(2)}$ are
row- and column-wise sample autocovariance matrices at lag $h$, and
$h_0$ is a user-specified upper lag. The construction in
equation~\eqref{eq:wang_M} exploits the fact that under the matrix factor
model, lagged autocovariances retain only the factor-driven signal: the
common factors are persistent so $\mathrm{Cov}(\bF_t, \bF_{t-h}) \neq 0$
for $h \geq 1$, while the idiosyncratic errors $\bE_t$ are assumed
serially uncorrelated and so contribute zero in expectation.
We follow the paper's default choice of $h_0 = 1$ throughout and report a
sensitivity analysis below.

The factor dimensions $(p_1, p_2)$ are selected by the eigenvalue-ratio
rule: $\widehat{p}_d = \arg\min_{i \leq P_d/2}
\lambda_{d,i+1}/\lambda_{d,i}$ for $d = 1, 2$, where
$\lambda_{d,1} \geq \lambda_{d,2} \geq \cdots$ are the eigenvalues of
$\widehat{\bM}_d$ and $P_d$ is the observed dimension along axis $d$
($P_1 = n = 19$, $P_2 = k = 10$ in our application). Loadings are then
estimated as the orthonormal eigenvectors corresponding to the leading
$\widehat{p}_d$ eigenvalues, and the factor matrix is recovered as
$\widehat{\bF}_t = \widehat{\bA}^\top \bY_t \widehat{\bB}$.

\subsection{Application to the OECD panel}

\begin{figure}[H]
	\centering
	\includegraphics[width=0.95\linewidth]{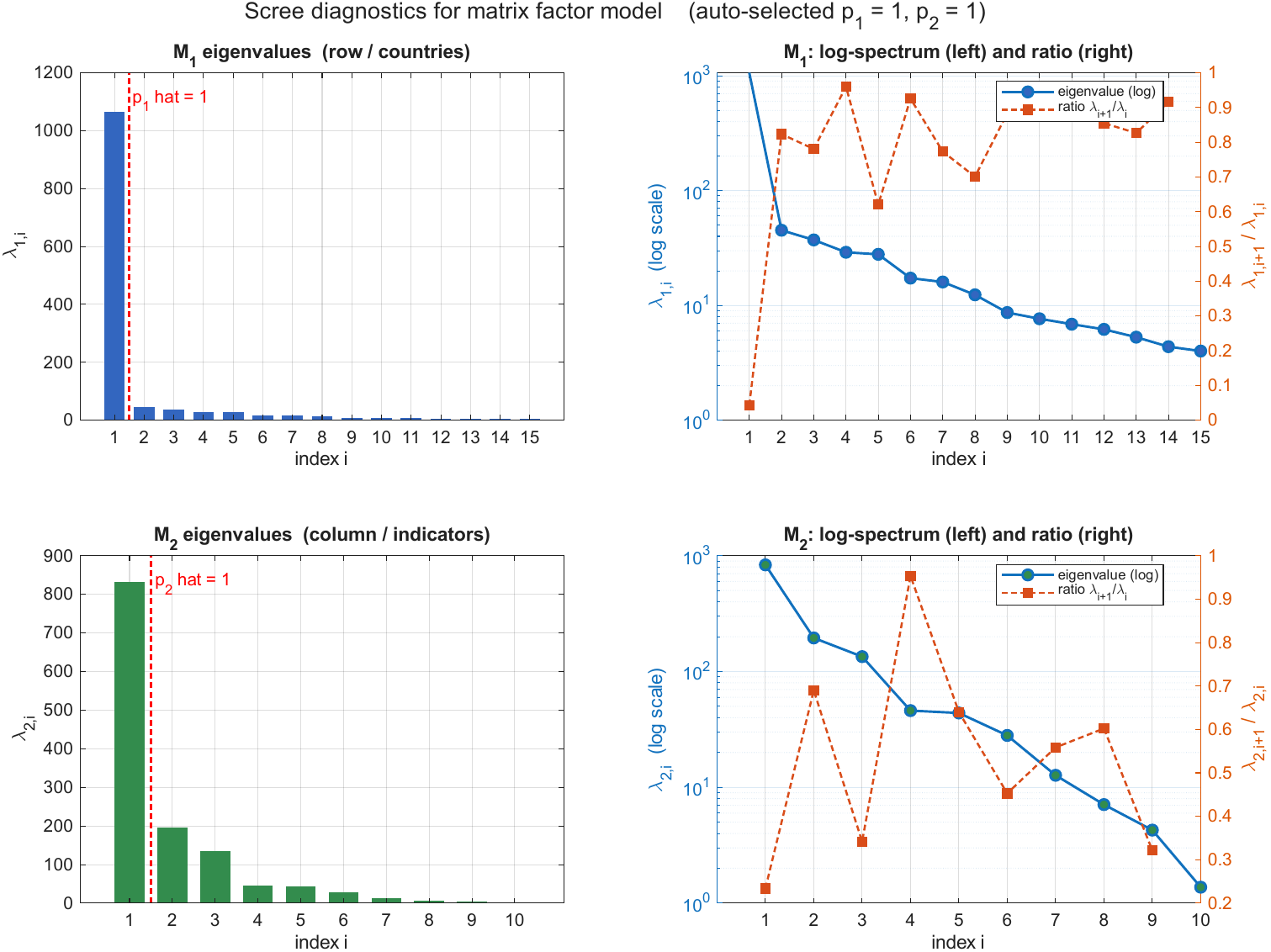}
	\caption{Scree diagnostics for the \citet{wang2019factor} estimator
		on the OECD panel. Top row: $\widehat{\bM}_1$ eigenvalues (row /
		countries) in linear scale (left) and log scale with the
		eigenvalue-ratio diagnostic (right). Bottom row: same diagnostics for
		$\widehat{\bM}_2$ (column / indicators). The vertical dashed lines
		mark the auto-selected $\widehat{p}_d$.}
	\label{fig:scree}
\end{figure}

On our OECD panel ($n = 19$ countries, $k = 10$ indicators, $T = 115$
quarters), the eigenvalue-ratio rule selects
$(\widehat{p}_1, \widehat{p}_2) = (1, 1)$, as shown in Figure \ref{fig:scree}. The selection of $\widehat{p}_1$ is decisive:
$\lambda_{1,2}/\lambda_{1,1} \approx 0.04$, three orders of magnitude
smaller than the next ratio in the sequence. The selection of
$\widehat{p}_2$ is less clear-cut: $\lambda_{2,2}/\lambda_{2,1} \approx
0.23$ versus $\lambda_{2,3}/\lambda_{2,2} \approx 0.69$ and
$\lambda_{2,4}/\lambda_{2,3} \approx 0.34$, with $\lambda_{2,2}$ and
$\lambda_{2,3}$ both clearly visible above the noise floor of the
$\widehat{\bM}_2$ spectrum. \citet{wang2019factor} document in their
Table~5 that the eigenvalue-ratio rule is unstable in moderate-$T$
panels, with $\widehat{p}_2 = 1$
selected with high frequency in simulations even when the true value is
$2$.

\subsection{Sensitivity to the lag parameter \texorpdfstring{$h_0$}{h0}}

Although $h_0 = 1$ is the canonical choice, we verify that the rank
selection and loading estimates are stable under alternative values.
Table~\ref{tab:h0_sensitivity} reports the auto-selected factor
dimensions, in-sample RMSE under both the auto-selected ranks and a
forced $(1, 2)$ specification, and subspace distances of the loading
estimates relative to the $h_0 = 1$ baseline.

\begin{table}[H]
	\centering
	\caption{Sensitivity of the \citet{wang2019factor} estimator to the
		lag parameter $h_0$ in equation~\eqref{eq:wang_M}. Columns report
		the auto-selected dimensions, in-sample RMSE under auto selection,
		in-sample RMSE under forced $(p_1, p_2) = (1, 2)$, and subspace
		distances of the row and column loading estimates relative to
		$h_0 = 1$.}
	\label{tab:h0_sensitivity}
	\begin{tabular}{cccccc}
		\toprule
		\toprule
		$h_0$ & $\widehat{p}_1$ & $\widehat{p}_2$ &
		RMSE auto & RMSE $(1, 2)$ & $D(\widehat{\bQ}_1) / D(\widehat{\bQ}_2)$ \\
		\midrule
		1 & 1 & 1 & 0.8280 & 0.7711 & 0.0000 / 0.0000 \\
		2 & 1 & 1 & 0.8272 & 0.7657 & 0.0297 / 0.0399 \\
		3 & 1 & 1 & 0.8273 & 0.7630 & 0.0378 / 0.0664 \\
		4 & 1 & 1 & 0.8264 & 0.7621 & 0.0452 / 0.0745 \\
		\bottomrule
		\bottomrule
	\end{tabular}
\end{table}

The auto-selected factor dimensions
$(\widehat{p}_1, \widehat{p}_2) = (1, 1)$ are invariant across $h_0$.
The disagreement with our marginal-likelihood selection is therefore not
an artifact of the choice of $h_0$. In addition, in-sample RMSE under auto
selection varies by less than $0.2\%$ across the grid, and the loading
subspace distances $D(\widehat{\bQ}_1)$ and $D(\widehat{\bQ}_2)$ remain
below $0.08$, indicating high stability of the loading estimates with
respect to $h_0$. Third, and most directly relevant, in-sample RMSE
under the forced $(1, 2)$ specification is uniformly lower than under
auto selection, by 6--7 percentage points, across all values of $h_0$.
This is an evidence that the OECD panel may exhibit a second column factor that the eigenvalue-ratio rule misses in a moderate-$T$ scenario.

\subsection{In-sample fit comparison}
\label{app:wang_insample}

\begin{figure}[H]
	\centering
	\includegraphics[width=0.9\linewidth]{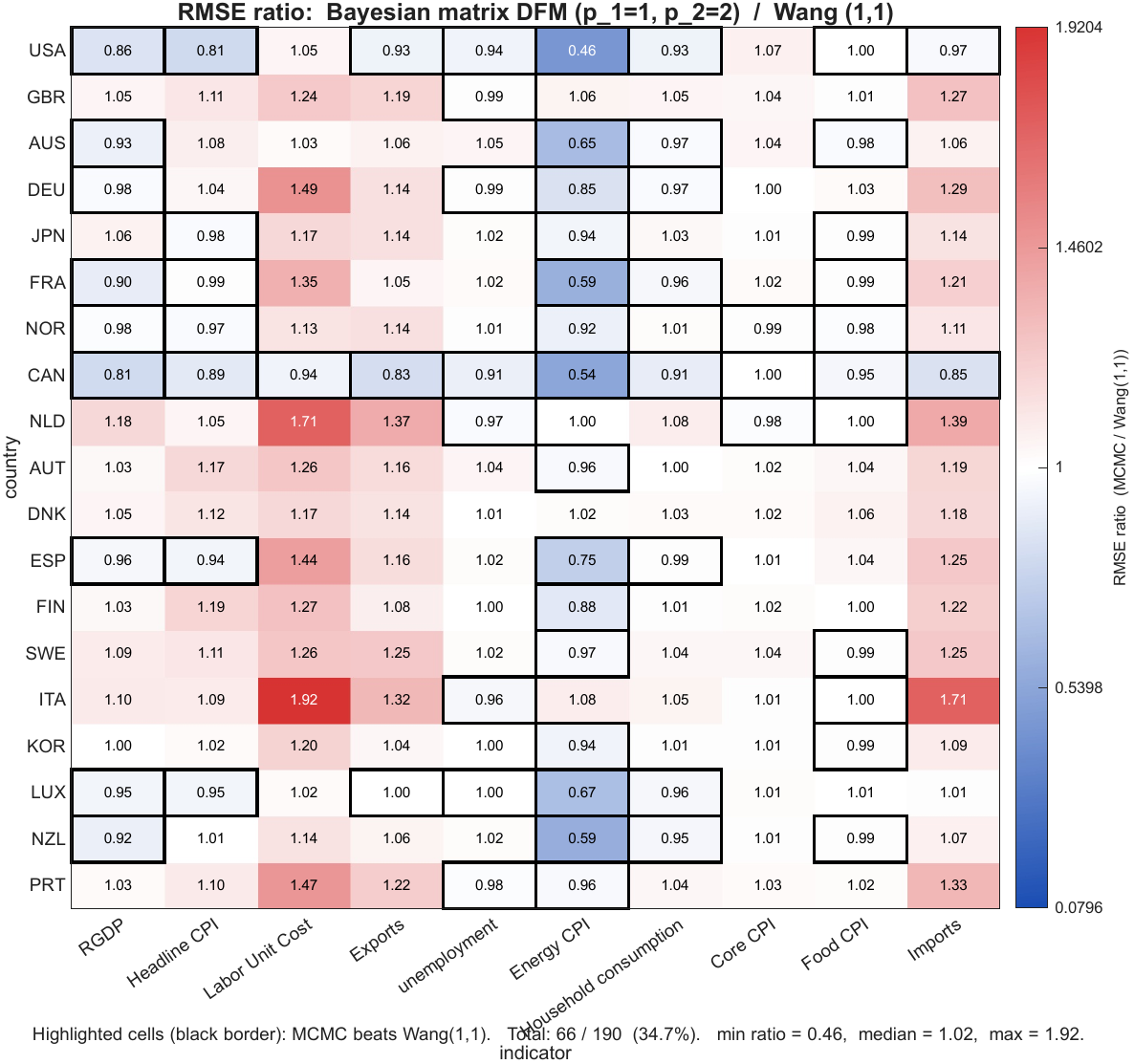}
	\caption{Per-cell ratio of in-sample RMSE, Bayesian MDFM-cross-sv
		with $(p_1, p_2) = (1, 2)$ relative to \citet{wang2019factor} with
		$(\widehat{p}_1, \widehat{p}_2) = (1, 1)$. Black borders mark cells
		where the Bayesian model achieves lower RMSE.}
	\label{fig:rmse_ratio}
\end{figure}

We compare the in-sample fit of the Bayesian MDFM-cross-sv with
$(p_1, p_2) = (1, 2)$ to the \citet{wang2019factor} benchmark with
$(\widehat{p}_1, \widehat{p}_2) = (1, 1)$. Figure~\ref{fig:rmse_ratio}
reports the per-cell ratio of in-sample RMSE between the two methods,
arranged as a $19 \times 10$ heatmap with countries on the rows and
indicators on the columns.

The MDFM achieves lower RMSE on 66 of 190 cells ($34.7\%$) with median ratio
$1.02$ — close to a tie. The in-sample comparison is structurally
unfavorable to the more flexible model: Wang's projection has
substantially fewer effective parameters than MDFM-cross-sv with its
Kronecker idiosyncratic covariance, AR(1) idiosyncratic dynamics, and
common stochastic volatility, and reusing the estimation sample for
evaluation flatters the parsimonious specification.

\subsection{Implementation of the forecasting benchmark}

For the out-of-sample comparison, the \citet{wang2019factor} estimator
is implemented as a direct projection benchmark in the spirit of
\citet{stock2002macroeconomic}. At each forecast origin $t$, we first
obtain the rank-$(1, 1)$ loading estimates $\widehat{\bQ}_1$ and
$\widehat{\bQ}_2$ from the in-sample data $\{\bY_s\}_{s=1}^{t}$, and
construct the scalar factor estimate
$\widehat{f}_s = \widehat{\bQ}_1^\top \bY_s \widehat{\bQ}_2$ for $s
\leq t$. For each forecast horizon $h$ and each (country, indicator)
pair $(i, j)$, we then estimate the horizon-specific direct projection
\begin{equation}
	Y_{i,j,s+h} \;=\; \alpha_{i,j}^{(h)} \;+\; \beta_{i,j}^{(h)}\,
	\widehat{f}_s \;+\; \varepsilon_{i,j,s+h}^{(h)},
	\quad s = 1, \ldots, t - h,
\end{equation}
by ordinary least squares, and produce the forecast
$\widehat{Y}_{i,j,t+h} = \widehat{\alpha}_{i,j}^{(h)} +
\widehat{\beta}_{i,j}^{(h)}\,\widehat{f}_t$.

This is a one-factor analogue of the standard direct projection
forecasting framework in the dynamic factor model literature. Direct
projection is the natural choice for the Wang benchmark because the
estimator does not specify dynamics for the latent factor matrix
$\bF_t$.

\section{Extra Forecasting Results}\label{app:extra}

\begin{figure}[H]
	\centering
	\includegraphics[width=0.8\linewidth]{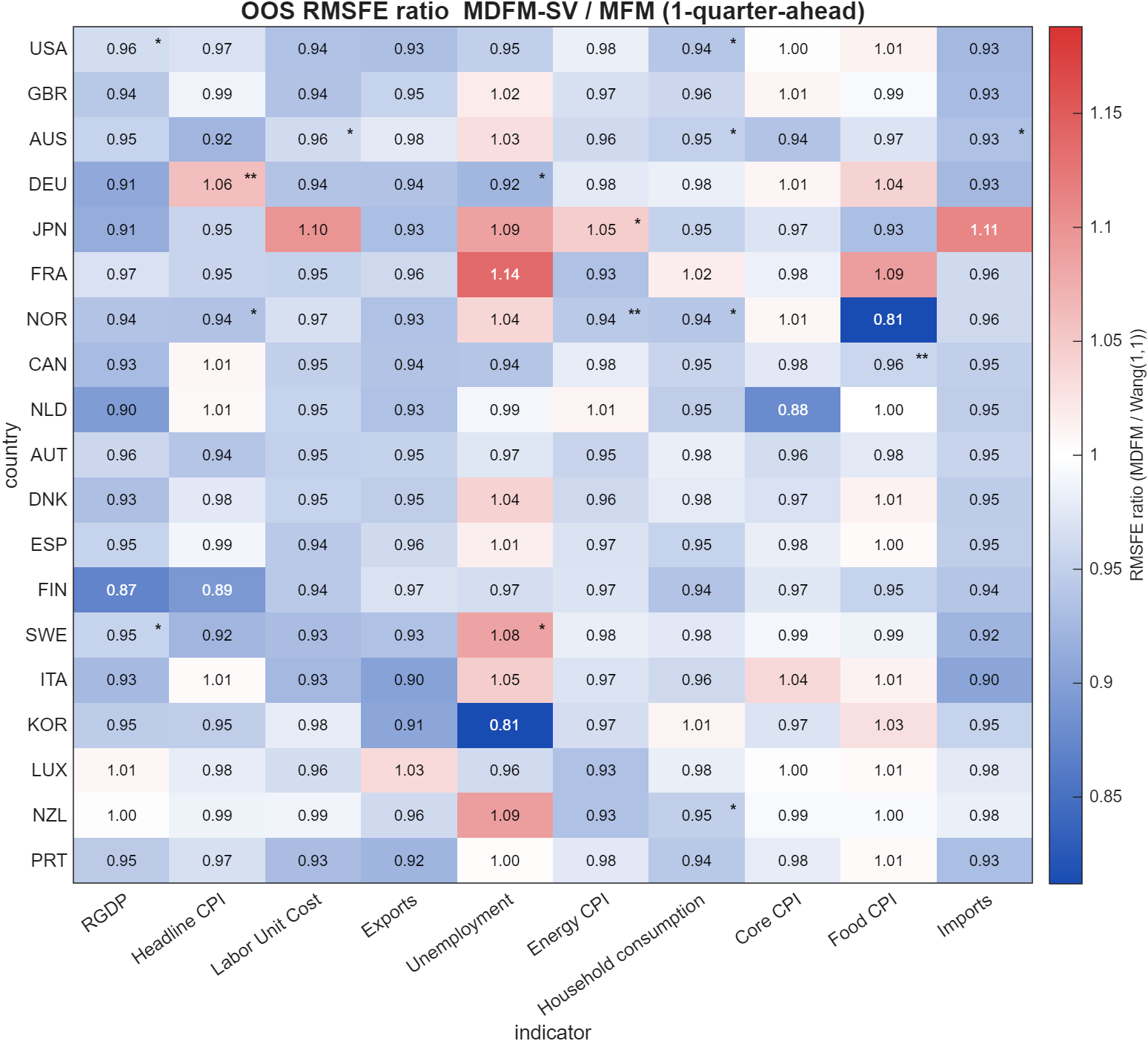}
	\caption{Per-cell RMSFE ratios of the MDFM-SV relative to the static
		matrix factor benchmark, by indicator and country, at the
		1-quarter-ahead horizon ($h=1$). Entries below one indicate that the
		MDFM-SV forecast is more accurate; asterisks denote rejection of the
		Diebold--Mariano test of equal predictive accuracy.}
	\label{fig:oos_rmsfe_appendix}
\end{figure}

\begin{figure}[H]
	\centering
	\includegraphics[width=0.8\linewidth]{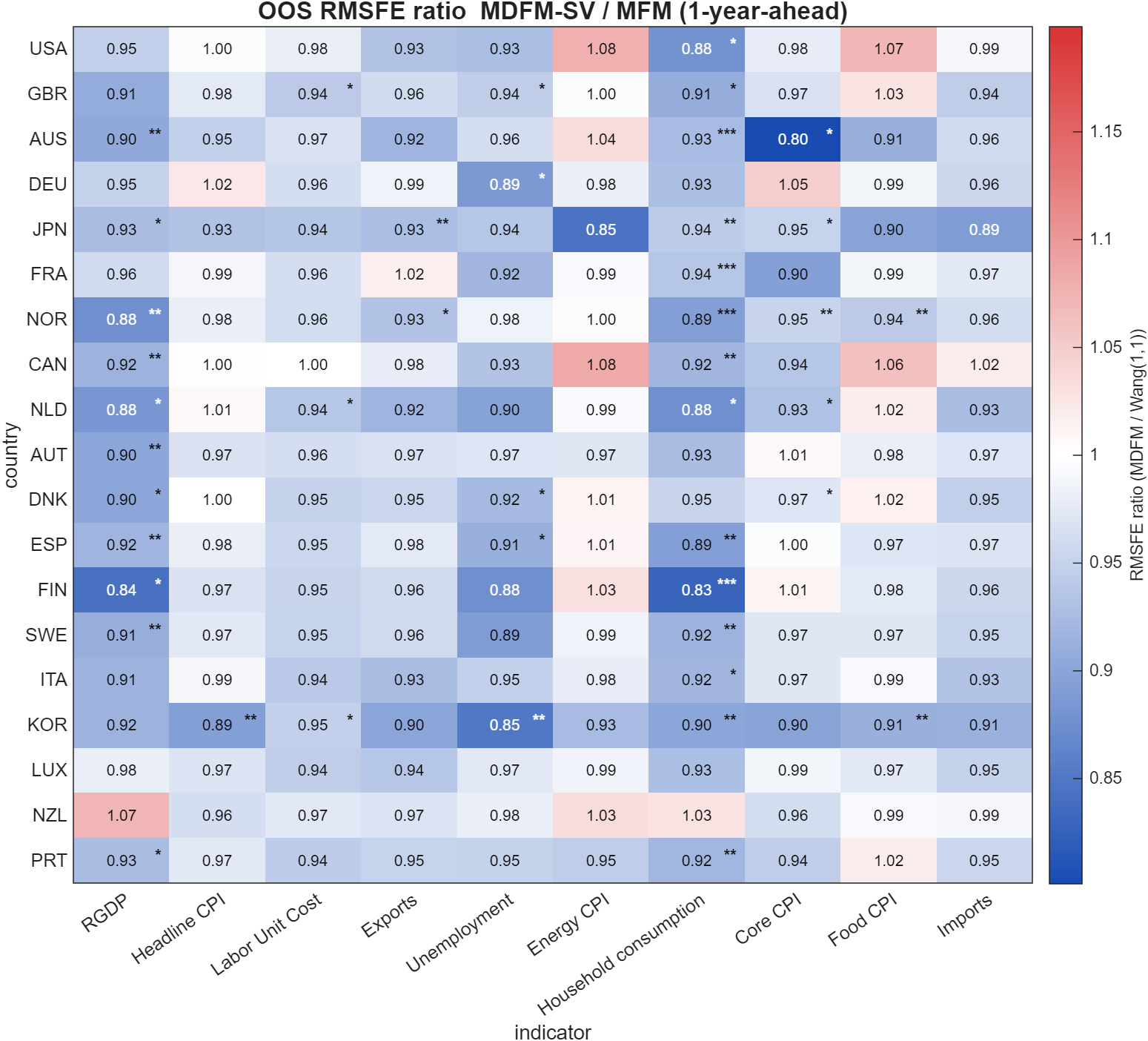}
	\caption{Per-cell RMSFE ratios of the MDFM-SV relative to the static
		matrix factor benchmark, by indicator and country, at the
		1-year-ahead horizon ($h=4$). See the note to
		Figure~\ref{fig:oos_rmsfe_appendix}.}
\end{figure}

\begin{figure}[H]
	\centering
	\includegraphics[width=0.8\linewidth]{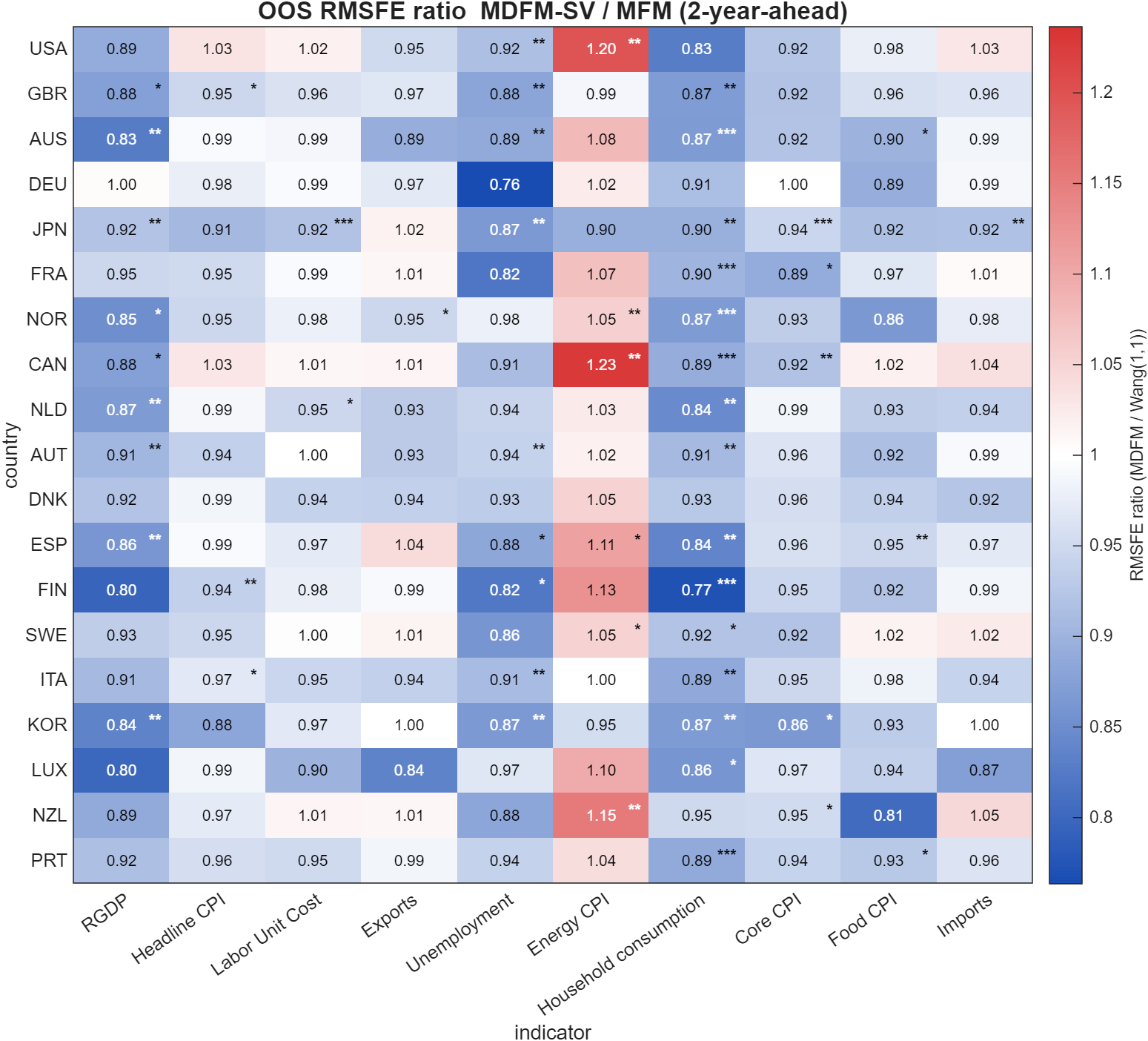}
	\caption{Per-cell RMSFE ratios of the MDFM-SV relative to the static
		matrix factor benchmark, by indicator and country, at the
		2-year-ahead horizon ($h=8$). See the note to
		Figure~\ref{fig:oos_rmsfe_appendix}.}
\end{figure}

\begin{figure}[H]
	\centering
	\includegraphics[width=\linewidth]{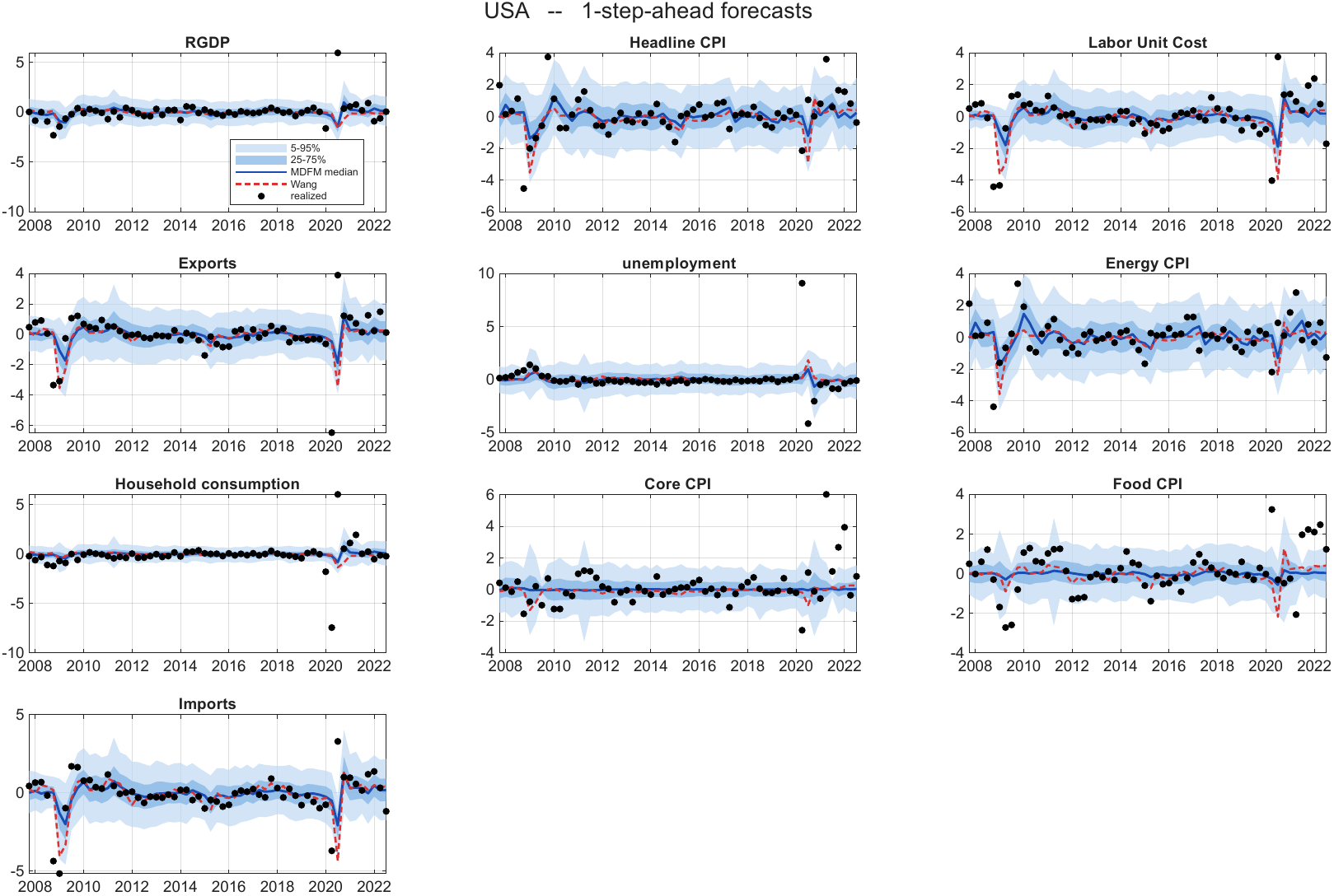}
	\caption{Posterior predictive fan charts for real GDP growth (annualized
		percent change) across the G7 economies. The top block shows
		1-quarter-ahead forecasts; the bottom block shows 4-quarter (1-year)
		average forecasts. Each panel reports the MDFM posterior median and
		5--95\% and 25--75\% predictive bands together with the realized
		series.}
	\label{fig:fancharts_g7}
\end{figure}


\section{Application: Fama-French \texorpdfstring{$10 \times 10$}{10x10} Panel}\label{app:fama}
\begin{figure}[H]
	\centering
	\includegraphics[width=\linewidth]{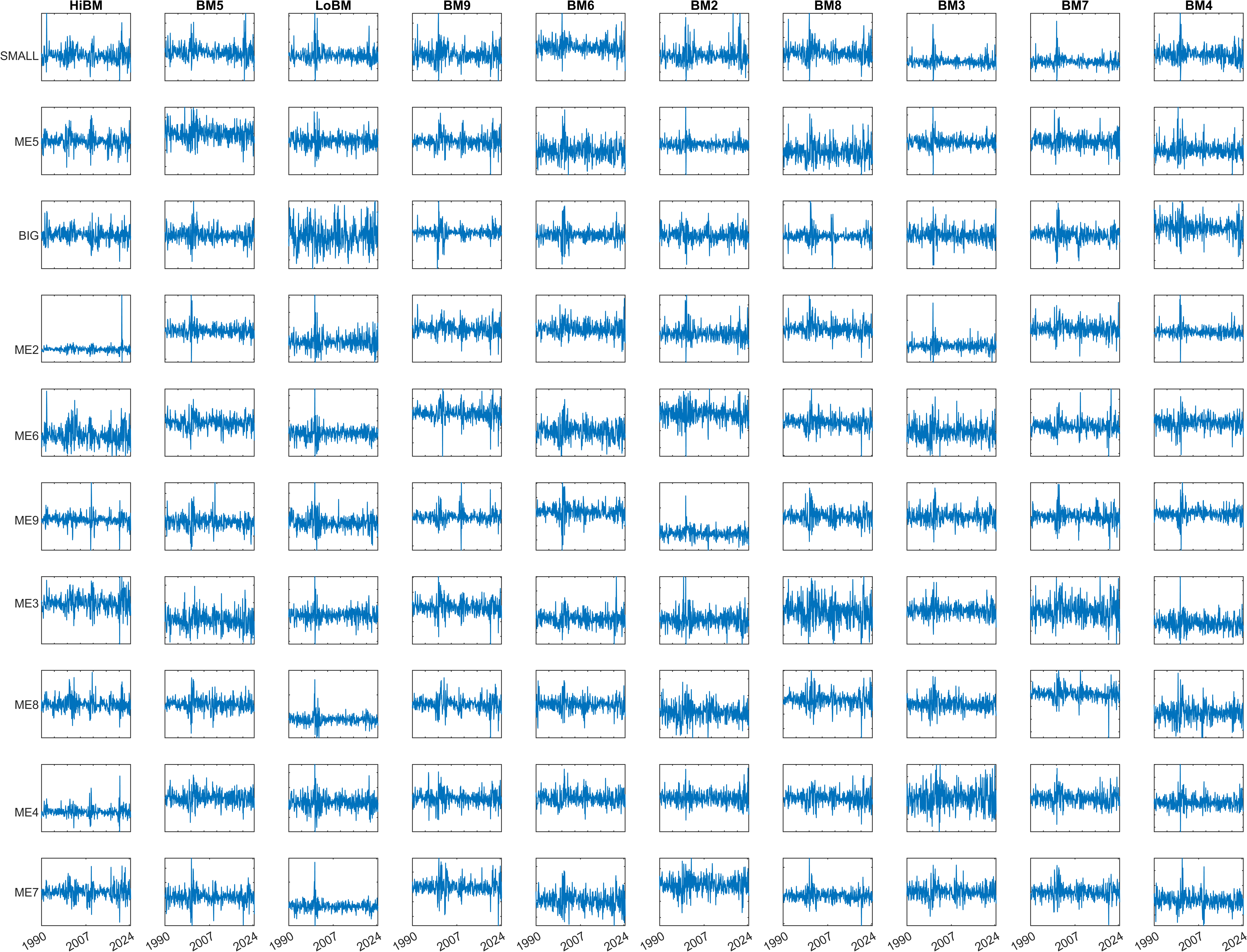}
	\caption{The return series of the portfolios structured by different levels of sizes (rows) and book equity to market equity ratio (columns). Note that we have rearranged the order of rows and columns. The horizontal axis represents time and vertical axis represents the standardized monthly returns. The ranges of the vertical values are not the same.}
	\label{fig:data_famma}
\end{figure}

In this application, we investigate the usefulness of the dynamic matrix factor model on the Fama-French return series, which was studied by \cite{wang2019factor}, \cite{yu2022projected} and \cite{he2024matrix}. The data include monthly returns of 100 portfolios, structured in a 10 by 10 matrix according to ten levels of sizes (market equity) and ten levels of ratio of book equity to market equity (BE/ME).\footnote{The data is available at \url{http://mba.tuck.dartmouth.edu/pages/faculty/ken.french/data_library.html}.} The return series span from January 1990 to June 2024 (414 observations).\footnote{We do not include data earlier than 1990 because there are many missing values in the early years.} Following \cite{chang2023modelling}, we impute the missing values by the weighted averages of the three previous months, i.e., set $y_{i,j,t} = 0.5y_{i,j,t-1} + 0.3y_{i,j,t-2}+0.2y_{i,j,t-3}$ for missing $y_{i,j,t}$. To account for market conditions, we follow \cite{wang2019factor}, \cite{yu2022projected}, and \cite{he2024matrix} and subtract the monthly excess market return from each series. We then standardize the data by subtracting the mean and dividing by the standard deviation. The standardized market-adjusted return series of the portfolios is shown in Figure \ref{fig:data_famma}.


Empirical evidence shows that small-cap stocks tend to earn higher average returns than large-cap stocks, while value stocks (high BE/ME) tend to outperform growth stocks (low BE/ME). To account for these cross-sectional return patterns, \cite{fama1992cross} introduce the Small Minus Big (SMB) and High Minus Low (HML) factors to explain return variations across stocks with varying size and book-to-market ratios. Motivated by this framework, we reorder the size and BE/ME ratios in the data matrix. Specifically, portfolios are ordered by size across rows, with small-cap (SMALL) portfolios listed first, followed by medium-cap (ME5), and then large-cap (BIG) portfolios. Similarly, columns are arranged by book-to-market ratio, with high BE/ME (HiBM) portfolios on the left, followed by medium (BM5), and then low BE/ME (LoBM) portfolios.

Table \ref{tab:ml_ldfm_fama}--\ref{tab:ml_mdfm_fama} report the log marginal likelihood estimates for both VDFMs and MDFMs. The results clearly indicate a strong preference for models with stochastic volatility, regardless of whether the VDFM or MDFM is used. Additionally, a comparison between the MDFM-exact and MDFM-cross reveals that accounting for cross-sectional correlation in idiosyncratic component further improves model performance. Among the MDFM specifications, the $2 \times 2$ MDFM with both cross-sectional correlation and stochastic volatility achieves the highest marginal likelihood (–42,951), although it remains slightly lower than that of the 4-factor VDFM (–42,943). The best overall performance is attained by the 5-factor VDFM with stochastic volatility, which yields a marginal likelihood of –42,877. These results suggest that the data favor the more flexible VDFM specification over the more structured MDFM, despite the marginal likelihood's built-in penalty for model complexity. For illustration, we use the MDFM with a factor matrix of dimensions $(p_1, p_2) = (2, 2)$ and stochastic volatility in the subsequent analysis.


\begin{table}[H]
	\centering
	\caption{Log marginal likelihood estimates using VDFM}\label{tab:ml_ldfm_fama}%
	\resizebox{0.78\textwidth}{!}{
	\begin{tabular}{ccccccccc}
		\toprule
		\toprule
		\multicolumn{4}{c}{VDFM-exact} &       & \multicolumn{4}{c}{VDFM-sv} \\
		\cmidrule{1-4}\cmidrule{6-9}    $k=1$   & $k=2$   & $k=3$   & $k=4$   &       & $k=1$   & $k=2$   & $k=3$   & $k=4$ \\
		-51647 & -47191 & -46217 & -45940 &       & -46096 & -43942 & -43069 & -42943 \\
		(0.1) & (0.2) & (0.4) & (0.5) &       & (0.4) & (0.4) & (0.5) & (0.8) \\
		\cmidrule{1-4}\cmidrule{6-9}    $k=5$   & $k=6$   & $k=7$   & $k = 8$   &       & $k=5$   & $k=6$   & $k=7$   & $k = 8$ \\
		-45787 & \textbf{-45729} & -45792 & -45886 &       & \textcolor[rgb]{ 1,  0,  0}{\textbf{-42877}} & -43007 & -43155 & -43328 \\
		(0.6) & (0.9) & (1.0) & (1.0) &       & (0.7) & (0.8) & (1.6) & (0.8) \\
		\bottomrule
		\bottomrule
	\end{tabular}}
	
\end{table}%

\begin{table}[H]
	\centering
	\caption{Log marginal likelihood estimates using MDFM}
	\label{tab:ml_mdfm_fama}
	\resizebox{\textwidth}{!}{
	\begin{tabular}{lccccccccccc}
		\toprule
		\toprule
		& \multicolumn{3}{c}{MDFM-exact} &       & \multicolumn{3}{c}{MDFM-cross} &       & \multicolumn{3}{c}{MDFM-cross-sv} \\
		\cmidrule{1-4}\cmidrule{6-8}\cmidrule{10-12}          & $p_2 =1$ & $p_2 =2$ & $p_2 =3$ &       & $p_2 =1$ & $p_2 =2$ & $p_2 =3$ &       & $p_2 =1$ & $p_2 =2$ & $p_2 = 3$ \\
		$p_1 = 1$ & -51867 & -49622 & -49382 &       & -47210 & -46809 & -46739 &       & -43527 & -43163 & -43161 \\
		& (0.2) & (0.3) & (0.4) &       & (0.2) & (0.4) & (0.4) &       & (1.3) & (1.4) & (1.4) \\
		$p_1 = 2$ & -49603 & -47130 & -46572 &       & -46897 & -46557 & -46164 &       & -43304 & \textbf{-42951} & -43024 \\
		& (0.3) & (0.4) & (0.4) &       & (0.4) & (0.4) & (0.7) &       & (2.0) & (0.9) & (2.9) \\
		$p_1 =3$ & -49285 & -46847 & \textbf{-46177} &       & -46928 & -46553 & \textbf{-46087} &       & -43406 & -43075 & -43027 \\
		& (0.4) & (0.5) & (0.9) &       & (0.6) & (0.9) & (0.8) &       & (1.3) & (1.2) & (1.4) \\
		\bottomrule
		\bottomrule
	\end{tabular}}
\end{table}%

\begin{figure}[H]
	\centering
	\includegraphics[width=\linewidth]{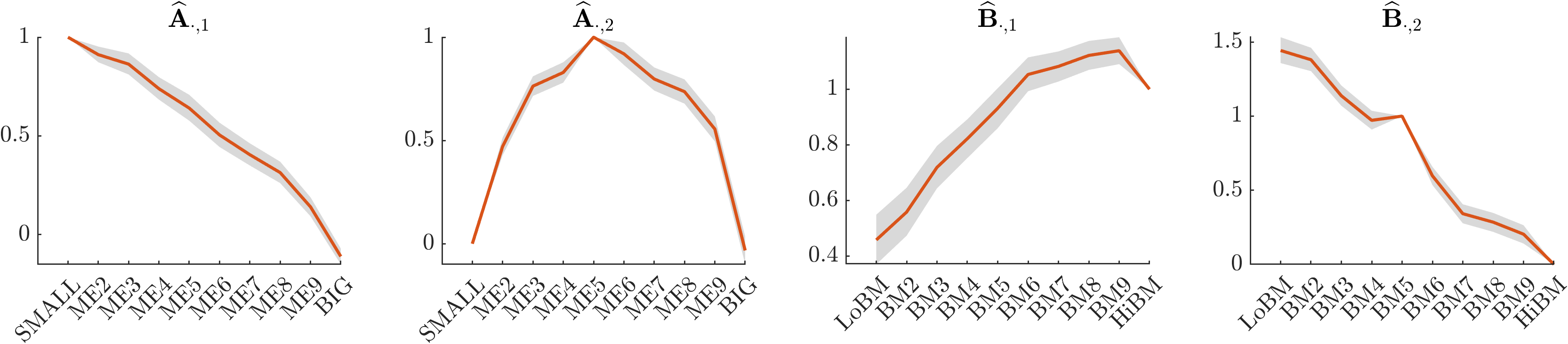}
	\caption{Estimates for row loading matrices (the first and second panel from left) and column loading matrices (the third and fourth panel). The gray area represents the $90\%$ credible interval.}
	\label{fig:fama_loading}
\end{figure}

Figure \ref{fig:fama_loading} shows the estimates for row loading matrices (the first and second panel from left) and column loading matrices (the third and fourth panel). In specific, the two subplots correspond to $\widehat{\bA}_{.,1}$ (first), $\widehat{\bA}_{.,2}$ (second), $\widehat{\bB}_{.,1}$ (third) and $\widehat{\bB}_{.,2}$ (fourth). Both the size (row) loadings and the BE/ME (column) loadings have shown cross-sectional patterns. Particularly, the small-cap factor exerts a strong influence on smaller portfolios, with its impact gradually decreasing as portfolio size increases, eventually turning negative for large-size portfolios. The medium-cap factor similarly influence portfolios with similar sizes, with its influence tapering off as the portfolio size shifts either smaller or larger. Similar patterns go for BE/ME factors.


\begin{figure}[H]
    \centering
    \includegraphics[width=\linewidth]{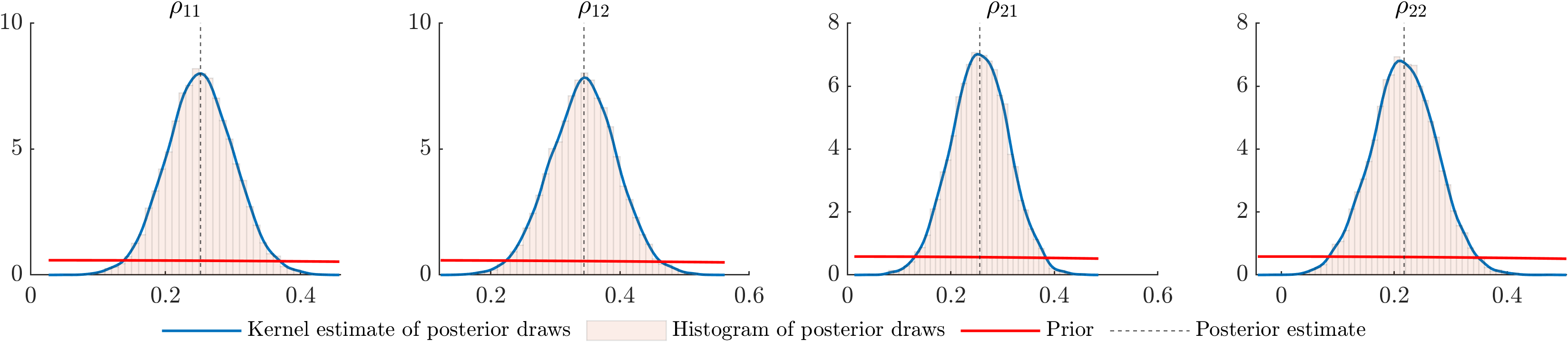}
    \caption{Posterior densities, histograms of posterior draws, priors and posterior estimates for autogressive coefficients $\vrho$}
    \label{fig:rho_famma}
\end{figure}

Figure \ref{fig:rho_famma} shows estimated posterior densities, histograms of posterior draws, priors, as well as the posterior estimates of autoregressive coefficients ($\vrho$) for the factor evolution process. All the six posterior densities have little mass on value 0, and the posterior estimates are around 0.2 or 0.3. This suggests that an AR process for factor evolution is supported by the data.

\begin{figure}[H]
    \centering
    \includegraphics[width=0.78\linewidth]{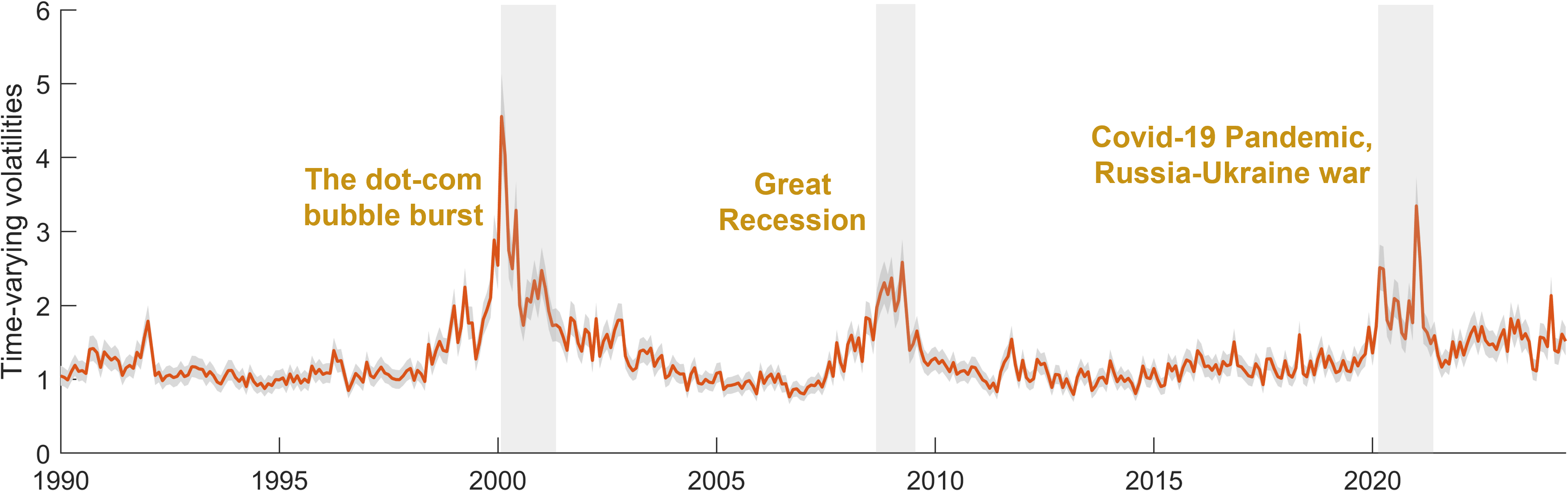}
    \caption{Estimates and standard errors of stochastic volatility: $\exp(\bh/2)$}
    \label{fig:volatilities_famma}
\end{figure}

Figure \ref{fig:volatilities_famma} shows the estimates and standard deviations of stochastic volatility for stock returns over time. Clearly, the volatility of stock returns exhibits considerable variation throughout the observed period. Notably, the volatility peaks around February 2000, about one month before the onset of the dot-com bubble burst. Additionally, significant spikes in volatility are observed during the 2008 financial crisis and the COVID-19 pandemic. 


\end{document}